\documentclass[rmp,aps,nofootinbib,twocolumn]{revtex4}

\usepackage{amsmath,graphicx,natbib}
\usepackage{hyperref}
\usepackage[varg]{txfonts}

\newcommand{\cvector}[1]{\left(\begin{array}{c}#1\end{array}\right)}
\renewcommand{\d}{\mathrm{d}}
\newcommand{\e}{\mathrm{e}}
\newcommand{\ii}{\mathrm{i}}
\newcommand{\nuc}[2]{\phantom{}^{#1}\mathrm{#2}}

\begin{document}

\title{The Dark Universe}

\author{Matthias Bartelmann}
\email{mbartelmann@ita.uni-heidelberg.de}
\affiliation{Zentrum f\"ur Astronomie der Universit\"at Heidelberg, Institut f\"ur Theoretische Astrophysik, Albert-\"Uberle-Str. 2, 69120 Heidelberg, Germany}
\date{draft version of \today}

\begin{abstract}

For a few years now, cosmology has a standard model. By this term, we mean a consistent theoretical background which is at the same time simple and broad enough to offer coherent explanations for the vast majority of cosmological phenomena. This review will briefly summarise the cosmological model, then proceed to discuss what we know from observations about the evolution of the Universe and its contents, and what we conclude about the origin and the future of the Universe and the structures it contains.

\end{abstract}

\maketitle

\section{The cosmological standard model}\label{sec:I}

\subsection{Historical outline}\label{sec:I-H}

This review will not follow the historical sequence of discoveries but rather present the arguments in a perhaps more logical order than history has allowed. Yet, let us begin with a brief and necessarily fragmental historical outline.

Modern cosmology began in 1915 with Einstein's theory of General Relativity, from which Friedmann in 1922 derived the homogeneous and isotropic class of world models which to this date form the foundation of the cosmological standard model. His discovery that static world models were impossible confirmed Einstein in his introduction of the cosmological constant, which he abandoned in the 1930s after Slipher, Hubble and Humason had discovered the cosmic expansion. Earlier, Lema{\^\i}tre had speculated about a primordial fireball which seemed the natural starting point of an expanding universe.

First evidence for dark matter was found by Zwicky in galaxy clusters, and by Babcock and Oort in galaxies, between 1933 and 1940. In 1946, Lifshitz worked out relativistic perturbation theory and started applying it to the linear growth of cosmic structures. In the late 1940s, Gamow, Alpher and Herman worked out how nuclear fusion may have proceeded in the early universe and predicted a cosmic radiation background with a temperature of a few Kelvin.

The cosmic microwave background (CMB) was discovered by Penzias and Wilson and explained by Dicke and collaborators in 1965. In 1970, Peebles and Yu as well as Sunyaev and Zel'dovich independently predicted structures in the CMB and found that they should have relative amplitudes near $10^{-4}$, where subsequent searches did not find them. Peebles suggested in 1982 that a lower CMB fluctuation level could be explained if the dark matter could not interact with light. It was quickly realised by means of the numerical simulations by Davis, Efstathiou, Frenk and White that cosmological structure formation could be explained within the emerging paradigm of cold dark matter. This paradigm experienced further strong support when temperature fluctuations at the revised level were finally found in the CMB with the COBE satellite in 1992.

It was suggested by Guth in 1981 that a phase of inflationary expansion could solve the horizon and flatness problems of Friedmann cosmology. Mukhanov and Chibisov immediately saw that inflated quantum fluctuations could be the origin of cosmic structures. At least in simple scenarios, inflation requires a spatially flat universe and thus a total energy density equal to its critical value, but it became obvious that matter alone cannot be as dense. The difference was tentatively attributed to the cosmological constant, which was confirmed by several groups in the late 1990s through the accelerated cosmic expansion inferred from type-Ia supernovae. The cosmological standard model could be considered complete when the recent measurements of the CMB temperature fluctuations confirmed that the universe is spatially flat.

Modern cosmology is thoroughly described in many textbooks. To mention only a few, see \cite{PA93.1,PE93.1,PE99.1,CO02.1,DO03.1,WE08.1} and the textbook on the CMB \cite{DU08.2}.

This review is structured as follows. The introduction proceeds by summarising the framework of the Friedmann models in \ref{sec:I-A} and the growth of cosmic structures in \ref{sec:I-B}. Section~\ref{sec:II} describes what we know about the age of the Universe from its ingredients through nuclear cosmo-chronology (\ref{sec:II-A}), stars (\ref{sec:II-B}) and the cooling of white dwarfs (\ref{sec:II-C}). In Sect.~\ref{sec:III}, measurements of the Hubble constant from Hubble's law (\ref{sec:III-A}) and gravitational lensing (\ref{sec:III-B}) are reviewed. Big-bang nucleosynthesis is the subject of Sect.~\ref{sec:IV}, with subsections on the production of light elements (\ref{sec:IV-A}) and the observed element abundances (\ref{sec:IV-B}). Section~\ref{sec:V} gives an overview of the matter density in the Universe, as constrained by galaxies (\ref{sec:V-A}), galaxy clusters as individual objects (\ref{sec:V-B}) and as an evolving population (\ref{sec:V-C}). Section~\ref{sec:VI} on the cosmic microwave background begins with the isotropic CMB (\ref{sec:VI-A}) and continues with its fluctuations (\ref{sec:VI-B}). Cosmic structures are reviewed in Sect.~\ref{sec:VII}, which starts with the quantification of structures (\ref{sec:VII-A}) and continues with their measurement and results obtained (\ref{sec:VII-B}). Section~\ref{sec:VIII} on cosmological weak lensing follows, in which gravitational light deflection is first summarised (\ref{sec:VIII-A}) before measurements are described (\ref{sec:VIII-B}). Supernovae of type~Ia are the subject of Sect.\ref{sec:IX}, where the principles of their cosmological application (\ref{sec:IX-A}), the reasons thereof and the observational results are described (\ref{sec:IX-B}. Section~\ref{sec:X} on the normalisation of the matter-fluctuation power spectrum begins with an introduction (\ref{sec:X-A}) and continues describing constraints from CMB fluctuations (\ref{sec:X-B}), cosmological weak lensing (\ref{sec:X-C}), galaxy clusters (\ref{sec:X-D}) and the Lyman-$\alpha$ forest (\ref{sec:X-E}). The final Sect.~\ref{sec:XI} discusses the motivation and the evidence for cosmological inflation (\ref{sec:XI-A}) and ends with remarks on dark energy (\ref{sec:XI-B}).

\subsection{Friedmann models}\label{sec:I-A}

\subsubsection{The metric}\label{sec:I-A-1}

Cosmology studies the physical properties of the Universe as a whole. The only of the four known interactions which can play a role on cosmic length scales is gravity. Electromagnetism, the only other interaction with infinite range, has sources of opposite charge which tend to shield each other on comparatively very small scales. Cosmic magnetic fields can perhaps reach coherence lengths on the order of $\gtrsim10\,\mathrm{Mpc}$, but they are far too weak for them to be important for the cosmic evolution. The weak and strong interactions, of course, have microscopic range and must thus be unimportant for cosmology as a whole.

The best current theory of gravity is Einstein's theory of general relativity, which relates the geometry of a four-dimensional space-time manifold to its material and energy content. Cosmological models must thus be constructed as solutions of Einstein's field equations. Symmetry assumptions greatly simplify this process. Guided by observations to be specified later, we assume that the Universe appears approximately identical in all directions of observation; in other words, it is assumed to be isotropic on average. While this assumption is obviously incorrect in our cosmological neighbourhood, it holds with increasing precision if observations are averaged on increasingly large scales. The assumption of isotropy can only be valid in a preferred reference frame which is at rest with respect to the mean cosmic motion. The motion of the Earth within this rest frame must be subtracted before any observation can be expected to appear isotropic.

The second assumption asserts that the Universe should appear equally isotropic about \emph{any} of its points. Then, it is homogeneous. Isotropic and homogeneous solutions for Einstein's field equations admit the Robertson-Walker metric whose line element is
\begin{equation}
  \d s^2=-c^2\d t^2+a^2(t)\left[
    \d w^2+f_K^2(w)\left(\d\theta^2+\sin^2\theta\,\d\phi^2\right)
  \right]\;,
\label{eq:01-1}
\end{equation}
with the radial coordinate $w$ and the radial function
\begin{equation}
  f_K(w)=\begin{cases}
    K^{-1/2}\sin(K^{1/2}w) & (K>0) \\
    w & (K=0) \\
    |K|^{-1/2}\sinh(|K|^{1/2}w) & (K<0) \\
         \end{cases}\;.
\label{eq:01-2}
\end{equation}
The curvature parameter $K$, which can be positive, negative or zero, has the dimension of an inverse squared length. The \emph{scale factor} $a(t)$ isotropically stretches or shrinks the three-dimensional spatial sections of the four-dimensional space-time. The scale factor is commonly normalised to $a_0=1$ at the present time. As usual, the line element $\d s$ gives the proper time measured by an observer moving by $(\d w, f_K(w)\d\theta, f_K(w)\sin\theta\d\phi)$ within the \emph{coordinate time} interval $\d t$. For light, $\d s=0$.

Observers attached to the coordinate grid $(w, \theta, \phi)$ on the spatial hypersurfaces for constant cosmological time $t$ are called comoving. The cosmic expansion does not change the comoving coordinates $\vec x$, but the physical coordinates $\vec r=a\vec x$.

\subsubsection{Redshift and expansion}\label{sec:I-A-2}

The changing scale of the Universe causes the cosmological redshift $z$. The wavelength of light from a distant comoving source seen by a comoving observer changes by the same amount as the Universe changes its scale while the light is travelling. Thus, the emitted and observed wavelengths $\lambda$ and $\lambda_0$, respectively, are related by
\begin{equation}
  \frac{\lambda_0}{\lambda}=\frac{a_0}{a}=\frac{1}{a}\;,
\label{eq:01-3}
\end{equation}
where $a$ is the scale factor at the time of emission. The \emph{relative} wavelength change is the redshift,
\begin{equation}
  z\equiv\frac{\lambda_0-\lambda}{\lambda}=\frac{1}{a}-1\;,
\label{eq:01-4}
\end{equation}
thus
\begin{equation}
  1+z=a^{-1}\;,\quad a=(1+z)^{-1}\;.
\label{eq:01-5}
\end{equation}

When the metric is inserted into Einstein's field equations, two ordinary differential equations result,
\begin{eqnarray}
  \left(\frac{\dot a}{a}\right)^2&=&\frac{8\pi G}{3}\rho-\frac{Kc^2}{a^2}+\frac{\Lambda}{3}
  \;,\label{eq:01-6}\\
  \frac{\ddot a}{a}&=&-\frac{4\pi G}{3}\left(\rho+\frac{3p}{c^2}\right)+\frac{\Lambda}{3}\;,
\label{eq:01-7}
\end{eqnarray}
relating $\dot a$ and $\ddot a$ to the density $\rho$, the pressure $p$, the cosmological constant $\Lambda$ and the curvature $K$. Equation~(\ref{eq:01-7}) can be eliminated and replaced by
\begin{equation}
  \frac{\d}{\d t}(\rho c^2a^3)+p\frac{\d(a^3)}{\d t}=0\;,
\label{eq:01-8}
\end{equation}
from which the change of $\rho$ with $a$ can be inferred once $p$ is given. Relativistic matter (``r'', ``radiation'') has $p=\rho c^2/3$ while nonrelativistic matter (``m'', ``dust'') is characterised by $\rho c^2\gg p\approx0$. Thus
\begin{equation}
  \rho_\mathrm{r}\propto a^{-4}\;,\quad\rho_\mathrm{m}\propto a^{-3}\;.
\label{eq:01-9}
\end{equation}

For a model universe containing only matter, radiation and the cosmological constant, Eq.~(\ref{eq:01-6}) can be brought into the form
\begin{eqnarray}
  H^2&=&H_0^2\left[
    \frac{\Omega_\mathrm{r0}}{a^4}+\frac{\Omega_\mathrm{m0}}{a^3}+
    \Omega_{\Lambda0}+
    \frac{1-\Omega_\mathrm{m0}-\Omega_\mathrm{r0}-\Omega_{\Lambda0}}
{a^2}\right]\nonumber\\&\equiv&
  H_0^2E^2(a)\;.
\label{eq:01-10}
\end{eqnarray}
This is commonly called Friedmann's equation, in which the relative expansion rate $\dot a/a$ is replaced by the \emph{Hubble function} $H(a)$ whose present value is the \emph{Hubble constant} $H_0\equiv H(a_0)=H(1)$, and the matter-energy content is described by the three density parameters $\Omega_\mathrm{r0}$, $\Omega_\mathrm{m0}$ and $\Omega_{\Lambda0}$.

$H_0$ is often expressed in dimension-less form by
\begin{equation}
  h\equiv\frac{H_0}{100\,\mathrm{km\,s^{-1}\,Mpc^{-1}}}
\label{eq:01-11}
\end{equation}
Since lengths in the Universe are typically measured with respect to the Hubble length, they are often given in units of $h^{-1}\mathrm{Mpc}$. Similarly, luminosities are typically obtained by multiplying fluxes with squared luminosity distances and are thus often given in units of $h^{-2}L_\odot$. We avoid this notation where possible in the following and insert $h=0.7$ where appropriate.

The dimension-less parameters $\Omega_\mathrm{m0}$ and $\Omega_\mathrm{r0}$ describe the densities of matter and radiation today in units of the critical density
\begin{equation}
  \rho_\mathrm{cr}\equiv\frac{3H^2}{8\pi G}\;,\quad
  \rho_\mathrm{cr0}\equiv\frac{3H_0^2}{8\pi G}\;.
\label{eq:01-12}
\end{equation}

A Robertson-Walker metric whose scale factor satisfies Friedmann's equation is called a Friedmann-Lema{\^\i}tre-Robertson-Walker metric. The cosmological standard model asserts that the Universe at large is described by such a metric, and is thus characterised by the four parameters $\Omega_\mathrm{m0}$, $\Omega_\mathrm{r0}$, $\Omega_{\Lambda0}$ and $H_0$. Since the critical density and the densities themselves evolve in time, so do the density parameters.

The remainder of this review article is devoted to answering two essential questions: (1) What are the values of the parameters characterising Friedmann's equation? (2) How can we understand the \emph{deviations} of the real universe from a purely homogeneous and isotropic space-time?

Table~\ref{tab:1}, adapted from \citep{KO08.1}, summarises the most important cosmological parameters and the values adopted below, unless stated otherwise.

\begin{table*}[ht]
\begin{center}
\begin{tabular}{|l|l|rcl|rcl|l|}
\hline
parameter & symbol &
  \multicolumn{6}{c|}{WMAP-5} & comment \\
& & \multicolumn{3}{c|}{alone} & \multicolumn{3}{c|}{$+$ BAO $+$ SNe} & \\
& & & & & \multicolumn{3}{c|}{(reference)} & \\
\hline
CMB temperature & $T_\mathrm{CMB}$ &
  $2.728$ & $\pm$ & $0.004\,\mathrm{K}$ & &--& & from \citep{FI96.1} \\
total energy density & $\Omega_\mathrm{tot}$ & 
  $1.099$ & $^+_-$ & $^{0.100}_{0.085}$ &
  $1.0052$ & $\pm$ & $0.0064$ & \\
matter density & $\Omega_\mathrm{m0}$ &
  $0.258$ & $\pm$& $0.03$ &
  $0.279$ & $\pm$ & $0.015$ & assuming spatial flatness \\
baryon density & $\Omega_\mathrm{b0}$ &
  $0.0441$ & $\pm$ & $0.0030$ &
  $0.0462$ & $\pm$ & $0.0015$ & here and below \\
cosmological constant & $\Omega_{\Lambda0}$ &
  $0.742$ & $\pm$ & $0.03$ &
  $0.721$ & $\pm$ & $0.015$ & \\
Hubble constant & $h$ &
  $0.719$ & $^+_-$ & $^{0.026}_{0.027}$ &
  $0.701$ & $\pm$ & $0.013$ & \\
power-spectrum normalisation & $\sigma_8$ &
  $0.796$ & $\pm$ & $0.036$ &
  $0.817$ & $\pm$ & $0.026$ & \\
age of the Universe in Gyr & $t_0$ &
  $13.69$ & $\pm$ & $0.13$ &
  $13.73$ & $\pm$ & $0.12$ & \\
decoupling redshift & $z_\mathrm{dec}$ &
  $1087.9$ & $\pm$ & $1.2$ &
  $1088.2$ & $\pm$ & $1.1$ & \\
reionisation optical depth & $\tau$ &
  $0.087$ & $\pm$ & $0.017$ &
  $0.084$ & $\pm$ & $0.016$ & \\
spectral index & $n_\mathrm{s}$ &
  $0.963$ & $^+_-$ & $^{0.014}_{0.015}$ &
  $0.960$ & $^+_-$ & $^{0.014}_{0.013}$ & \\
\hline
\end{tabular}
\end{center}
\caption{Cosmological parameters obtained from the 5-year data release of WMAP \citep{KO08.1}, without and with the additional constraints imposed by baryonic acoustic oscillations (BAO, \S~\ref{sec:VII-A-6}) and type-Ia supernovae (SNe, \S~\ref{sec:IX}). The dimension-less Hubble constant $h$ is defined in (\ref{eq:01-11}), the normalisation parameter $\sigma_8$ in (\ref{eq:10-1}) and the spectral index in (\ref{eq:01-28}). Reionisation is briefly discussed in \S~\ref{sec:X-B-2}. Note that spatial flatness ($K=0$) was assumed in deriving most of these values, as mentioned in last column.}
\label{tab:1}
\end{table*}

\subsubsection{The radiation-dominated phase}\label{sec:I-A-3}

It is an empirical fact that the Universe is expanding. Earlier in time, therefore, the scale factor must have been smaller than today, $a<1$. In principle, it is possible for Friedmann models to never reach a vanishing scale, $a=0$, within finite time into the past. However, a few crucial observational results suffice to rule out such ``bouncing'' models. This implies that a Universe like ours which is expanding today must have started from $a=0$ a finite time ago; thus, there must have been a Big Bang.

Equation (\ref{eq:01-10}) shows that the radiation density increases like $a^{-4}$ as the scale factor decreases, while the matter density increases with one power of $a$ less. Even though the radiation density is very much smaller today than the matter density, there must have been a period in the early evolution of the Universe when radiation dominated the energy density. The scale factor when both densities were equal is called $a_\mathrm{eq}$. This \emph{radiation-dominated era} is very important for several observational aspects of the cosmological standard model. Since the radiation retains the Planckian spectrum which it acquired in the very early Universe in the intense interactions with charged particles, its energy density is fully characterised by its temperature $T$. As the energy density is both proportional to $T^4$ and $a^{-4}$, its temperature falls like $T\propto a^{-1}$.

At the times relevant for our purposes, only photons and neutrinos need to be considered as relativistic species. Since photons are heated by electron-proton annihilation after neutrinos decoupled from the cosmic fluid, the photons are heated above the neutrinos by a factor $(11/4)^{1/3}$, which follows directly from entropy conservation. If $T$ is the temperature of the cosmic microwave background after electron-positron annihilation, the energy densities in the photons and the three neutrino species are
\begin{equation}
  \rho_\mathrm{r, CMB}=\frac{\pi^2}{15}\frac{(kT)^4}{(\hbar c)^3}\;,\quad
  \rho_{\mathrm{r}, \nu}=3\times\frac{7}{8}\times\left(\frac{4}{11}\right)^{4/3}\rho_\mathrm{r, CMB}
\label{eq:01-13}
\end{equation}
or, as long as all neutrino species are relativistic,
\begin{equation}
  \rho_\mathrm{r}=\rho_\mathrm{r, CMB}\left[1+\frac{21}{8}\left(\frac{4}{11}\right)^{4/3}\right]
  \approx1.68\,\rho_\mathrm{r, CMB}\;.
\label{eq:01-14}
\end{equation}

\subsubsection{Age, distances and horizons}\label{sec:I-A-4}

\begin{figure}[ht]
  \includegraphics[width=\hsize]{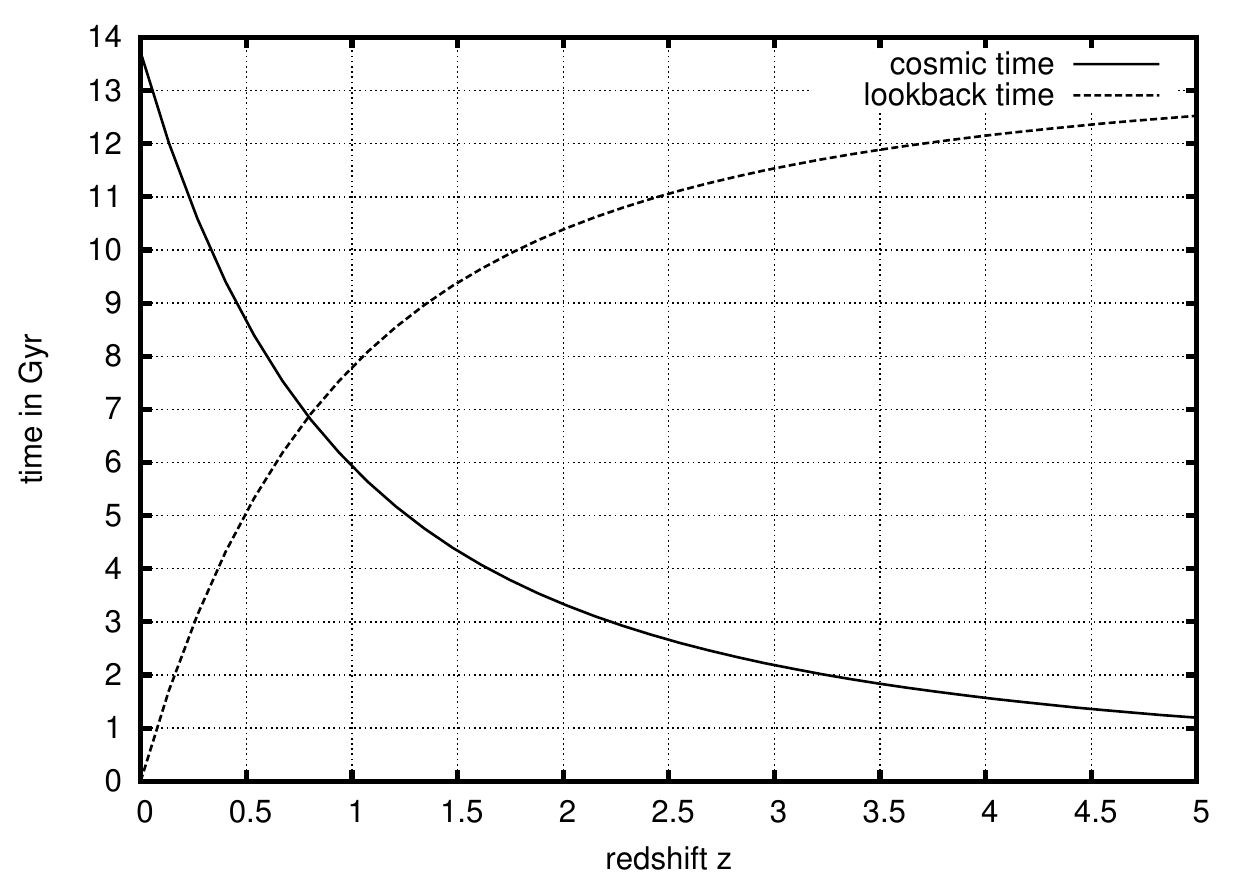}
\caption{Cosmic time (solid curve) and lookback time (present age of the universe minus cosmic time) in Gyr as functions of redshift. Here and below, the parameters labelled as ``reference'' in Tab.~\ref{tab:1} are used.}
\label{fig:1-1}
\end{figure}

Since Friedmann's equation gives the relative expansion rate $\dot a/a$, we can use it to infer the cosmic time,
\begin{equation}
  t=\int_0^t\d t'=\int_0^{a(t)}\frac{\d a}{\dot a}=
  \int_0^{a(t)}\frac{\d a}{aH(a)}=
  \frac{1}{H_0}\int_0^{a(t)}\frac{\d a}{aE(a)}\;,
\label{eq:01-15}
\end{equation}
which illustrates that the age scale is the inverse Hubble constant $H_0^{-1}$. A simple example is given by the Einstein-de Sitter model, which (unrealistically) assumes $\Omega_\mathrm{m0}=1$, $\Omega_\mathrm{r0}=0$ and $\Omega_\Lambda=0$. Then, from Eq.~(\ref{eq:01-10}), $E(a)=a^{-3/2}$ and
\begin{equation}
  t=\frac{1}{H_0}\int_0^1\sqrt{a}\d a=\frac{2}{3H_0}\;.
\label{eq:01-16}
\end{equation}
The cosmic age is $t_0=t(a_0)=t(1)$. The cosmic time and the lookback time $t_0-t$ are shown in Fig.~\ref{fig:1-1} as functions of the redshift.

\begin{figure}[ht]
  \includegraphics[width=\hsize]{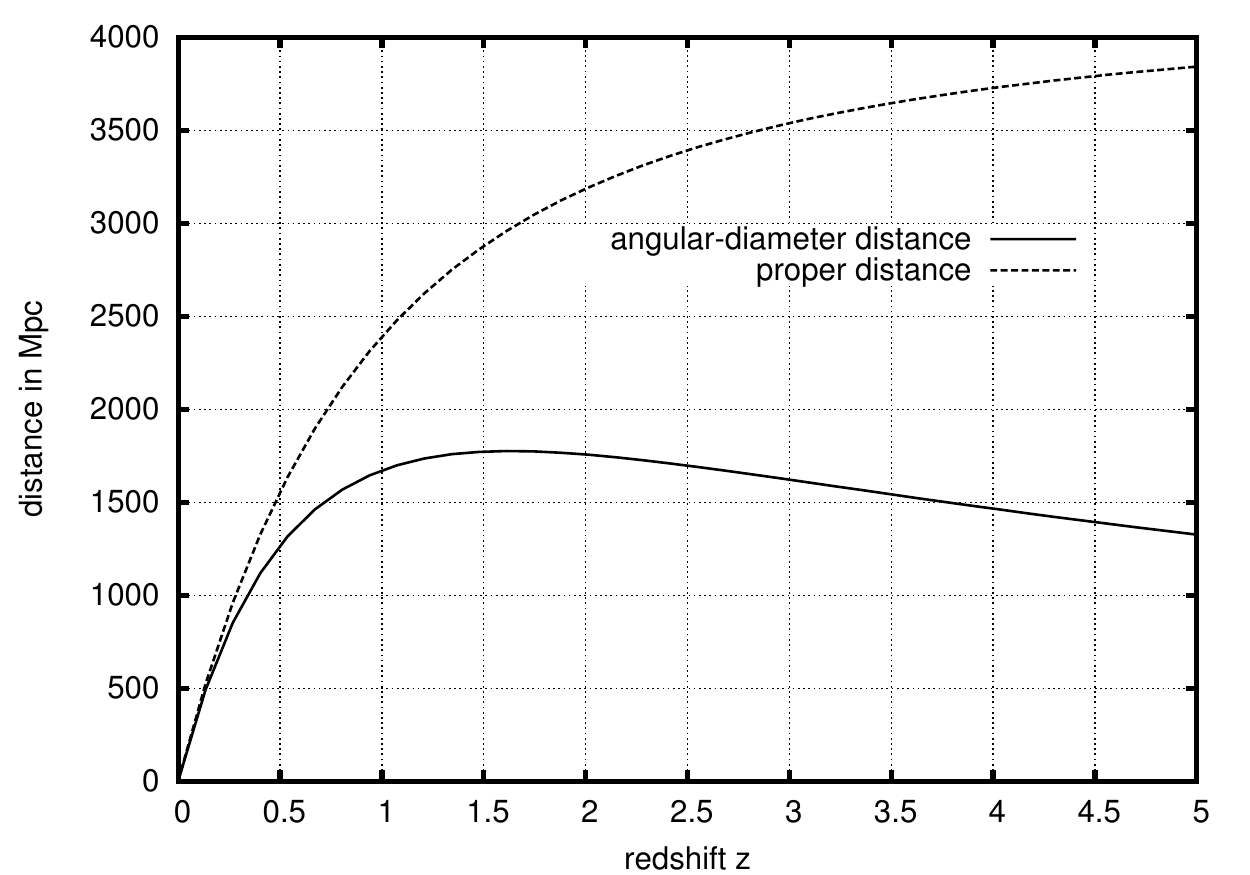}
\caption{Angular-diameter distance (solid curve) and proper distance in Mpc as functions of redshift.}
\label{fig:1-2}
\end{figure}

Distances can be defined in many ways which typically lead to different expressions. The \emph{proper distance} $D_\mathrm{prop}$ is the distance measured by the light-travel time, thus
\begin{equation}
  \d D_\mathrm{prop}=c\d t\quad\Rightarrow\quad
  D_\mathrm{prop}=\frac{c}{H_0}\int\frac{\d a}{aE(a)}\;,
\label{eq:01-17}
\end{equation}
where the integral has to be evaluated between the scale factors of emission and observation of the light signal. The \emph{comoving radial coordinate} $w$ is the comoving distance measured along a radial light ray. Since light rays propagate with zero proper time, $\d s=0$, this gives
\begin{equation}
  \d w=\frac{c\d t}{a}\quad\Rightarrow\quad
  w=c\int\frac{\d a}{a\dot a}=\frac{c}{H_0}\int\frac{\d a}{a^2E(a)}\;.
\label{eq:01-18}
\end{equation}
The \emph{angular-diameter distance} $D_\mathrm{ang}$ is defined such that the same relation as in Euclidean space holds between the physical size of an object and its angular size. It turns out to be
\begin{equation}
  D_\mathrm{ang}(a)=af_K[w(a)]\;,
\label{eq:01-19}
\end{equation}
with $f_K(w)$ given by (\ref{eq:01-2}). The \emph{luminosity distance} $D_\mathrm{lum}$ is analogously defined to reproduce the Euclidean relation between the bolometric luminosity of an object and its observed flux. This results in
\begin{equation}
  D_\mathrm{lum}(a)=\frac{D_\mathrm{ang}(a)}{a^2}=\frac{f_K[w(a)]}{a}\;.
\label{eq:01-20}
\end{equation}
Figure~\ref{fig:1-2} shows the angular-diameter and the proper distance as functions of redshift.

These distance measures rapidly diverge for scale factors $a<1$. For small distances, i.e.~for $a\approx1$, they all reproduce the linear relation
\begin{equation}
  D(z)=\frac{cz}{H_0}\;.
\label{eq:01-21}
\end{equation}

Since time is finite in a universe with a Big Bang, any particle can only be influenced by, and can only influence, events within finite regions, called \emph{horizons}. Several different definitions of horizons exist. They are typically characterised by some speed, e.g.~the light speed, times the inverse Hubble function which sets the time scale.

\subsection{Structures}\label{sec:I-B}

\subsubsection{Structure growth}\label{sec:I-B-1}

\begin{figure}[ht]
  \includegraphics[width=\hsize]{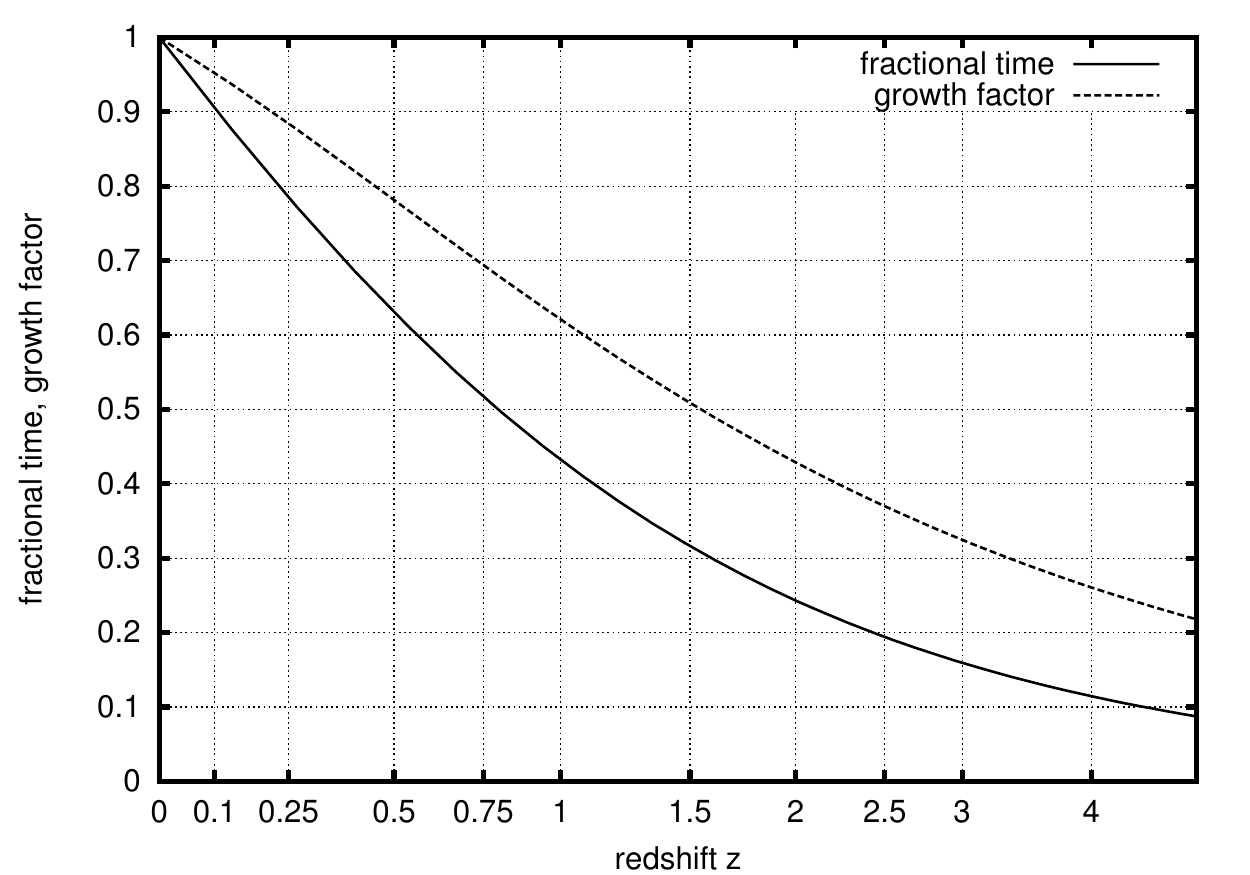}
\caption{Fractional cosmic age $t/t_0$ (solid curve) and structure growth as functions of redshift.}
\label{fig:1-3}
\end{figure}

The hierarchy of cosmic structures is assumed to have grown from primordial seed fluctuations in the process of gravitational collapse: overdense regions attract material and grow. They are described by the \emph{density contrast} $\delta(\vec x,t )$ as a function of the comoving coordinates $\vec x$, which is the density fluctuation relative to the mean density $\bar\rho(t)$,
\begin{equation}
  \delta(\vec x, t)\equiv\frac{\rho(\vec x, t)-\bar\rho(t)}{\bar\rho(t)}\;.
\label{eq:01-22}
\end{equation}
Linear perturbation theory shows that, during the matter-dominated era, the density contrast $\delta$ of sub-horizon perturbations is described by the second-order differential equation
\begin{equation}
  \ddot\delta+2H\dot\delta-4\pi G\bar\rho\delta=0
\label{eq:01-23}
\end{equation}
if the dark matter is cold, i.e.~if dark-matter flows have negligible velocity dispersion. Equation (\ref{eq:01-23}) has two solutions, one growing and one decaying. While the latter is irrelevant for structure growth, the growing mode is described by the \emph{growth factor} $D_+(a)$, defined such that the density contrast at the scale factor $a$ is related to the density contrast today $\delta_0$ by $\delta(a)=\delta_0D_+(a)$. In most cases of practical relevance, the growth factor is accurately described by
\begin{equation}
  D_+(a)=\frac{G(a)}{G(1)}
\label{eq:01-24}
\end{equation}
with the fitting formula
\begin{equation}
  G(a)\equiv a\Omega_\mathrm{m}\left[
    \Omega_\mathrm{m}^{4/7}-\Omega_\Lambda+\left(
      1+\frac{\Omega_\mathrm{m}}{2}
    \right)\left(
      1+\frac{\Omega_\Lambda}{70}
    \right)\right]^{-1}\;,
\label{eq:01-25}
\end{equation}
where the density parameters have to be evaluated at the scale factor $a$. For a standard cosmological model, (\ref{eq:01-25}) deviates from the accurate solution by $\le0.2\,\%$ for $a\in[0.01,1]$. The linear growth factor is shown in Fig.~\ref{fig:1-3} together with the fractional cosmic age $t/t_0$.

A very important length scale for cosmic structure growth is set by the horizon size at the end of the radiation-dominated phase. Structures smaller than that became causally connected while radiation was still dominating. The fast expansion due to the radiation density inhibited further growth of such structures until the matter density started dominating. Small structures are thus suppressed compared to large structures, which became causally connected only after radiation domination. The horizon radius $w_\mathrm{eq}$ at the end of the radiation-dominated era thus separates larger structures which could grow without inhibition, from smaller structures which were suppressed during radiation domination. In comoving coordinates,
\begin{equation}
  w_\mathrm{eq}=2\left(\sqrt{2}-1\right)\frac{c}{H_0}
  \left(\frac{a_\mathrm{eq}}{\Omega_\mathrm{m0}}\right)^{1/2}\;.
\label{eq:01-26}
\end{equation}

\subsubsection{The power spectrum}\label{sec:I-B-2}

It is physically plausible that the density contrast in the Universe is a Gaussian random field, i.e.~that the probability for finding a density contrast between $\delta$ and $\delta+\d\delta$ is given by a Gaussian distribution. The principal reason for this is the central limit theorem. A Gaussian random process is characterised by two numbers, the mean and the variance. By construction, the mean of the density contrast vanishes, such that the variance defines it completely.

In linear approximation, density perturbations grow in place, as Eq.~(\ref{eq:01-23}) shows because the density contrast at one comoving position $\vec x$ does not depend on the density contrast at another. As long as structures evolve linearly, their scale will be preserved, which implies that it is advantageous to study structure growth in Fourier rather than real space. The variance of the density contrast $\hat\delta(\vec k)$ at the comoving wave vector $\vec k$ in Fourier space is called the \emph{power spectrum}
\begin{equation}
  \left\langle
    \hat\delta(\vec k)\hat\delta^*(\vec k')
  \right\rangle\equiv(2\pi)^3P_\delta(k)\,
  \delta_\mathrm{D}(\vec k-\vec k')\;,
\label{eq:01-27}
\end{equation}
where the Dirac function $\delta_\mathrm{D}$ ensures that modes with different wave vectors are independent.

\begin{figure}[ht]
  \includegraphics[width=\hsize]{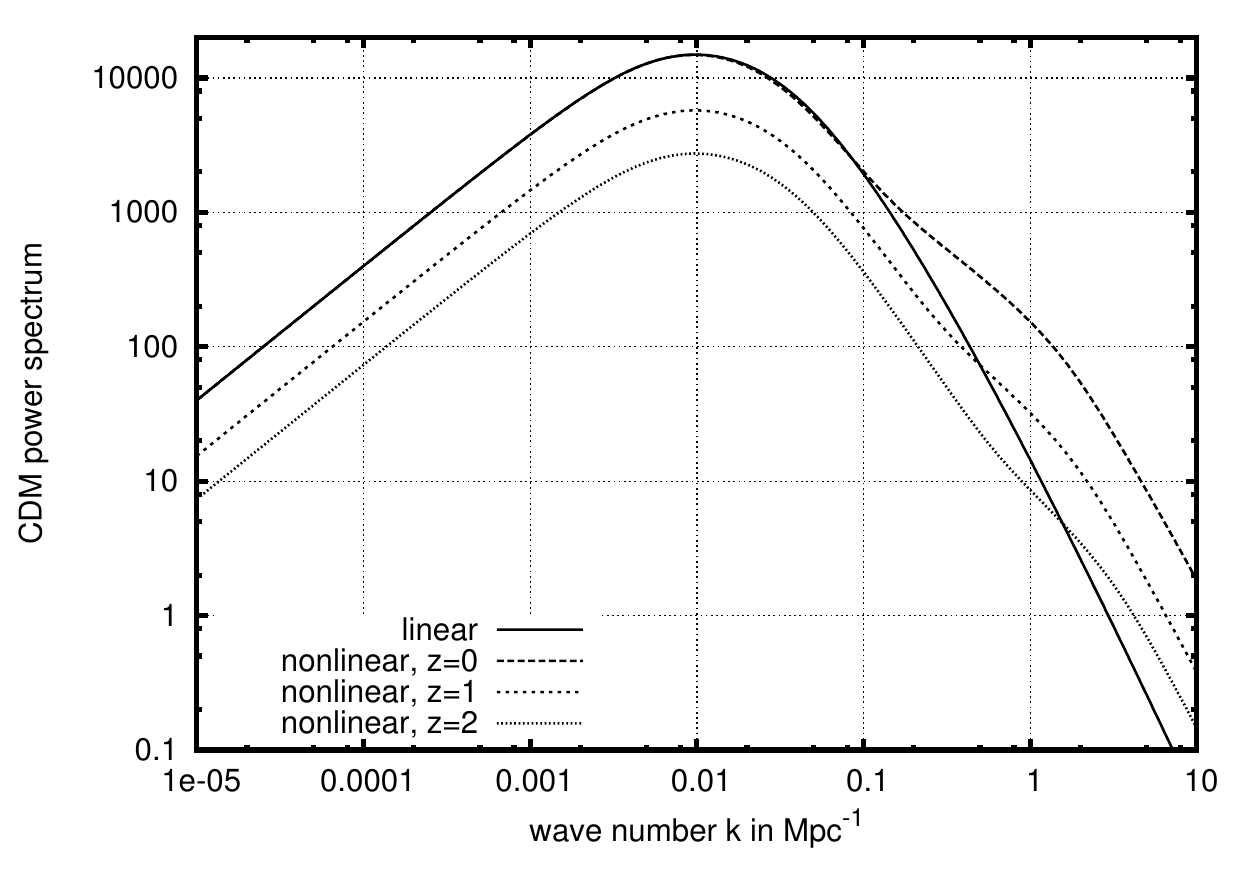}
\caption{Linearly and non-linearly evolved cold dark matter (CDM) power spectra. The linear CDM spectrum is shown for $a=1$ and scales like $D_+^2(a)$.}
\label{fig:1-4}
\end{figure}

Once the power spectrum is known, the statistical properties of the linear density contrast are completely specified. It is a remarkable fact that two simple assumptions about the nature of the cosmic structures and the dark matter constrain the shape of $P_\delta(k)$ completely. If the \textit{rms} mass fluctuation enclosed by the horizon is independent of time, and if the dark matter is cold, the power spectrum will behave as
\begin{equation}
  P_\delta(k)\propto\begin{cases}
    k^{n_\mathrm{s}}      & (k\ll k_\mathrm{eq}) \\
    k^{n_\mathrm{s}-4} & (k\gg k_\mathrm{eq})
                    \end{cases}\;,
\label{eq:01-28}
\end{equation}
with the \textit{spectral index} $n_\mathrm{s}=1$ \citep{HA70.1,ZE72.1,PE70.1,PE82.1,WE08.1}. The comoving wave number $k_\mathrm{eq}=0.01\,\mathrm{Mpc}^{-1}$ of the peak location in the power spectrum is set by the comoving horizon radius at matter-radiation equality, $w_\mathrm{eq}$, given in (\ref{eq:01-26}). The steep decline for structures smaller than $w_\mathrm{eq}$ reflects the suppression of structure growth during radiation domination. Figure~\ref{fig:1-4} shows the linearly and the nonlinearly evolved power spectrum of cold dark matter (CDM).

\subsubsection{Non-linear evolution}\label{sec:I-B-3}

As the density contrast approaches unity, its evolution becomes non-linear. In the course of non-linear evolution, overdensities contract, causing matter to flow from larger to smaller scales. Power in the density-fluctuation field is thus transported towards smaller modes, or towards larger wave numbers $k$. This \emph{mode coupling} process deforms the power spectrum on small scales, i.e.~for large $k$. Detailed studies of the non-linear evolution of cosmic structures require numerical simulations, which need to cover large scales and to resolve small scales well at the same time. Much progress has been achieved in this field within the last two decades due to the fortunate combination of increasing computer power with highly sophisticated numerical algorithms, such as particle-mesh and tree codes, and adaptive mesh refinement techniques \citep{EF85.1,BA86.1,HO88.1,SP02.1,MO05.1}.

The onset of non-linear evolution can be described by the so-called \emph{Zel'dovich approximation}, which approximately describes particle trajectories \citep{ZE70.1}. Although the Zel'dovich approximation breaks down as the non-linear evolution proceeds, it is remarkable for two applications. First, it allows a computation of the shapes of collapsing dark-matter structures and arrives at the conclusion that the collapse must be anisotropic, leading to the formation of sheets and filaments \citep{DO70.1}. Filamentary structures thus appear as a natural consequence of gravitational collapse in a Gaussian random field. Second, it provides an explanation for the origin of the angular momentum of cosmic structures.

\section{The age of the Universe}\label{sec:II}

\subsection{Nuclear cosmo-chronology}\label{sec:II-A}

\subsubsection{The age of the Earth}\label{sec:II-A-1}

How old is the Universe? We have no direct way to measure how long ago the Big Bang happened, but there are various ways to set lower limits to the age of the Universe. They are all based on the same principle: since the Universe cannot be younger than any of its parts, it must be older than the oldest objects it contains. Three methods for age determination have been developed. One is based on the radioactive decay of long-lived isotopes, another constrains the age of globular clusters, and the third is based on the age of white dwarfs.

Nuclear cosmo-chronology compares the measured abundances of certain long-lived radioactive isotopes with their initial abundances, which are eliminated by comparing abundances in different probes.

To give a specific example, consider the two uranium isotopes $\nuc{235}{U}$ and $\nuc{238}{U}$. They both decay into stable lead isotopes, $\nuc{235}{U}\to\nuc{207}{Pb}$ through the actinium series and $\nuc{238}{U}\to\nuc{206}{Pb}$ through the radium series. The abundance of any of these two lead isotopes is the sum of the initial abundance, plus the amount produced by the uranium decay, e.g.
\begin{equation}
  N_{207}=N_{207,0}+N_{235}\left(\e^{\lambda_{235}t}-1\right)
\label{eq:02-1}
\end{equation}
for $\nuc{207}{Pb}$, where $N_{235}$ is the abundance of $\nuc{235}{U}$ nuclei today. A similar equation with 235 replaced by 238 and 207 replaced by 206 holds for the decay of $\nuc{238}{U}$ to $\nuc{206}{Pb}$. The decay constants for the two uranium isotopes are measured as
\begin{equation}
  \lambda_{235}=(1.015\,\mathrm{Gyr})^{-1}\;,\quad
  \lambda_{238}=(6.45\,\mathrm{Gyr})^{-1}\;.
\label{eq:02-2}
\end{equation}

The idea is now that ores on Earth or meteorites formed during a period very short compared to the age of the Earth $t_\mathrm{e}$, so that their abundances can be assumed to have been locked up instantaneously and simultaneously a time $t_\mathrm{e}$ ago. Chemical fractionation has given different abundances to different samples, but could not distinguish between different isotopes of the same element. Thus, we expect different samples to show different isotope \emph{abundances}, but identical \emph{abundance ratios} between different isotopes.

The unstable lead isotope $\nuc{204}{Pb}$ has no long-lived parents and is therefore a measure for the primordial lead abundance. Thus, the \emph{abundance ratios} between $\nuc{206}{Pb}$ and $\nuc{207}{Pb}$ to $\nuc{204}{Pb}$ calibrate the abundances in different samples. Suppose we have two independent samples $a$ and $b$, in which the abundance ratios
\begin{equation}
  R_{206}\equiv\frac{N_{206}}{N_{204}}\quad\mbox{and}\quad
  R_{207}\equiv\frac{N_{207}}{N_{204}}
\label{eq:02-3}
\end{equation}
are measured. They are given by
\begin{equation}
  R_i=R_{i,0}+\frac{N_j}{N_{204}}\left(\e^{\lambda_jt_\mathrm{e}}-1\right)\;,
\label{eq:02-4}
\end{equation}
with $(i,j)=(206, 238)$ or $(207, 235)$. The present lead abundance ratios $R_{206,0}$ and $R_{207,0}$ should be the same in the two samples and cancel when the difference between the ratios in the two samples is taken. Then, the ratio of differences can be written as
\begin{equation}
  \frac{R_{207}^a-R_{207}^b}{R_{206}^a-R_{206}^b}=
  \frac{N_{235}}{N_{238}}
  \frac{\e^{\lambda_{235}t_\mathrm{e}}-1}
{\e^{\lambda_{238}t_\mathrm{e}}-1}\;.
\label{eq:02-5}
\end{equation}
Once the lead abundance ratios have been measured in the two samples, and the present uranium isotope ratio
\begin{equation}
  \frac{N_{235}}{N_{238}}=0.00725
\label{eq:02-6}
\end{equation}
is known, the age of the Earth $t_\mathrm{e}$ is the only unknown in (\ref{eq:02-5}). This method yields \citep{PA56.1,WA77.1}
\begin{equation}
  t_\mathrm{e}=(4.6\pm0.1)\,\mathrm{Gyr}\;.
\label{eq:02-7}
\end{equation}

\subsubsection{The age of the Galaxy}\label{sec:II-A-2}

A variant of this method can be used to estimate the age of the Galaxy, but this requires a model for how the radioactive elements were formed during the lifetime of the Galaxy until they were locked up in samples where we can measure their abundances today. Again, we can assume that the Galaxy formed quickly compared to its age $t_\mathrm{g}$.

Suppose there was an instantaneous burst of star formation and subsequent supernova explosions a time $t_\mathrm{g}$ ago and no further production thereafter. Then, the radioactive elements found on Earth today decayed for the time $t_\mathrm{g}-t_\mathrm{e}$ until they were locked up when the Solar System formed. If we can infer from supernova theory the initially produced abundance ratio $N_{235}/N_{238}$, we can conclude from its present value (\ref{eq:02-6}) and the age of the Earth what the age of the Galaxy must be.

The situation is slightly more complicated because element production did not stop after the initial burst. Suppose that a fraction $f$ of the total number $N_\mathrm{p}$ of the heavy nuclei locked up in the Solar System was produced in a burst at $t=0$, and the remaining fraction $1-f$ was added at a steady rate until $t=t_\mathrm{g}-t_\mathrm{e}$ when the Earth was formed. Given the constant production rate $p$, the abundance of a radioactive element with decay constant $\lambda$ is
\begin{equation}
  N=C\e^{-\lambda t}+\frac{p}{\lambda}
\label{eq:02-8}
\end{equation}
before $t_\mathrm{g}-t_\mathrm{e}$, with a constant $C$ to be suitably chosen, and
\begin{equation}
  N=N_0\e^{-\lambda[t-(t_\mathrm{g}-t_\mathrm{e})]}
\label{eq:02-9}
\end{equation}
thereafter, where $N_0$ is the abundance of the element locked up in the Solar System, as before. The initial conditions then require
\begin{equation}
  N(0)=C+\frac{p}{\lambda}=fN_\mathrm{p}\;,
\label{eq:02-10}
\end{equation}
after the burst at $t=0$, thus
\begin{equation}
  N(t_\mathrm{g}-t_\mathrm{e})=
  \e^{-\lambda(t_\mathrm{g}-t_\mathrm{e})}\left[
    fN_\mathrm{p}+\frac{p}{\lambda}\left(
      \e^{\lambda(t_\mathrm{g}-t_\mathrm{e})}-1
    \right)
  \right]
\label{eq:02-11}
\end{equation}
when the Earth formed, and
\begin{equation}
  N(t_\mathrm{g})=\e^{-\lambda t_\mathrm{g}}\left[
    fN_\mathrm{p}+\frac{p}{\lambda}\left(
      \e^{\lambda(t_\mathrm{g}-t_\mathrm{e})}-1
    \right)
  \right]
\label{eq:02-12}
\end{equation}
today, when the galaxy reaches its age $t_\mathrm{g}$. Since the production rate must be
\begin{equation}
  p=\frac{(1-f)N_\mathrm{p}}{t_\mathrm{g}-t_\mathrm{e}}\;,
\label{eq:02-13}
\end{equation}
the present abundance is
\begin{equation}
  N=N_\mathrm{p}\e^{-\lambda t_\mathrm{g}}\left[
    f+\frac{(1-f)}{\lambda(t_\mathrm{g}-t_\mathrm{e})}\left(
      \e^{\lambda(t_\mathrm{g}-t_\mathrm{e})}-1
    \right)
  \right]
\label{eq:02-14}
\end{equation}
in terms of the produced abundance $N_\mathrm{p}$. Supernova theory says that the \emph{produced} abundance ratio of the isotopes $\nuc{235}{U}$ and $\nuc{238}{U}$ is
\citep{CO87.1,CO91.1}
\begin{equation}
  \frac{N_\mathrm{235,p}}{N_\mathrm{238,p}}=1.4\pm0.2\;.
\label{eq:02-15}
\end{equation}
Taking the ratio of (\ref{eq:02-14}) for the present abundances of $\nuc{235}{U}$ and $\nuc{238}{U}$, inserting the decay constants from (\ref{eq:02-2}), the abundance ratios from (\ref{eq:02-6}) and (\ref{eq:02-15}), and the age of the Earth $t_\mathrm{e}$ from (\ref{eq:02-7}) yields an equation which contains only the age of the Galaxy $t_\mathrm{g}$ in terms of the assumed fraction $f$. This gives
\begin{equation}
  t_\mathrm{g}=\begin{cases}
    (6.3\pm0.2)\,\mathrm{Gyr} & f=1 \\
    (8.0\pm0.6)\,\mathrm{Gyr} & f=0.5 \\
    (12\pm2)\,\mathrm{Gyr}    & f=0 \\
  \end{cases}\;.
\label{eq:02-16}
\end{equation}
More detailed models \citep{ME86.1,CO87.1,FO89.1,CO91.1} yield comparable results. Common assumptions and results from galaxy-formation theory assert that at least $1\,\mathrm{Gyr}$ is necessary before galactic disks could have been assembled. Therefore, nuclear cosmochronology constrains the age of the Galaxy to fall approximately within
\begin{equation}
  7\,\mathrm{Gyr}\lesssim t_\mathrm{g}\lesssim 13\,\mathrm{Gyr}\;.
\label{eq:02-17}
\end{equation}

Interestingly, nuclear cosmochronology has also been applied to metal-poor and therefore presumably very old stars, in which heavy elements are nonetheless overabundant. For example, \cite{CA01.1} constrain the age of such a star to $(12.5\pm3)\,\mathrm{Gyr}$.

\subsection{Stellar ages}\label{sec:II-B}

\begin{figure}[ht]
  \includegraphics[width=\hsize]{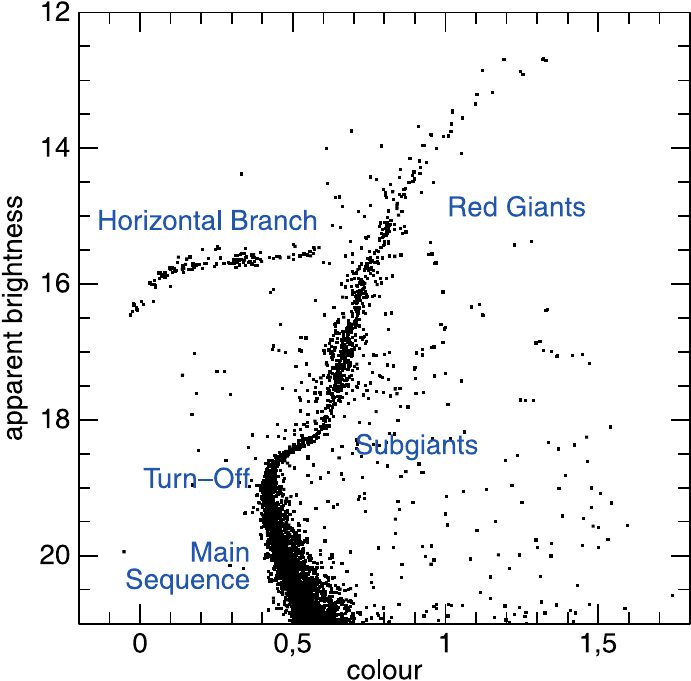}
\caption{Hertzsprung-Russell or colour-magnitude diagram of the globular cluster M~68. The magnitude is a logarithmic measure of the luminosity, colours are defined as magnitude differences. Stars populate the main sequence while burning hydrogen, then turn away from it towards the giant branch. The main-sequence turn-off is clearly visible (courtesy of Achim Wei{\ss}).}
\label{fig:02-1}
\end{figure}

Another method for measuring the age of the Universe caused much trouble for cosmologists for a long time. It is based on stellar evolution and exploits the fact that the time spent by stars on the main sequence of the Hertzsprung-Russell or colour-magnitude diagram (see Fig.~\ref{fig:02-1}) depends sensitively on their mass and thus on their colour.

Stars are described by the stellar-structure equations, which relate the mass $M$, the density $\rho$ and the pressure $p$ to the radius $r$ and specify the temperature $T$ and the luminosity $L$. They are
\begin{equation}
  \frac{\d p}{\d r}=-\frac{GM\rho}{r^2}\;,\quad
  \frac{\d M}{\d r}=4\pi r^2\rho\;,
\label{eq:02-18}
\end{equation}
which simply state hydrostatic equilibrium and mass conservation, and
\begin{equation}
  \frac{\d T}{\d r}=\frac{3L\kappa\rho}{4\pi r^2a_\mathrm{SB}cT^3}\;,\quad
  \frac{\d L}{\d r}=4\pi r^2\rho\epsilon\;,
\label{eq:02-19}
\end{equation}
which describe radiative energy transport and energy production. $\kappa$ is the opacity of the stellar material, $\epsilon$ is the energy production rate per unit mass, and $a_\mathrm{SB}$ is the Stefan-Boltzmann constant. Assuming $\kappa$ is independent of temperature, the energy-transport equation, mass conservation, hydrostatic equilibrium and the equation-of-state for an ideal gas yield the scaling relations
\begin{equation}
  L\sim\frac{RT^4}{\rho}\;,\quad
  M\sim\rho R^3\;,\quad
  p\sim\frac{\rho M}{R}\;,\quad
  p\sim\rho T\;.
\label{eq:02-20}
\end{equation}
The second pair of equations gives $T\sim M/R$, which yields $L\sim M^3$ when inserted into the first pair.

The total lifetime $\tau$ of a star on the main sequence must scale as $L\tau\sim M$ because the total energy radiated, $L\tau$, must be a fraction of the total rest-mass energy. Together with the earlier result, we find
\begin{equation}
  \tau\sim\frac{M}{L}\sim M^{-2}\sim L^{-2/3}\;.
\label{eq:02-21}
\end{equation}
According to the Stefan-Boltzmann law, the star's luminosity must be
\begin{equation}
  L\sim R^2T^4\quad\Rightarrow\quad R^2\sim\frac{M^3}{T^4}\;,
\label{eq:02-22}
\end{equation}
but we also know from above that $T\sim M/R$. Thus
\begin{equation}
  R^2\sim\frac{M^3R^4}{M^4}\sim\frac{R^4}{M}\quad\Rightarrow\quad
  R\sim M^{1/2}\;,\quad T\sim M^{1/2}\;,
\label{eq:02-23}
\end{equation}
and the lifetime $\tau$ on the main sequence turns out to scale as $\tau\sim T^{-1}$, by Eq.~(\ref{eq:02-21}). Afterwards, stars move away from the main sequence towards the giant branch. Thus, as a coeval stellar population ages, the point in its Hertzsprung-Russell diagram up to which the main sequence remains populated moves towards lower luminosities and temperatures along $(L,T)\sim(\tau^{-3/2},\tau^{-1})$. Old, coeval stellar populations exist: they are the globular clusters which surround the centre of the Galaxy in an approximately spherical halo. Therefore, the main-sequence turn-off points can be used to derive lower limits to the age of the Galaxy and the Universe.

In practice, such age determinations proceed by adapting simulated stellar-evolution tracks to the Hertzsprung-Russell diagrams of globular clusters and assigning the age of the best-fitting stellar-evolution model to the cluster \citep{RE88.1,ST91.1,VA00.1}. The simulated stellar-evolution tracks depend on the assumed metalicity of the stellar material, which changes the opacity and thus the energy transport through the stars.

Since observations cannot tell the luminosity of the turn-off point on the main sequence, but only its apparent brightness, age determinations from globular clusters require observations measuring both the apparent brightness of the main sequence turn-off, and the distance to the globular cluster. There are several ways for estimating cluster distances. One uses the period-luminosity relation of a class of variable stars, the RR~Lyrae stars, which are similar to the Cepheids that will play a central role in the next section. Another method exploits that stars on the horizontal branch have a typical luminosity and can thus be used to calibrate the cluster distance.

Therefore, uncertainties in the distance determinations directly translate to uncertainties in the age determinations. If the distance is overestimated, so is the luminosity, which implies that the age is underestimated, and vice versa. Globular clusters typically appeared to have ages well above estimates based on the cosmological parameters assumed \citep{VA83.1,SA90.0}. In the past decade or so, this has changed because improvements in stellar-evolution theory and direct distance measurements with the \textit{Hipparcos} satellite have lowered the globular-cluster ages, while recently determined cosmological parameters yield a higher age for the Universe as assumed before \citep{SA97.2,CH98.1}. Now, globular-cluster ages imply
\begin{equation}
  t_0\gtrsim(12.5\pm1.3)\,\mathrm{Gyr}
\label{eq:02-24}
\end{equation}
for the age of the Universe, allowing for a Gyr between the Big Bang and the formation of the globular clusters \citep{CH98.2}. The uncertainty in this result may still be considerable, though \cite[see, e.g.~][]{GR00.1}.

\subsection{Cooling of white dwarfs}\label{sec:II-C}

A third method for cosmic age determinations is based on the cooling of white dwarfs (see \citealt{HA03.2} for a review). They are the end products in the evolution of low-mass stars and form when the nuclear fuel in the stellar cores is exhausted. This happens typically when a core of carbon and oxygen has formed and temperature necessary for further fusion cannot be reached. The C-O core then shrinks until it is stabilised by electron degeneracy pressure. At this point, the mean free path of the electrons becomes effectively infinite and the core turns isothermal.

The pressure of a degenerate electron gas is independent of temperature and depends only on the density. In the non-relativistic limit, $p\propto\rho^{5/3}$ or
\begin{equation}
  p\propto\frac{M^{5/3}}{R^5}\quad\hbox{and}\quad
  \frac{p}{R}\sim\frac{GM\rho}{R^2}\sim\frac{GM^2}{R^5}\;,
\label{eq:02-25}
\end{equation}
where the latter two equations express hydrostatic equilibrium. Together, these relations imply $R\propto M^{-1/3}$, showing that more massive white dwarfs have smaller radii. Their surface gravity,
\begin{equation}
  g\propto\frac{M}{R^2}\propto M^{5/3}\;,
\label{eq:02-26}
\end{equation}
which can be determined from their spectra, is thus a direct measure for their mass. It turns out that the vast majority of white dwarfs are born with similar masses, $M_\mathrm{wd}\approx(0.55\pm0.05)\,M_\odot$ \citep{CO96.1}, which is a conspiracy of stellar evolution: more massive stars lose more mass through stellar winds before they turn into white dwarfs.

Except for the latent heat of crystallisation, white dwarfs cool passively by radiating the thermal energy stored by their mass. Being born with essentially the same mass, they start with very similar amounts of thermal energy. Their temperatures and luminosities are determined by the radiation processes and the opacities of their atmospheres, which in turn depend on their metalicities. Age inferences from white dwarfs thus require that their metalicities be known and the energy transport through their atmospheres be modelled.

After an initial short phase of rapid neutrino cooling, the energy loss slows down. This change in cooling time causes the white dwarfs to pile up at the temperature and the luminosity where the energy loss due to neutrinos falls below that due to radiative cooling. Thus, as time proceeds, their luminosity distribution develops a peak which is later slowly depopulated. Models for the white-dwarf cooling sequence allow the construction of time-dependent, theoretical luminosity distributions from which, by comparison with the observed luminosity distribution, the age of a white-dwarf population can be determined. This methods yields
\begin{equation}
  t_\mathrm{wd}\approx(9.5\pm1)\,\mathrm{Gyr}
\label{eq:02-27}
\end{equation}
for the age of white dwarfs in the Galactic disk \citep{OS96.1,SA97.3,CH98.2}. If we assume that massive spiral disks form at redshifts below $z\lesssim3$, the implied age of the Universe is approximately
\begin{equation}
  t_0\approx(11\pm1.4)\,\mathrm{Gyr}\;.
\label{eq:02-28}
\end{equation}
The age of the white-dwarf population in the globular cluster M~4 has been measured to be $\sim12.1\,\mathrm{Gyr}$ \citep{HA04.1}.

\subsection{Summary}\label{sec:II-D}

Combining results, we see that the age of the Universe, as measured by its supposedly oldest parts, is at least $\gtrsim11\,\mathrm{Gyr}$, and this places serious cosmological constraints. In the framework of the Friedmann-Lema{\^\i}tre models, they can be interpreted as limits on the cosmological parameters. Suppose we lived in an Einstein-de Sitter universe with $\Omega_\mathrm{m0}=1$ and $\Omega_\mathrm{\Lambda0}=0$. Then, we know from (\ref{eq:01-16}) that
\begin{equation}
  t_0=\frac{2}{3H_0}\gtrsim11\,\mathrm{Gyr}\quad\Rightarrow\quad
  H_0\lesssim2\times10^{-18}\,\mathrm{s}\;,
\label{eq:02-29}
\end{equation}
or $H_0\lesssim61\,\mathrm{km\,s^{-1}\,Mpc^{-1}}$ in conventional units. The Hubble constant is measured to be larger than this, which can immediately be interpreted as an indication that we are \emph{not} living in an Einstein-de Sitter universe.

\section{The Hubble Constant}\label{sec:III}

\subsection{Hubble constant from Hubble's law}\label{sec:III-A}

\subsubsection{Hubble's law}\label{sec:III-A-1}

Slipher discovered preceding 1920 that distant galaxies typically move away from us \citep{SL27.1}. Hubble found that their recession velocity grows with distance,
\begin{equation}
  v=H_0\,D\;,
\label{eq:03-1}
\end{equation} 
and determined the constant of proportionality as $H_0\approx559\,\mathrm{km\,s^{-1}\,Mpc^{-1}}$ \citep{HU31.1}. We had seen in (\ref{eq:01-21}) that all distance measures in a Friedmann-Lema{\^\i}tre universe follow the linear relation
\begin{equation}
  D=\frac{cz}{H_0}
\label{eq:03-2}
\end{equation}
to first order in $z\ll1$. Since $cz=v$ is the velocity according to the linearised relation between redshift and velocity, (\ref{eq:03-2}) is exactly the relation Hubble found.

The problem with any measurement of the Hubble constant from (\ref{eq:03-1}) is that, while (\ref{eq:03-2}) holds for an idealised, homogeneous and isotropic universe, real galaxies have peculiar motions on top of their Hubble velocity which are caused by the attraction from local density inhomogeneities. For instance, galaxies in our neighbourhood feel the gravitational pull of a cosmologically nearby supercluster called the Great Attractor and accelerate towards it. The galaxy M~31 in Andromeda is moving within the Local Group of galaxies and approaches the Milky Way.

Thus, the peculiar velocities of the galaxies must either be known well enough, for which a model for the velocity field is necessary, or they must be observed at so large distances that any peculiar motion is unimportant compared to their Hubble velocity. Requiring that the typical peculiar velocities, of order $(300\ldots600)\,\mathrm{km\,s^{-1}}$, be less than 10\% of the Hubble velocity, galaxies with redshifts
\begin{equation}
  z\gtrsim10\times\frac{(300\ldots600)\,\mathrm{km\,s^{-1}}}{c}\approx
  0.01\ldots0.02
\label{eq:03-4}
\end{equation} 
need to be observed. This is already so distant that it is hard or impossible to apply simple geometric distant estimators. This illustrates why accurate measurements of the Hubble constant are so difficult: nearby galaxies, whose distances are more accurately measurable, do not follow the Hubble expansion well, but the distances to galaxies far enough to follow the Hubble law are very hard to measure (see \cite{SA74.1} for an interesting account of the difficulties).

\subsubsection{The distance ladder}\label{sec:III-A-2}

Measurements of the Hubble constant from Hubble's law thus require accurate distance measurements out to cosmologically relevant distance scales. Since this is impossible in one step, the so-called distance ladder must be applied, in which each step calibrates the next.

The most fundamental distance measurement in the astrophysical toolbox is the trigonometric parallax caused by the annual motion of the Earth around the Sun. By definition, a star at a distance of a parsec perpendicular to the Earth's orbital plane has a parallax of one arc second. Astrometric measurement accuracies of order $10^{-5}\,''$ are thus necessary to measure distances of order $10\,\mathrm{kpc}$. Although the astrometric satellite \emph{Hipparcos} could measure distances only to $\le1\,\mathrm{kpc}$, its results were used to calibrate the distance ladder with nearby stars. In this way, it was possible to determine the distance to the Large Magellanic Cloud as $D_\mathrm{LMC}=(54.0\pm2.5)\,\mathrm{kpc}$ \cite{FE97.1}. An independent distance estimate based on the elliptical ring that appeared around the supernova 1987A gave $D_\mathrm{LMC}=(46.77\pm0.04)\,\mathrm{kpc}$ (\citealt{GO95.1}; see also \citealt{PA91.1}).

The next important step in the distance ladder is formed by the Cepheids. These are stars in late evolutionary stages which undergo periodic variability. The underlying instability is driven by the temperature dependence of the atmospheric opacity in these stars, which is caused by the transition between singly and doubly ionised Helium. The cosmologically important aspect of the Cepheids is that their variability period $\tau$ and their luminosity $L$ are related,
\begin{equation}
  L\propto\tau^{1.1}
\label{eq:03-5}
\end{equation}
in the $V$ band \citep{FE87.1,MA91.1,UD99.1}, hence their luminosity can be inferred from their variability period, and their distance from the ratio of their luminosity to the flux $S$ observed from them,
\begin{equation}
  D=\sqrt{\frac{L}{4\pi S}}\;.
\label{eq:03-6}
\end{equation}
At the relevant distances, any distinction between differently defined distance measures is irrelevant.

It is crucially important here that the calibration of the period-luminosity relation depends on the metalicity of the pulsating variables used, and thus on the stellar population they belong to. Hubble's originally much too high result for $H_0$ was corrected when Baade realised that he had confused a metal-poor class of variable stars with the metal-rich Cepheids \citep{BA56.1}.

Measuring the periods of Cepheids and comparing their apparent brightnesses in different galaxies thus allows the determination of the relative distances to the galaxies. For example, comparisons between Cepheids in the LMC and the Andromeda galaxy M~31 show \citep{FE97.1}
\begin{equation}
  \frac{D_\mathrm{M~31}}{D_\mathrm{LMC}}=16.4\pm1.1\;,
\label{eq:03-7}
\end{equation} 
while Cepheids in the member galaxies of the Virgo cluster yield \citep{FR01.1}
\begin{equation}
  \frac{D_\mathrm{Virgo}}{D_\mathrm{LMC}}=305\pm16\;.
\label{eq:03-8}
\end{equation}

Of course, for the Cepheid method to be applicable, it must be possible to resolve at least the outer parts of distant galaxies into individual stars and to reliably identify Cepheids among them. This was one reason why the \textit{Hubble Space Telescope} was proposed \citep{FR94.1}, to apply the superb resolution of an orbiting telescope to the measurement of $H_0$. Cepheid distance measurements are possible to distances $\lesssim30\,\mathrm{Mpc}$.

Scaling relations within classes of galaxies provide additional distance indicators. In the three-dimensional parameter space spanned by the velocity dispersion $\sigma_v$ of their stars, the radius $R_\mathrm{e}$ enclosing half the luminosity, and the surface brightness $I_\mathrm{e}$ within $R_\mathrm{e}$, elliptical galaxies populate the tight \emph{fundamental plane} \citep{DR87.1,DJ87.1} defined by
\begin{equation}
  R_\mathrm{e}\propto\sigma_v^{1.4}I_\mathrm{e}^{-0.85}\;.
\label{eq:03-9}
\end{equation} 
Since the luminosity must be
\begin{equation}
  L\propto I_\mathrm{e}R_\mathrm{e}^2\;,
\label{eq:03-10}
\end{equation} 
the fundamental-plane relation implies
\begin{equation}
  L\propto\sigma_v^{2.8}I_\mathrm{e}^{-0.7}\;.
\label{eq:03-11}
\end{equation}
Such a relation follows directly from the virial theorem if the mass-to-light ratio in elliptical galaxies increases gently with mass,
\begin{equation}
  \frac{M}{L}\propto M^{0.2}\;.
\label{eq:03-12}
\end{equation}
Thus, if it is possible to measure the surface brightness $I_\mathrm{e}$ and the velocity dispersion $\sigma_v$ of an elliptical galaxy, the fundamental-plane relation gives the luminosity, which can be compared to the flux to find the distance.

An interesting way for determining distances to galaxies uses the fluctuations in their surface brightness \citep{TO88.1,TO97.1,TO00.1}. The idea behind this method is that the fluctuations in the surface brightness will be dominated by the rare brightest stars, and that the optical luminosity of the entire galaxy will be proportional to the number $N$ of such stars. Assuming Poisson statistics, the fluctuation level will be proportional to $\sqrt{N}$, from which $N$ and $L\propto N$ can be determined once the method has been calibrated with galaxies whose distance is known otherwise. Again, the distance is then found by comparing the flux to the luminosity.

Planetary nebulae, which are late stages in the evolution of massive stars, have a luminosity function with a steep upper cut-off \citep{JA80.1,HU93.1}. Moreover, their spectra are dominated by sharp nebular emission lines which facilitate their detection even at large distances because they appear as bright objects in narrow-band filters tuned to the emission lines. Since the cut-off luminosity is known, it can be converted to a distance.

An important class of distance indicators are supernovae of type Ia. As will be described in \S~\ref{sec:IX-B}, their luminosities can be standardised, allowing their distances to be inferred from their fluxes.

Although they are not standard (or standardisable) candles, core-collapse supernovae of type II can also be used as distance indicators through the \emph{Baade-Wesselink method} \citep{BA26.1,WE46.1}. Suppose the spectrum of the supernova photosphere can be approximated by a Planck curve whose temperature can be determined from the spectral lines. Then, the Stefan-Boltzmann law fixes the total luminosity,
\begin{equation}
  L=\sigma_\mathrm{SB}R^2T^4\;.
\label{eq:03-14}
\end{equation}
The photospheric radius, however, can be inferred from the expansion velocity of the photosphere, which is measurable by the Doppler shift in the emission lines, times the time after the explosion. When applied to the supernova SN~1987A in the Large Magellanic Cloud, the Baade-Wesselink method yields a distance of
\begin{equation}
  D_\mathrm{LMC}=(49\pm3)\,\mathrm{kpc}
\label{eq:03-15}
\end{equation}
based on infrared photometry \citep{SC92.2} which agrees with the direct distance measurements \citep{GO95.1}. Note that the Bade-Wesselink method gives a one-step geometrical distance indicator bypassing the distance ladder.

\subsubsection{The HST Key Project}\label{sec:III-A-3}

All these distance indicators were used in the \emph{HST Key Project} \citep{FR01.1} to determine accurate distances to 26 galaxies between $(3\ldots25)\,\mathrm{Mpc}$, and five very nearby galaxies for testing and calibration.

Double-blind photometry was applied to the identified distance indicators. Since Cepheids tend to lie in star-forming regions and are thus attenuated by dust, and since their period-luminosity relation depends on metalicity, respective corrections had to be carefully applied. Then, the measured velocities had to be corrected by the peculiar velocities, which were estimated by a model for the flow field. The estimated luminosities of the distance indicators could then be compared with the extinction-corrected fluxes to determine distances, whose proportionality with the velocities corrected by the peculiar motions finally gave the Hubble constant. A weighed average over all distance indicators is \citep{FR01.1}
\begin{equation}
  H_0=(72\pm8)\,\mathrm{km\,s^{-1}\,Mpc^{-1}}\;,
\label{eq:03-16}
\end{equation} 
where the error is the square root of the systematic and statistical errors summed in quadrature. An alternative interpretation of the data is summarised in \cite{SA06.1}, who find a $14\,\%$ smaller value for $H_0$ because of a different calibration.

\subsection{Gravitational Lensing}\label{sec:III-B}

A completely different method for determining the Hubble constant uses gravitational lensing. Masses bend passing light paths towards themselves and therefore act similarly to convex glass lenses. As in ordinary geometrical optics, this effect can be described applying Fermat's principle to a medium with an index of refraction \citep{SC92.1}
\begin{equation}
  n=1-\frac{2\Phi}{c^2}\;,
\label{eq:03-17}
\end{equation} 
where $\Phi$ is the Newtonian gravitational potential.

If it is strong enough, the bending of the light paths causes multiple images to appear from single sources. Compared to the straight light paths in absence of the deflecting mass distribution, the curved paths are geometrically longer, and they have to additionally propagate through a medium whose index of refraction is $n>1$. This gives rise to a \emph{time delay} \citep{BL86.1} which has a geometrical and a gravitational component,
\begin{equation}
  \tau=\frac{1+z_\mathrm{d}}{c}\frac{D_\mathrm{d}D_\mathrm{ds}}{D_\mathrm{s}}\left[
    \frac{1}{2}\left(\vec\theta-\vec\beta\right)^2-\psi(\vec\theta)
  \right]\;,
\label{eq:03-18}
\end{equation}
where $\vec\theta$ are angular coordinates on the sky and $\vec\beta$ is the angular position of the source. $\psi$ is the appropriately scaled Newtonian gravitational potential of the deflector, projected along the line-of-sight. According to Fermat's principle, images occur where $\tau$ is extremal, i.e.~$\vec\nabla_\theta\tau=0$. The prefactor contains the distances $D_\mathrm{d, s, ds}$ from the observer to the deflector, to the source, and from the deflector to the source, respectively, and the redshift $z_\mathrm{d}$ of the lens.

The projected \emph{lensing potential} $\psi$ is related to the surface-mass density $\Sigma$ of the deflector by
\begin{equation}
  \vec\nabla^2\psi=2\frac{\Sigma}{\Sigma_\mathrm{cr}}\equiv2\kappa\;,
\label{eq:03-19}
\end{equation}
where the \emph{critical surface-mass density} is
\begin{equation}
  \Sigma_\mathrm{cr}\equiv\frac{c^2}{4\pi G}
  \frac{D_\mathrm{s}}{D_\mathrm{d}D_\mathrm{ds}}\;.
\label{eq:03-20}
\end{equation}

The dimension-less time delay $\tau$ from (\ref{eq:03-18}) is related to the true physical time delay $t$ by
\begin{equation}
  t\propto\frac{\tau}{H_0}\;,
\label{eq:03-21}
\end{equation}
where the proportionality constant is a dimension-less combination of the distances $D_\mathrm{d,s,ds}$ with the Hubble radius $cH_0^{-1}$ and the deflector redshift $1+z_\mathrm{d}$. Equation (\ref{eq:03-21}) shows that the true time delay is proportional to the Hubble time, as it can intuitively be expected \citep{RE64.1}.

Time delays are measurable in multiple images of a variable source. The variable signal arrives after different times in the images seen by the observer. If the deflector is a galaxy, time delays are typically of order days to months and therefore observable with a reasonable monitoring effort.

Interestingly, it can be shown in an elegant, but lengthy calculation \citep{KO02.1} that measured time delays can be inverted to find the Hubble constant from the approximate equation
\begin{equation}
  H_0\approx A(1-\langle\kappa\rangle)+
  B\langle\kappa\rangle(\eta-1)\;,
\label{eq:03-22}
\end{equation} 
where $A$ and $B$ are constants depending on the measured image positions and time delays, $\langle\kappa\rangle$ is the mean scaled surface-mass density of the deflector averaged within an annulus bounded by the image positions, and $\eta\approx2$ is the logarithmic slope of the deflector's density profile.

Therefore, if a model exists for the gravitationally-lensing galaxy, the Hubble constant can be found from the positions and time delays of the images. Applying this technique to five different lens systems, \cite{KO02.1} found
\begin{equation}
  H_0=(73\pm8)\,\mathrm{km\,s^{-1}}
\label{eq:03-23}
\end{equation}
assuming that the lensing galaxies have radially constant mass-to-light ratios.

This result is remarkable because it was obtained in one step without any reference to the extragalactic distance ladder. On the other hand, it is also problematic because it is obtained only if significantly more concentrated density profiles of the lensing galaxies are assumed than obtained e.g.~from the kinematics of the stars in galaxies. If less concentrated lenses are adopted which agree with the kinematic constraints, $H_0$ derived from gravitational time delays drops substantially to values near $50\,\mathrm{km\,s^{-1}}$ \citep{KO03.1}. This hints at an as yet unexplained discrepancy between measurements of $H_0$ and the measured time delays within the CDM framework.

\subsection{Summary}\label{sec:III-C}

The results given so far are in excellent agreement with the value
\begin{equation}
  H_0=(70.1\pm1.3)\,\mathrm{km\,s^{-1}\,Mpc^{-1}}\;,
\label{eq:03-24}
\end{equation}
derived from CMB measurements (see \S~\ref{sec:VI} and Tab.~\ref{tab:1}) assuming spatial flatness, $K=0$. adopting it, we can calibrate several important numbers that scale with some power of the Hubble constant. First, in cgs units, the Hubble constant is
\begin{equation}
  H_0=(2.26\pm0.04)\times10^{-18}\,\mathrm{s}\;,
\label{eq:03-25}
\end{equation}
which implies the Hubble time
\begin{equation}
  H_0^{-1}=(14.01\pm0.26)\,\mathrm{Gyr}
\label{eq:03-26}
\end{equation} 
and the Hubble radius
\begin{equation}
  \frac{c}{H_0}=(1.327\pm0.025)\times10^{28}\,\mathrm{cm}=
  (4.299\pm0.08)\,\mathrm{Gpc}\;.
\label{eq:03-27}
\end{equation}
The critical density of the Universe turns out to be
\begin{equation}
  \rho_\mathrm{cr0}=\frac{3H_0^2}{8\pi G}=
  (9.15\pm0.3)\times10^{-30}\,\mathrm{g\,cm^{-3}}\;.
\label{eq:03-28}
\end{equation}
It corresponds to five protons per $\mathrm{m}^3$, or approximately one galaxy per $\mathrm{Mpc}^3$.

\section{Big-Bang Nucleosynthesis}\label{sec:IV}

\subsection{The origin and abundance of Helium-4}\label{sec:IV-A}

\subsubsection{Elementary considerations}\label{sec:IV-A-1}

Stellar spectra show that the abundance of $\nuc{4}{He}$ in stellar atmospheres ranges between $0.2\lesssim Y\lesssim0.3$ by mass (the Sun has $Y=0.263$), i.e.~about a quarter of the baryonic mass in the Universe is composed of $\nuc{4}{He}$. It is produced in stars in the course of hydrogen burning. Per $\nuc{4}{He}$ nucleus, the amount of energy released corresponds to $0.7\%$ of the masses involved, or
\begin{eqnarray}
  \Delta E&=&\Delta mc^2=0.007\,(2m_\mathrm{p}+2m_\mathrm{n})c^2\approx
  0.028\,m_\mathrm{p}c^2\nonumber\\
  &\approx&26\,\mathrm{MeV}\approx4.2\times10^{-5}\mathrm{erg}\;.
\label{eq:04-1}
\end{eqnarray} 
Suppose a galaxy such as ours, the Milky Way, shines with a luminosity of $L\approx10^{10}\,L_\odot\approx3.8\times10^{43}\,\mathrm{erg\,s^{-1}}$ for a good fraction of the age of the Universe, say for $\tau=5\times10^9\,\mathrm{yr}\approx1.5\times10^{17}\,\mathrm{s}$. Then, it releases a total energy of
\begin{equation}
  E_\mathrm{tot}\approx L\tau\approx5.7\times10^{60}\,\mathrm{erg}\;.
\label{eq:04-2}
\end{equation}
The number of $\nuc{4}{He}$ nuclei required to produce this energy is
\begin{equation}
  \Delta N=\frac{E}{\Delta E}\approx
  \frac{5.7\times10^{60}}{4.2\times10^{-5}}\approx
  1.4\times10^{65}\;,
\label{eq:04-3}
\end{equation}
which amounts to a $\nuc{4}{He}$ mass of
\begin{equation}
  M_\mathrm{He}\approx4m_\mathrm{p}\Delta N\approx
  9.3\times10^{41}\,\mathrm{g}\;.
\label{eq:04-4}
\end{equation} 
Assume further that the galaxy's stars were all composed of pure hydrogen initially, and that they are all more or less similar to the Sun. Then, the mass in hydrogen was $M_\mathrm{H}\approx10^{10}\,M_\odot\approx2\times10^{43}\,\mathrm{g}$
initially, and the final $\nuc{4}{He}$ abundance by mass expected from the energy production is
\begin{equation}
  Y_*\approx\frac{9.3\times10^{41}}{2\times10^{43}}\approx5\%\;,
\label{eq:04-5}
\end{equation}
much less than the $\nuc{4}{He}$ abundance actually observed. This discrepancy is exacerbated by the fact that $\nuc{4}{He}$ is destroyed in later stages of the evolution of massive stars, a process affected also by mixing in stellar interiors.

We thus see that the amount of $\nuc{4}{He}$ observed in stars was highly unlikely produced by these stars themselves under reasonable assumptions during the lifetime of the galaxies. We must therefore consider that most of the $\nuc{4}{He}$ which is now observed
may have existed already well before the galaxies formed.

Nuclear fusion of $\nuc{4}{He}$ and similar light nuclei in the early Universe is possible only if the Universe was hot enough for a sufficiently long period during its early evolution. The nuclear binding energies of order $\sim\mathrm{MeV}$ imply that temperatures had to have fallen below $T\sim10^6\,\mathrm{MeV}\times1.16\times10^4\,\mathrm{K}/\mathrm{MeV}=1.16\times10^{10}\,\mathrm{K}$ before nucleosynthesis could begin. On the other hand, temperatures needed to be still high enough for charged nuclei to surmount the Coulomb barrier. Typical temperatures during primordial nucleosynthesis were of order $kT\lesssim100\,\mathrm{keV}$. Since the temperature of the (photon background in the) Universe is now $T_0\sim3\,\mathrm{K}$ as we shall see later, this corresponds to times when the scale factor of the Universe was \begin{equation}
  a_\mathrm{nuc}\gtrsim\frac{3}{1.16\times10^9}\approx2.59\times10^{-9}\;.
\label{eq:04-6}
\end{equation}

At times so early, the actual mass density and a possible cosmological constant are entirely irrelevant for the expansion of the Universe, which is only driven by the radiation density. Thus, the expansion function can be simplified to read $E(a)=\Omega_\mathrm{r0}^{1/2}a^{-2}$, and we find for the cosmic time according to (\ref{eq:01-15})
\begin{equation}
  t(a)=\frac{1}{\Omega_\mathrm{r0}^{1/2}H_0}\int_0^aa'\d a'=
  \frac{a^2}{2\Omega_\mathrm{r0}^{1/2}H_0}\approx
  2.40\times10^{19}a^2\,\mathrm{s}\;,
\label{eq:04-7}
\end{equation}
where we have inserted the Hubble constant from (\ref{eq:03-24}) and the radiation-density parameter today, $\Omega_\mathrm{r0}\approx8.51\times10^{-5}$, which will be justified later [see (\ref{eq:06-3})]. It contains the contributions from photons and three neutrino species, which are the particles relativistic at the relevant time.

Inserting $a_\mathrm{nuc}$ from (\ref{eq:04-6}) into (\ref{eq:04-7}) yields a time scale for nucleosynthesis of order a few minutes. It is instructive for later purposes to establish a relation between time and temperature based on (\ref{eq:04-7}). Since $T=T_0/a$,
\begin{equation}
  t=2.40\times10^{19}\left(\frac{T_0}{T}\right)^2\,\mathrm{s}\approx
  0.89\left(\frac{T}{\mathrm{MeV}}\right)^{-2}\,\mathrm{s}\;.
\label{eq:04-8}
\end{equation}

\subsubsection{The Gamow criterion}\label{sec:IV-A-2}

A crucially important step in the fusion of $\nuc{4}{He}$ is the fusion of deuterium $d$,
\begin{equation}
  p+n\to d+\gamma
\label{eq:04-9}
\end{equation} 
because the direct fusion of $\nuc{4}{He}$ from two neutrons and two protons is extremely unlikely. The conditions in the early universe must have been such that deuterium could be formed efficiently enough for the subsequent fusion of $\nuc{4}{He}$, but not too efficiently because otherwise too much deuterium would be left over after the end of nucleosynthesis. Realising this, \cite{GA48.1} suggested that the amount of deuterium produced had to be ``just right'', which he translated into the intuitive criterion
\begin{equation}
  n_\mathrm{B}\langle\sigma v\rangle t\approx1\;,
\label{eq:04-10}
\end{equation}
where $n_\mathrm{B}$ is the baryon number density, $\langle\sigma v\rangle$ is the velocity-averaged cross section for the reaction (\ref{eq:04-9}), and $t$ is the time available for the fusion, which we have seen in (\ref{eq:04-8}) to be set by the present temperature of the cosmic radiation background, $T_0$, and the temperature $T$ required for deuterium fusion.

Thus, from an estimate of the baryon density $n_\mathrm{B}$ in the Universe, from the known velocity-averaged cross section $\langle\sigma v\rangle$, and from the known temperature required for deuterium fusion, Gamow's criterion enables an estimate of the present temperature $T_0$ of the cosmic radiation background. Based on this, \cite{AL49.1} were able to predict $T_0\approx(1\ldots5)\,\mathrm{K}$! After these remarkably simple and far-reaching conclusions, we shall now study primordial nucleosynthesis and consequences thereof in more detail.

\subsubsection{Elements produced}\label{sec:IV-A-3}

The fusion of deuterium (\ref{eq:04-9}) is the crucial first step. Since the photodissociation cross section of $d$ is large, destruction of $d$ is very likely because of the intense photon background until the temperature has dropped way below the binding energy of $d$, which is only $2.2\,\mathrm{MeV}$, corresponding to $2.6\times10^{10}\,\mathrm{K}$. In fact, substantial $d$ fusion is delayed until the temperature falls to $T=9\times10^8\,\mathrm{K}$ or $kT\approx78\,\mathrm{keV}$! As (\ref{eq:04-8}) shows, this happens $t\approx150\,\mathrm{s}$ after the Big Bang.

From deuterium, $\nuc{3}{He}$ and tritium $t$ can be built, which can both be processed to $\nuc{4}{He}$. These reactions are now fast, immediately converting the newly formed $d$. In detail, these reactions are
\begin{eqnarray}
  d+p&\to&\nuc{3}{He}+\gamma\;,\nonumber\\
  d+d&\to&\nuc{3}{He}+n\;,\nonumber\\
  d+d&\to&t+p\;,\quad\mbox{and}\nonumber\\
  \nuc{3}{He}+n&\to&t+p\;,
\label{eq:04-11}
\end{eqnarray}
followed by
\begin{eqnarray}
  \nuc{3}{He}+d&\to&\nuc{4}{He}+p\quad\mbox{and}\nonumber\\
  t+d&\to&\nuc{4}{He}+n\;.
\label{eq:04-12}
\end{eqnarray}

Fusion reactions with neutrons are irrelevant because free neutrons are immediately locked up in deuterons once deuterium fusion begins, and passed on to $t$, $\nuc{3}{He}$ and $\nuc{4}{He}$ in the further fusion steps.

Since there are no stable elements with atomic weight $A=5$, addition of protons to $\nuc{4}{He}$ is unimportant. Fusion with $d$ is unimportant because its abundance is very low due to the efficient follow-up reactions. We can therefore proceed only by fusing $\nuc{4}{He}$ with $t$ and $\nuc{3}{He}$ to build up elements with $A=7$,
\begin{eqnarray}
  t+\nuc{4}{He}&\to&\nuc{7}{Li}+\gamma\;,\nonumber\\
  \nuc{3}{He}+\nuc{4}{He}&\to&\nuc{7}{Be}+\gamma\;,\quad
  \mbox{followed by}\nonumber\\
  \nuc{7}{Be}+e^-&\to&\nuc{7}{Li}+\nu_e\;.
\label{eq:04-13}
\end{eqnarray}
Some $\nuc{7}{Li}$ is destroyed by
\begin{equation}
  \nuc{7}{Li}+p\to2\,\nuc{4}{He}\;.
\label{eq:04-14}
\end{equation} 
The fusion of two $\nuc{4}{He}$ nuclei leads to $\nuc{8}{Be}$, which is unstable. Further fusion of $\nuc{8}{Be}$ in the reaction
\begin{equation}
  \nuc{8}{Be}+\nuc{4}{He}\to\nuc{12}{C}+\gamma
\label{eq:04-15}
\end{equation}
is virtually impossible because the low density of the reaction partners essentially excludes that a $\nuc{8}{Be}$ nucleus meets a $\nuc{4}{He}$ nucleus during its lifetime. While the reaction (\ref{eq:04-15}) is possible and extremely important in stars, it is suppressed below any importance in the early Universe. The absence of stable elements with $A=8$ thus prohibits any primordial element fusion beyond $\nuc{7}{Li}$.

\subsubsection{Helium abundance}\label{sec:IV-A-4}

Once stable hadrons can form from the quark-gluon plasma in the very early universe, neutrons and protons are kept in thermal equilibrium by the weak interactions
\begin{equation}
  p+e^-\leftrightarrow n+\nu_e\;,\quad n+e^+\leftrightarrow p+\bar\nu_e
\label{eq:04-16}
\end{equation}
until the interaction rate falls below the expansion rate of the Universe. While equilibrium is maintained, the abundances $n_n$ and $n_p$ are controlled by the Boltzmann factor
\begin{equation}
  \frac{n_n}{n_p}=\left(\frac{m_n}{m_p}\right)^{3/2}
  \exp\left(-\frac{Q}{kT}\right)\approx
  \exp\left(-\frac{Q}{kT}\right)\;,
\label{eq:04-17}
\end{equation}
where $Q=1.3\,\mathrm{MeV}$ is the energy equivalent of the mass difference between the neutron and the proton.

The weak interaction freezes out when $T\approx10^{10}\,\mathrm{K}$ or $kT\approx0.87\,\mathrm{MeV}$, which is reached $t\approx2\,\mathrm{s}$ after the Big Bang. At this time, the $n$ abundance by mass is
\begin{equation}
  X_n(0)\equiv\frac{n_nm_n}{n_nm_n+n_pm_p}\approx
  \frac{n_n}{n_n+n_p}=\left[
    1+\exp\left(\frac{Q}{kT_n}\right)
  \right]^{-1}\approx0.17\;.
\label{eq:04-18}
\end{equation}
Afterwards, the free neutrons undergo $\beta$ decay with a half life\footnote{Particle Data Group, http://pdg.lbl.gov/} of $\tau_n=(885.7\pm0.8)\,\mathrm{s}$, thus
\begin{equation}
  X_n=X_n(0)\e^{-t/\tau_n}\;.
\label{eq:04-19}
\end{equation}
When $d$ fusion finally sets in at $t_d\approx150\,\mathrm{s}$ after the Big Bang, the neutron abundance has dropped to
\begin{equation}
  X_n(t_d)\approx X_n(0)\e^{-t_d/\tau_n}\approx0.14\;.
\label{eq:04-20}
\end{equation}
Now, essentially all these neutrons are collected into $\nuc{4}{He}$ because the abundances of the other elements can be neglected to first order. This simple estimate yields a $\nuc{4}{He}$ abundance by mass of
\begin{equation}
  Y\approx2X_n(t_d)\approx0.28
\label{eq:04-21}
\end{equation}
because the neutrons are locked up in pairs to form $\nuc{4}{He}$ nuclei. The Big-Bang model thus allows the prediction that $\nuc{4}{He}$ must have been produced such that its abundance is approximately $28\%$ by mass, which is in remarkable agreement with the observed abundance and thus a strong confirmation of the Big-Bang model.

\subsubsection{Expected abundances and abundance trends}\label{sec:IV-A-5}

Precise abundances of the light elements as produced by the primordial fusion must be calculated solving rate equations based on the respective fusion cross sections. Uncertainties involved concern the exact values of the cross sections and their energy dependence, and the precise life time of the free neutrons. Since primordial nucleosynthesis happens during the radiation-dominated era, the expansion rate is exclusively set by the radiation density. Then, the only other parameter controlling the primordial fusion processes is the baryon density, if the neutron lifetime is taken for granted.

An excellent recent review of these matters is \cite{ST07.1}, to which we refer for further detail.

Ignoring possible modifications by non-standard temperature evolution, the only relevant parameter defining the primordial abundances is the ratio between the number densities of baryons and photons. Since both densities scale like $a^{-3}$ or, equivalently, like $T^3$, their \emph{ratio} $\eta$ is constant. Anticipating the photon number density to be determined from the temperature of the CMB,
\begin{equation}
  \eta=\frac{n_\mathrm{B}}{n_\gamma}=
  10^{-10}\eta_{10}\;,\quad
  \eta_{10}\equiv273\Omega_\mathrm{B}h^2\;.
\label{eq:04-22}
\end{equation} 
Thus, once we know the photon number density, and once we can determine the parameter $\eta$ from the primodial element abundances, we can infer the baryon number density. Typical 2-$\sigma$ uncertainties at a fiducial $\eta$ parameter of $\eta_{10}=5$ are 0.4\% for $\nuc{4}{He}$, 15\% for $d$ and $\nuc{3}{He}$, and 42\% for $\nuc{7}{Li}$. The $\nuc{4}{He}$ abundance depends only very weakly on $\eta$ because the largest fraction of free neutrons is swept up into $\nuc{4}{He}$ without strong sensitivity to the detailed conditions.

The principal effects determining the abundances of $d$, $\nuc{3}{He}$ and $\nuc{7}{Li}$ are the following: with increasing $\eta$, they can more easily be burned to $\nuc{4}{He}$, and so their abundances drop as $\eta$ increases. At low $\eta$, an increase in the proton density causes $\nuc{7}{Li}$ destruction in the reaction (\ref{eq:04-14}), while the precursor nucleus $\nuc{7}{Be}$ is more easily produced if the baryon density increases further. This creates a characteristic ``valley'' of the predicted $\nuc{7}{Li}$ abundance near $\eta\approx(2\ldots3)\times10^{-10}$.

\subsection{Observed element abundances}\label{sec:IV-B}

\subsubsection{Principles}\label{sec:IV-B-1}

Of course, the main problem with any comparison between light-element abundances predicted by primordial nucleosynthesis and their observed values is that much time has passed since the primordial fusion ceased, and further fusion processes have happened since. Seeking to determine the primordial abundances, observers must therefore either select objects in which little or no contamination by later nucleosynthesis can reasonably be expected, in which the primordial element abundance may have been locked up and separated from the surroundings, or whose observed element abundances can be corrected for their enrichment during cosmic history in some way.

Deuterium can be observed in cool, neutral hydrogen gas (HI regions) via resonant UV absorption from the ground state, or in radio wavebands via the hyperfine spin-flip transition, or in the sub-millimetre regime via $d$H molecule lines. These methods all employ the fact that the heavier $d$ nucleus causes small changes in the energy levels of electrons bound to it. $\nuc{3}{He}$ is observed through the hyperfine transition in its ion $\nuc{3}{He}^+$ in radio wavebands, or through its emission and absorption lines in HII regions. $\nuc{4}{He}$ is of course most abundant in stars, but the fusion of $\nuc{4}{He}$ in stars is virtually impossible to be corrected precisely. Rather, $\nuc{4}{He}$ is probed via the emission from optical recombination lines in HII regions \citep{IZ04.1}. Measurements of $\nuc{7}{Li}$ must be performed in old, local stellar populations. This restricts observations to cool, low-mass stars because of their long lifetime, and to stars in the Galactic halo to allow precise spectra to be taken despite the low $\nuc{7}{Li}$ abundance.

\subsubsection{Evolutionary corrections}\label{sec:IV-B-2}

Stars brooded heavy elements as early as $z\sim6$ or even higher \citep{CI05.1}. Any attempts at measuring primordial element abundances must therefore concentrate on gas with a metal abundance as low as possible. The dependence of the element abundances on metalicity allows extrapolations to zero enrichment.

Such evolutionary corrections are low for $d$ because it is observed in the Lyman-$\alpha$ forest lines, which arise from absorption in low-density, cool gas clouds at high redshift. Likewise, they are low for the measurements of $\nuc{4}{He}$ because it is observed in low-metalicity, extragalactic HII regions. Probably, little or no correction is required for the $\nuc{7}{Li}$ abundances determined from the spectra of very metal-poor halo stars \citep{CH05.1}, but a lively discussion is going on \citep{AS06.2,KO06.1}.

Inferences from $\nuc{3}{He}$ are different because $\nuc{3}{He}$ is produced from $d$ in stars during the pre-main sequence evolution. It is burnt to $\nuc{4}{He}$ during the later phases of stellar evolution in stellar cores, but conserved in stellar exteriors. Observations indicate that a net destruction of $\nuc{3}{He}$ must happen, possibly due to extra mixing in stellar interiors. For these uncertainties, $\nuc{3}{He}$ is commonly excluded from primordial abundance measurements.

\subsubsection{Specific results}\label{sec:IV-B-3}

Due to the absence of strong evolutionary effects and its steep monotonic abundance decrease with increasing $\eta$, $d$ is perhaps the most trustworthy baryometer. Since it is produced in the early Universe and destroyed by later fusion in stars, all $d$ abundance determinations are lower bounds to its primordial abundance. Measurements of the $d$ abundance at high redshift are possible through absorption lines in QSO spectra, which are likely to probe gas with primordial element composition or close to it. Such measurements are challenging in detail because the tiny isotope shift in the $d$ lines needs to be distinguished from velocity-shifted hydrogen lines, H abundances from saturated H lines need to be corrected by comparison with higher-order lines, and high-resolution spectroscopy is required for accurate continuum subtraction.

A deuterium abundance of
\begin{equation}
  \frac{n_d}{n_\mathrm{H}}=\left(2.68^{+0.27}_{-0.25}\right)\times10^{-5}
\label{eq:04-23}
\end{equation}
in high-redshift QSOs relative to hydrogen appears consistent with most relevant QSO spectra, although substantially lower values have been derived \cite[e.g.][]{PE01.1}. A substantial depletion from the primordial value is unlikely because any depletion should be caused by $d$ fusion and thus be accompanied by an increase in metal abundances, which should be measurable.

Some spectra which were interpreted as having $\lesssim10$ times the $d$ abundance from (\ref{eq:04-24}) may be due to lack of spectral resolution. The $d$ abundance in the local interstellar medium is typically lower,
\begin{equation}
  \frac{n_d}{n_\mathrm{H}}\sim1.5\times10^{-5}\;,
\label{eq:04-24}
\end{equation}
which is consistent with $d$ consumption due to fusion processes. Conversely, the $d$ abundance in the Solar System is higher because $d$ is locked up in the ice on the giant planets.

In low-metalicity systems, $\nuc{4}{He}$ should be near its primordial abundance, and a metalicity correction can be applied. Possible systematic uncertainties are due to modifications by underlying stellar absorption, collisional excitation of observed recombination lines, and the exact regression towards zero metalicity. Values between $Y_\mathrm{p}=0.240\pm0.006$ and $Y_\mathrm{p}=0.2477\pm0.0029$ are considered realistic and likely \citep{PE07.1}.

Observations of the $\nuc{7}{Li}$ abundance aim at the oldest stars in the Galaxy, which are halo (Pop-II) stars with very low metalicity. They should have locked up very nearly primordial gas, but may have processed it \citep{SA01.1,PI02.1}. Cool stellar atmospheres are difficult to model, and $\nuc{7}{Li}$ may have been produced by cosmic-ray spallation on the interstellar medium \citep{ST92.1}.

In the limit of low stellar metalicity, the observed $\nuc{7}{Li}$ abundances group around an approximately flat line (the so-called Spite plateau \citep{SP82.1}), which is asymptotically independent of metalicity,
\begin{equation}
  A(\nuc{7}{Li})=12+\log(n_\mathrm{Li}/n_\mathrm{H})=2.2\pm0.1\;,
\label{eq:04-25}
\end{equation} 
and shows very little dispersion. Stellar rotation is important here because it enhances mixing in stellar interiors \citep{PI02.1}.

The Spite plateau is unlikely to reflect the primordial $\nuc{7}{Li}$ abundance, but corrections are probably moderate. A possible increase of $\nuc{7}{Li}$ with the iron abundance indicates low production of $\nuc{7}{Li}$, but the probable net effect is a depletion with respect to the primordial abundance by no more than $\sim0.2$~dex. A conservative estimate yields
\begin{equation}
  2.1\le A(\nuc{7}{Li})\le2.5\;.
\label{eq:04-26}
\end{equation} 
In absence of depletion, this value falls into the valley expected in the primordial $\nuc{7}{Li}$ at the boundary between destruction by protons and production from $\nuc{8}{Be}$. However, if $\nuc{7}{Li}$ was in fact depleted, its primordial abundance was higher than the value (\ref{eq:04-26}), and then two values for $\eta_{10}$ are possible.

\subsubsection{Summary of results}\label{sec:IV-B-4}

Through the relation (\ref{eq:04-22}), the density of \emph{visible} baryons alone implies $\eta_{10}\ge1.5$. The deuterium abundance derived from absorption systems in the spectra of high-redshift QSOs indicates $\eta_{10}=5.65\ldots6.38$. The $\nuc{7}{Li}$ abundance \emph{predicted} from this value of $\eta$ is then $A(\nuc{7}{Li})_\mathrm{p}=3.81\ldots4.86$, which is at least in mild conflict with the observed value $A(\nuc{7}{Li})=2.1-2.3$, even if a depletion by 0.2~dex due to stellar destruction is allowed.

The \emph{predicted} primordial abundance of $\nuc{4}{He}$ is then $Y_\mathrm{p}=0.2477\ldots0.2489$, which overlaps with the measured value $Y_\mathrm{P}=0.228\ldots0.248$. Thus, the light-element abundances draw a consistent picture starting from the observed deuterium abundance.

\begin{figure}[ht]
  \includegraphics[width=\hsize]{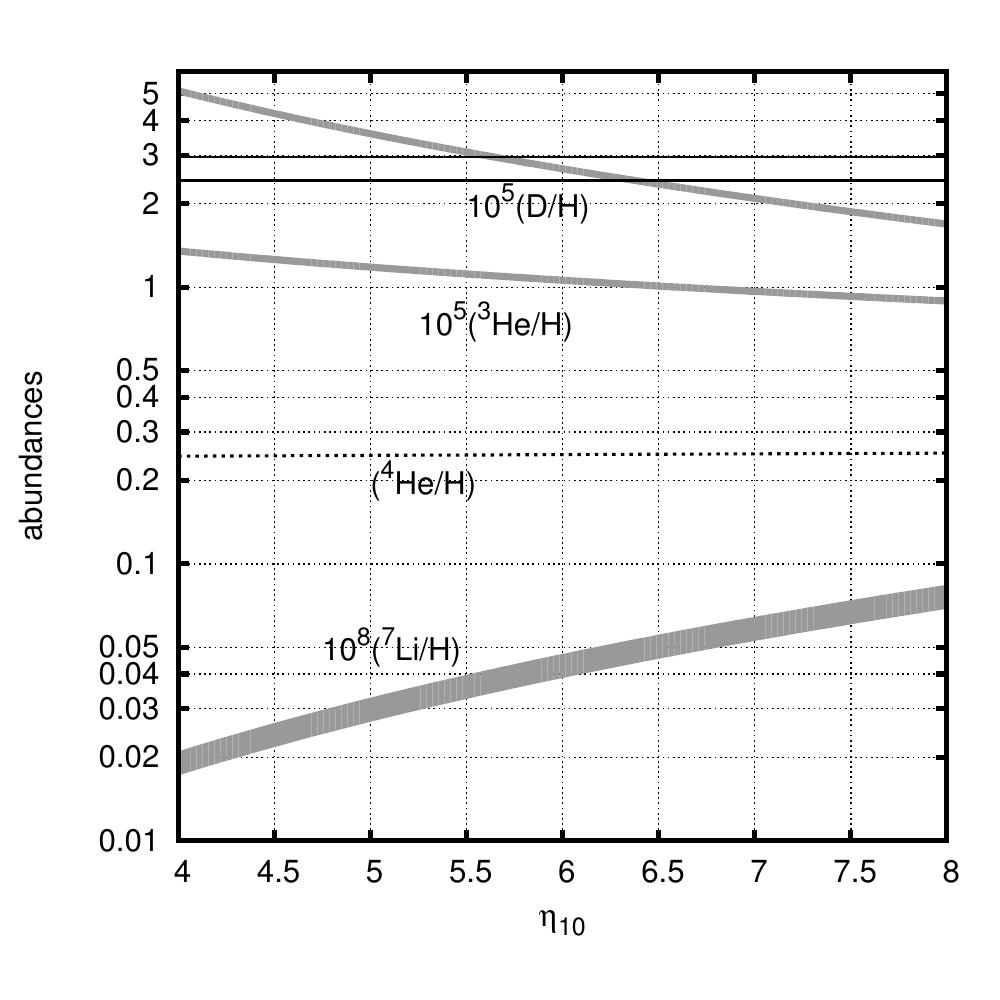}
\caption{Theoretically expected abundances of the light elements deuterium, $^3$He, $^4$He and $^7$Li as functions of $\eta_{10}$ (following \cite{ST07.1}). The observed deuterium abundance is marked by the two horizontal solid lines.}
\end{figure}

We thus find that Big-Bang nucleosynthesis \emph{alone} implies
\begin{equation}
  0.0207\le\Omega_\mathrm{B}h^2\le0.0234\quad\mbox{or}\quad
  0.0399\le\Omega_\mathrm{B}\le0.045
\label{eq:04-27}
\end{equation}
if conclusions are predominantly based on the deuterium abundance in high-redshift absorption systems. We shall later see that this result is in fantastic agreement with independent estimates of the baryon density obtained from the analysis of structures in the CMB.

At this point, it is important to note that primordial nucleosynthesis depends on the number of relativistic particle species, and thus on the number of light neutrino families. The theoretical abundances given here assume three lepton families in the standard model of the electroweak interaction. In fact, light-element abundances had been used to constrain the number of lepton families before it was measured in particle detectors \cite{ST77.1,YA84.1}.

\section{The Matter Density in the Universe}\label{sec:V}

\subsection{Mass in galaxies}\label{sec:V-A}

\subsubsection{Stars}\label{sec:V-A-1}

Given the luminosity of a stellar population, what is its mass? If all stars were like the Sun, the answer was trivial, but this is not the case. We shall focus the discussion here on stars which fall on the main sequence of the colour-magnitude diagram. Stars are formed with an initial mass distribution, the ``initial mass function'', which is for our purposes reasonably approximated by the Salpeter form \citep{SA55.1}
\begin{equation}
  \frac{\d N}{\d\ln M}\propto M^{-1.35}\;.
\label{eq:05-1}
\end{equation}
Expressing the mass $M$ in solar units, $m\equiv M/M_\odot$, and normalising the mass distribution to unity in the mass range $m_0\le m<\infty$ yields
\begin{equation}
  \frac{\d N}{\d m}=1.35\left(\frac{m_0}{m}\right)^{2.35}\,
  \frac{\d m}{m_0}\;.
\label{eq:05-2}
\end{equation}

The accepted lower mass limit for a star is $m_0=0.08$ because nuclear fusion cannot set in in objects of lower mass. However, we are interested in stars that are able to produce visible or near-infrared light so that we can translate the respective measured luminosities into mass. For a simple estimate, assume that stars have approximate Planck spectra, for which Wien's law holds, relating the wavelength $\lambda_\mathrm{max}$ of the peak in the Planck curve to the temperature $T$,
\begin{equation}
  \lambda_\mathrm{max}=0.29\,\left(\frac{\mathrm{K}}{T}\right)\,
  \mathrm{cm}\;.
\label{eq:05-3}
\end{equation}
The effective Solar temperature $T_\odot=5780\,\mathrm{K}$ implies $\lambda_\mathrm{max}=5.0\times10^{-5}\,\mathrm{cm}$. Stars releasing the majority of their energy in the optical and near-infrared regime should have $\lambda_\mathrm{max}\lesssim1\mu\mathrm{m}=10^{-4}\,\mathrm{cm}$ and thus $T\gtrsim2900\,\mathrm{K}\approx0.5\,T_\odot$. We saw in (\ref{eq:02-23}) that the temperature scales as $T\sim M^{1/2}$; thus, $T\gtrsim0.5\,T_\odot$ implies $m_0\approx0.25$.

We also saw following (\ref{eq:02-20}) that the luminosity scales as $L\sim M^3$. With $l\equiv L/L_\odot$, the mean mass-to-light ratio of the visible stellar population is thus expected to be
\begin{equation}
  \left\langle\frac{m}{l}\right\rangle=
  \int_{m_0}^\infty\frac{\d N}{\d m}\left(\frac{m}{l}\right)\d m=
  \int_{m_0}^\infty\frac{\d N}{\d m}\frac{\d m}{m^2}=
  \frac{1.35}{3.35\,m_0^2}\approx6.4\;.
\label{eq:05-4}
\end{equation}
This shows that an average stellar population visible in the optical and near-infrared spectral ranges is expected to require $\approx6.4$ solar masses for one solar luminosity. In order to produce, say, $10^{10}\,L_\odot$, a galaxy thus needs to have a mass of at least $\approx6.4\times10^{10}\,M_\odot$. This estimate assumes a homogeneous stellar population at the beginning of its evolution, neglects corrections from the exact spectral-energy distribution of the stars and the emission of giant stars, but $\langle m/l\rangle\gtrsim6.4$ still seems appropriate for evolved populations.

\subsubsection{Galaxies}\label{sec:V-A-2}

The rotation velocities of stars orbiting in spiral galaxies are observed to rise quickly with radius and then to remain roughly constant \citep{SO01.1}. If measurements are continued with neutral hydrogen beyond the radii out to which stars can be seen, these \emph{rotation curves} are observed to continue at an approximately constant level \citep{KR04.1}.

In a spherically-symmetric mass distribution, test particles on circular orbits have orbital velocities of
\begin{equation}
  v_\mathrm{rot}^2(r)=\frac{GM(r)}{r}\;.
\label{eq:05-5}
\end{equation} 
Flat rotation curves thus imply that $M(r)\propto r$, requiring the density to drop like $\rho(r)\propto r^{-2}$. This is much flatter than the light distribution, which shows that spiral galaxies are characterised by an increasing amount of dark matter as the radius increases.

A mass distribution with $\rho\propto r^{-2}$ has formally infinite mass, which is physically implausible. However, at finite radius, the density of the galaxy falls below the mean density of the surrounding universe. The spherical collapse model \citep{GU72.1} often invoked in cosmology shows that a spherical mass distribution can be considered in dynamical equilibrium if its mean overdensity is approximately 200 times higher than the mean density $\bar\rho$. Let $R$ be the radius enclosing this overdensity, and $M$ the mass enclosed, then
\begin{equation}
  \frac{M}{V}=\frac{3M}{4\pi R^3}=200\bar\rho\quad\Rightarrow\quad
  \frac{M}{R}=\frac{800\pi\bar\rho R^2}{3}\;.
\label{eq:05-6}
\end{equation}
At the same time, (\ref{eq:05-5}) needs to be satisfied, hence
\begin{equation}
  \frac{800\pi\bar\rho R^2}{3}=\frac{v_\mathrm{rot}^2}{G}
  \quad\Rightarrow\quad
  R=\left(
    \frac{3v_\mathrm{rot}^2}{800\pi G\bar\rho}
  \right)^{1/2}\;.
\label{eq:05-7}
\end{equation}
Inserting typical numbers yields
\begin{equation}
  R=290\,\mathrm{kpc}\,\left(
    \frac{v_\mathrm{rot}}{200\,\mathrm{km\,s^{-1}}}
  \right)\;.
\label{eq:05-8}
\end{equation} 
With (\ref{eq:05-5}), this implies
\begin{equation}
  M=\frac{Rv_\mathrm{rot}^2}{G}=2.7\times10^{12}\,M_\odot\,\left(
    \frac{v_\mathrm{rot}}{200\,\mathrm{km\,s^{-1}}}
  \right)^3\;.
\label{eq:05-9}
\end{equation}

Typical luminosities of spiral galaxies are given by the Tully-Fisher relation \citep{TU77.1},
\begin{equation}
  L=L_*\left(
    \frac{v_\mathrm{rot}}{220\,\mathrm{km\,s^{-1}}}
  \right)^{3\ldots4}\;,
\label{eq:05-10}
\end{equation}
with $L_*\approx2.4\times10^{10}\,L_\odot$. Thus, the mass-to-light ratio of a massive spiral galaxy is found to be
\begin{equation}
  \frac{m}{l}\approx150
\label{eq:05-11}
\end{equation}
in solar units, where it is assumed that the galaxy extends out to the virial radius of $\approx290\,\mathrm{kpc}$ with the same density profile $r^{-2}$. Evidently, this exceeds the \emph{stellar} mass-to-light ratio by far. Of course, the mass-to-light ratio of galaxies depends on the limiting radius assumed. Values of $m/l\approx30$ are often quoted, which are typically based on the largest radii out to which rotation curves can be measured.

\subsubsection{The galaxy population}\label{sec:V-A-3}

Galaxy luminosities are observed to be distributed according to the Schechter function \citep{SC76.1},
\begin{equation}
  \frac{\d N}{\d L}=\frac{\Phi_*}{L_*}
  \left(\frac{L}{L_*}\right)^{-\alpha}
  \exp\left(-\frac{L}{L_*}\right)\;,
\label{eq:05-12}
\end{equation} 
where the normalising factor is $\Phi_*\approx3.7\times10^{-3}\,\mathrm{Mpc}^{-3}$, the scale luminosity is $L_*$ given under (\ref{eq:05-10}) above, and the power-law exponent is $\alpha\approx1$ \citep{MA02.1}. Irrespective of which physical processes this distribution originates from, its functional form turns out to characterise mixed galaxy populations very well, even in galaxy clusters.

The luminosity density in galaxies is easily found to be
\begin{eqnarray}
  \mathcal{L}_\mathrm{g}&=&\int_0^\infty L\frac{\d N}{\d L}\,\d L=
  \Phi_*L_*\int_0^\infty l^{1-\alpha}\e^{-l}\,\d l\nonumber\\&=&
  \Gamma(2-\alpha)\Phi_*L_*\approx\Phi_*L_*\approx
  7.4\times10^7\,\frac{L_\odot}{\mathrm{Mpc}^3}\;.
\label{eq:05-13}
\end{eqnarray}
The average mass-to-light ratio (\ref{eq:05-11}) then allows converting this number into a mass density,
\begin{equation}
  \langle\rho_\mathrm{g}\rangle=\left\langle\frac{m}{l}\right\rangle\,
  \mathcal{L}_\mathrm{g}\approx
  1.1\times10^{-4}\,\frac{M_\odot}{\mathrm{Mpc}^3}\approx
  7.5\times10^{-31}\,\mathrm{g\,cm^{-3}}
\label{eq:05-14}
\end{equation}
and thus, with
$\rho_\mathrm{cr0}=9.15\times10^{-30}\,\mathrm{g\,cm^{-3}}$ from Eq.~(\ref{eq:03-28}),
\begin{equation}
  \Omega_\mathrm{g0}\approx0.08\;.
\label{eq:05-15}
\end{equation}
Of course, estimates based on the more conservative mass-to-light ratio $m/l\approx30$ find values which are lower by a factor of $\sim5$. In summary, this shows that the total mass expected to be contained in the dark-matter halos hosting galaxies contributes of order $8\%$ to the critical density in the Universe.

\subsection{Mass in galaxy clusters}\label{sec:V-B}

\subsubsection{Kinematic masses}\label{sec:V-B-1}

The next step upward in the cosmic hierarchy are galaxy clusters, which were first identified as significant galaxy overdensities in relatively small areas of the sky \citep{HE86.1,HE89.1}. Although the majority of galaxies is not in galaxy clusters, rich galaxy clusters contain several hundred galaxies, which by themselves contain a total amount of mass
\begin{equation}
  M_\mathrm{g}\lesssim10^2L_*\left\langle\frac{m}{l}\right\rangle\approx
  3\times10^{14}\,M_\odot\;.
\label{eq:05-16}
\end{equation} 
The mass in stars is of course considerably lower. With the mean stellar mass-to-light ratio of $m/l\approx6.4$ from (\ref{eq:05-4}), the same luminosity implies
\begin{equation}
  M_*\lesssim1.3\times10^{13}\,M_\odot\;.
\label{eq:05-17}
\end{equation}
The stellar mass of the Coma cluster, for instance, is inferred to be $M_{*,\mathrm{Coma}}\approx(1.4\pm0.3)\times10^{13}\,M_\odot$ \citep{WH93.1}.

The galaxies in rich galaxy clusters move with typical velocities of order $\lesssim10^3\,\mathrm{km\,s^{-1}}$, measured from the redshifts in their spectra. Therefore, only one component of the galaxy velocity is observed. Its distribution is characterised by the velocity dispersion $\sigma_v$.

If these galaxies were not gravitationally bound to the clusters, the clusters would disperse within $\lesssim1\,\mathrm{Gyr}$. Since they exist over cosmological time scales, clusters need to be (at least in some sense) gravitationally stable. Assuming virial equilibrium, the potential energy of the cluster galaxies should be minus two times the kinetic energy, or
\begin{equation}
  \frac{GM}{R}\approx3\sigma_v^2\;,
\label{eq:05-18}
\end{equation} 
where $M$ and $R$ are the total mass and the virial radius of the cluster, and the factor three arises because the velocity dispersion represents only one of three velocity components. We combine (\ref{eq:05-18}) with (\ref{eq:05-6}) to find
\begin{equation}
  R=\left(
    \frac{9\sigma_v^2}{800\pi G\bar\rho}
  \right)^{1/2}\approx2.5\,\mathrm{Mpc}\,
  \left(\frac{\sigma_v}{1000\,\mathrm{km\,s^{-1}}}\right)\;,
\label{eq:05-19}
\end{equation}
and, with (\ref{eq:05-18}),
\begin{equation}
  M\approx1.7\times10^{15}\,\mathrm{M_\odot}\,
  \left(\frac{\sigma_v}{1000\,\mathrm{km\,s^{-1}}}\right)^3\;.
\label{eq:05-20}
\end{equation}
Hence, the mass required to keep cluster galaxies bound despite their high velocities exceeds the mass in galaxies by about an order of magnitude, even if the entire virial mass of the galactic halos is accounted for \citep{ZW33.1,ZW37.1}. The stellar mass apparently contributes just about one per cent to the total cluster mass.

\subsubsection{Mass in the hot intracluster gas}\label{sec:V-B-2}

Galaxy clusters are diffuse sources of thermal X-ray emission. Their X-ray continuum is caused by thermal \emph{bremsstrahlung}, whose bolometric volume emissivity is
\begin{equation}
  j_\mathrm{X}=Z^2g_\mathrm{ff}C_\mathrm{X}\,n^2\,\sqrt{T}
\label{eq:05-21}
\end{equation}
in cgs units, where $Z$ is the ion charge, $g_\mathrm{ff}$ is the Gaunt factor, $n$ is the ion number density, $T$ is the gas temperature, and
\begin{equation}
  C_\mathrm{X}=2.68\times10^{-24}
\label{eq:05-22}
\end{equation}
in cgs units, if $T$ is measured in keV.

A common simple, axisymmetric model for the gas-density distribution in clusters, adapted to X-ray observations, is
\begin{equation}
  n(x)=\frac{n_0}{(1+x^2)^{3\beta/2}}\;,\quad
  x\equiv\frac{r}{r_\mathrm{c}}=\frac{\theta}{\theta_\mathrm{c}}\;,
\label{eq:05-23}
\end{equation}
where $r_\mathrm{c}$ and $\theta_\mathrm{c}$ are the physical and angular core radii. The line-of-sight projection of the X-ray emissivity yields the X-ray surface brightness as a function of the angular radius $\theta$,
\begin{equation}
  S_\mathrm{X}(\theta)=\int_{-\infty}^\infty j_\mathrm{X}\d z=
  \frac{\sqrt{\pi}\Gamma(3\beta-1/2)}{\Gamma(3\beta)}\,
  \frac{Z^2g_\mathrm{ff}C_\mathrm{X}r_\mathrm{c}n_0^2\sqrt{T}}
{\left[1+x^2\right]^{3\beta-1/2}}\;,
\label{eq:05-24}
\end{equation}
where we have combined (\ref{eq:05-21}) and (\ref{eq:05-23}) and assumed for simplicity that the cluster is isothermal, so $T$ does not change with radius. The latter equation shows that two parameters of the density profile (\ref{eq:05-23}), namely the slope $\beta$ and the (angular) core radius $\theta_\mathrm{c}$, can be read off the observable surface-brightness profile.

The missing normalisation constant can be obtained from the X-ray luminosity,
\begin{equation}
  L_\mathrm{X}=4\pi\int_0^\infty j_\mathrm{X} r^2\d r=
  4\pi r_\mathrm{c}^3Z^2g_\mathrm{ff}C_\mathrm{X}\sqrt{T}n_0^2
  \frac{\sqrt{\pi}\Gamma(3\beta-3/2)}{4\Gamma(3\beta)}\;,
\label{eq:05-25}
\end{equation}
and a spectral determination of the temperature $T$. Finally, the total mass of the X-ray gas enclosed in spheres of radius $R$ is
\begin{equation}
  M_\mathrm{X}(R)=4\pi\int_0^Rn(r)r^2\d r\;,
\label{eq:05-26}
\end{equation}
which is a complicated function for general $\beta$. For $\beta=2/3$, which is a commonly measured value,
\begin{equation}
  M_\mathrm{X}(R)=4\pi r_\mathrm{c}^3n_0\left(
    \frac{R}{r_\mathrm{c}}-\arctan\frac{R}{r_\mathrm{c}}
  \right)\;,
\label{eq:05-27}
\end{equation}
which is of course formally divergent for $R\to\infty$ because the density falls off $\propto r^{-2}$ for $\beta=2/3$ and $r\to\infty$.

Inserting typical numbers, we set $Z=1=g_\mathrm{ff}$, $\beta=2/3$ and assume a hypothetic cluster with $L_\mathrm{X}=10^{45}\,\mathrm{erg\,s^{-1}}$, a temperature of $kT=10\,\mathrm{keV}$ and a core radius of $r_\mathrm{c}=250\,\mathrm{kpc}=7.75\times10^{23}\,\mathrm{cm}$. Then, (\ref{eq:05-25}) yields the central ion density
\begin{equation}
  n_0=5\times10^{-3}\,\mathrm{cm^{-3}}
\label{eq:05-28}
\end{equation} 
and thus the central gas mass density
\begin{equation}
  \rho_0=m_\mathrm{p}n_0=8.5\times10^{-27}\,\mathrm{g\,cm^{-3}}\;.
\label{eq:05-29}
\end{equation}

Based on the virial radius (\ref{eq:05-19}) and on the mass (\ref{eq:05-27}), we find the total gas mass
\begin{equation}
  M_\mathrm{X}\approx1.0\times10^{14}\,M_\odot\;.
\label{eq:05-30}
\end{equation}
This is of the same order as the cluster mass in galaxies, and approximately one order of magnitude less than the total cluster mass.

It is reasonable to believe that clusters are closed systems in the sense that there cannot have been much material exchange between their interior and their surroundings \citep{WH93.1}. If this is indeed the case, and the mixture between dark matter and baryons in clusters is typical for the entire universe, the matter-density parameter should be
\begin{equation}
  \Omega_\mathrm{m0}\approx
  \Omega_\mathrm{b0}\,\frac{M}{M_*+M_\mathrm{X}}\approx
  10\Omega_\mathrm{b0}\approx0.4\;,
\label{eq:05-31}
\end{equation}
given the determination (\ref{eq:04-27}) of $\Omega_\mathrm{b0}$ from primordial nucleosynthesis. More precise estimates based on detailed investigations of individual clusters yield
\begin{equation}
  \Omega_\mathrm{m0}\approx0.3\;.
\label{eq:05-32}
\end{equation}

\subsubsection{Alternative cluster mass estimates}\label{sec:V-B-3}

Cluster masses can be estimated in several other ways. One of them is directly related to the X-ray emission discussed above. The hydrostatic Euler equation for a gas sphere in equilibrium with the spherically-symmetric gravitational potential of a mass $M(r)$
requires
\begin{equation}
  \frac{1}{\rho}\frac{\d p}{\d r}=-\frac{GM(r)}{r^2}\;,
\label{eq:05-33}
\end{equation} 
where $\rho$ and $p$ are the gas density and pressure, respectively. Assuming an ideal gas, the equation of state is $p=nkT$, where $n=\rho/(\mu m_\mathrm{p})$ is the particle density and $T$ the temperature. $\mu$ is the mean particle mass in units of the proton mass $m_\mathrm{p}$. A mixture of hydrogen and helium with helium fraction $Y$ has $\mu=4/(8-5Y)$, or $\mu\approx0.59$ for $Y=0.25$. If we further simplify the problem assuming an isothermal gas distribution, we can write
\begin{equation}
  \frac{kT}{\mu m_\mathrm{p}\rho}\frac{\d\rho}{\d r}=-\frac{GM(r)}{r^2}
\label{eq:05-34}
\end{equation}
or, solving for the mass
\begin{equation}
  M(r)=-\frac{rkT}{G\mu m_\mathrm{p}}\frac{\d\ln\rho}{\d\ln r}\;.
\label{eq:05-35}
\end{equation}

For the $\beta$ model introduced in (\ref{eq:05-23}), the logarithmic density slope is
\begin{equation}
  \frac{\d\ln\rho}{\d\ln r}=\frac{\d\ln n}{\d\ln r}=
  -3\beta\frac{x^2}{1+x^2}\;,
\label{eq:05-36}
\end{equation}
thus the cluster mass is determined from the slope of the X-ray surface brightness and the cluster temperature,
\begin{equation}
  M(r)=\frac{3\beta rkT}{G\mu m_\mathrm{p}}\frac{x^2}{1+x^2}\;.
\label{eq:05-37}
\end{equation}
With the typical numbers used before, i.e.~$R\approx2.5\,\mathrm{Mpc}$, $\beta\approx2/3$ and $kT=10\,\mathrm{keV}$, the X-ray mass estimate gives
\begin{equation}
  M(R)\approx1.8\times10^{15}\,M_\odot\;,
\label{eq:05-38}
\end{equation} 
in reassuring agreement with the mass estimate (\ref{eq:05-20}) from galaxy kinematics, and again an order of magnitude more than either the gas mass or the stellar mass.

A third, completely independent way of measuring cluster masses is provided by gravitational lensing. Without going into any detail, we mention here that it can generate image distortions from which the projected lensing mass distribution can be reconstructed. Mass estimates obtained in this way by and large confirm those from X-ray emission and galaxy kinematics, although interesting discrepancies exist in detail \citep{CY04.1,HO07.1}.

None of the cluster mass estimates is particularly reliable because they are all to some degree based on stability and symmetry assumptions. For mass estimates based on galaxy kinematics, for instance, assumptions have to be made on the shape of the galaxy orbits, the symmetry of the gravitational potential and the mechanical equilibrium between orbiting galaxies and the body of the cluster. Numerous assumptions also enter X-ray based mass determinations, such as hydrostatic equilibrium, spherical symmetry and, in some cases, isothermality of the intracluster gas. Gravitational lensing does not need any equilibrium assumption, but inferences from strongly distorted images depend very sensitively on the assumed symmetry of the mass distribution, and measure only the mass enclosed by the lensed images.

\subsection{Mass density from cluster evolution}\label{sec:V-C}

A conceptually very interesting constraint on the cosmic mass density is based on the evolution of cosmic structures. Abell's cluster catalogue \citep{AB89.1} covers the redshift range $0.02\lesssim z\lesssim0.2$, which encloses a volume of $\approx9\times10^8\,\mathrm{Mpc}^3$. There are 1894 clusters in that volume, which yields an estimate for the spatial cluster density of
\begin{equation}
  n_\mathrm{c}\approx2\times10^{-6}\,\mathrm{Mpc}^{-3}\;.
\label{eq:05-39}
\end{equation}

It is a central assumption in cosmology that structures formed from Gaussian random density fluctuations. The spherical collapse model then asserts that gravitationally bound objects of mass $M$ form where the linear density contrast, smoothed within spheres of radius $R$ enclosing the average mass $M$, exceeds a critical threshold of $\delta_\mathrm{c}\approx1.686$, quite independent of cosmology. The probability for this to happen in a Gaussian random field with a standard deviation $\sigma_R(z)$ is
\begin{equation}
  p_\mathrm{c}(z)=\frac{1}{2}\,\mathrm{erfc}\left(
    \frac{\delta_\mathrm{c}}{\sqrt{2}\sigma_R(z)}
  \right)\;,
\label{eq:05-40}
\end{equation} 
where
\begin{equation}
  \sigma_R(z)=\sigma_{R0}D_+(z)
\label{eq:05-41}
\end{equation}
because the linear growth of the density contrast is determined by the growth factor, for which a fitting formula was given in (\ref{eq:01-25}).

Now, the present-day standard deviation $\sigma_{R0}$ must be chosen such as to reproduce the observed number density of clusters given in (\ref{eq:05-39}). The fraction of cosmic matter contained in clusters is approximated by
\begin{equation}
  p_\mathrm{c}'=\frac{Mn_\mathrm{c}}{\rho_\mathrm{cr0}\Omega_\mathrm{m0}}
  \approx3\times10^{-3}\Omega_\mathrm{m0}^{-1}\;.
\label{eq:05-42}
\end{equation}
The standard deviation $\sigma$ in (\ref{eq:05-40}) must now be chosen such that this number is reproduced, which yields
\begin{equation}
  \sigma_{R0}\approx\begin{cases}
    0.61 & \Omega_\mathrm{m0}=1.0 \\
    0.72 & \Omega_\mathrm{m0}=0.3 \\
                 \end{cases}\;.
\label{eq:05-43}
\end{equation}

Equations (\ref{eq:05-40}) and (\ref{eq:05-41}) can now be used to estimate how the cluster abundance should change with redshift. Simple evaluation reveals that the comoving cluster abundance is expected to drop very rapidly with increasing redshift if $\Omega_\mathrm{m0}$ is high, and much more slowly if $\Omega_\mathrm{m0}$ is low. Qualitatively, this behaviour is easily understood. If, in a low-density universe, clusters do not form early, they cannot form at all because the rapid expansion due to the low matter density prevents them from growing late in the cosmic evolution. From the observed slow evolution of the cluster population as a whole, it can be concluded that the matter density must be low. Estimates arrive at
\begin{equation}
  \Omega_\mathrm{m0}\approx0.3\;,
\label{eq:05-44}
\end{equation}
in good agreement with the preceding determinations.

\subsection{Conclusions}\label{sec:V-D}

What does it all mean? The preceding discussion should have demonstrated that the matter density in the Universe is considerably less than its critical value, approximately one third of it. Since only a small fraction of this matter is visible, we call the invisible majority \emph{dark matter}.

What is this dark matter composed of? Can it be baryons? Tight limits are set by primordial nucleosynthesis, which predicts that the density in baryonic matter should be $\Omega_\mathrm{B}\approx0.04$, cf.~(\ref{eq:04-27}). In the framework of the Friedmann-Lema{\^\i}tre models, the baryon density in the Universe can be higher than this only if baryons are locked up in some way before nucleosynthesis commences. They could form black holes before, but their mass is limited by the mass enclosed within the horizon at, say, up to one minute after the Big Bang. According to (\ref{eq:04-7}), the scale factor at this time was $a\approx10^{-10}$, and thus the matter density was of order $\rho_\mathrm{m}\approx10^{30}\rho_\mathrm{cr0}\approx 10\,\mathrm{g\,cm^{-3}}$. The horizon size is $r_\mathrm{H}\approx ct\approx1.8\times10^{12}\,\mathrm{cm}$, thus the mass enclosed by the horizon is $\approx3\times10^4M_\odot$, which limits possible black-hole masses from above.

Gravitational microlensing was used to constrain the amount of dark, compact objects of subsolar mass in the halo of the Milky Way. Although they were found to contribute part of the mass, they can certainly not account for all of it \citep{AL00.1,LA00.1}. Black holes with masses of the order $10^6\,M_\odot$ should have been found through their microlensing effects \citep{WA92.1}. The abundance of lower-mass black holes is limited by stellar-dynamics arguments, in particular by the presence of cold disks \citep{KL96.1,RI93.1}.

We are thus guided to the conclusion that the dark matter is most probably \textit{not baryonic} and \textit{not composed of compact dark objects}. We shall see in \S~\ref{sec:VI-B-2} that and why the most favoured hypothesis now holds that it is composed of weakly interacting massive particles. 

Neutrinos, however, are ruled out because their total mass has been measured to be way too low ($<2\,\mathrm{eV}$\footnote{Particle Data Group, http://pdg.lanl.gov/}). Their correspondingly high velocities require objects more massive than galaxies to keep them bound. The formation of galaxies would be much delayed in this case until larger-scale objects could fragment into smaller pieces. This is the opposite of the observed hierarchy of cosmic structure formation, which clearly shows that galaxies appear substantially earlier than galaxy clusters. Therefore, the conclusion seems inevitable that the dark matter must be cold, i.e.~consisting of particles moving non-relativistically \citep{PE82.1}. A sequence of numerical simulations has shown that structure formation in a universe filled with cold dark matter can be brought into agreement with the observed large-scale cosmic structures \citep{DA85.1,WH87.1}.

\section{The Cosmic Microwave Background}\label{sec:VI}

\subsection{The isotropic CMB}\label{sec:VI-A}

\subsubsection{Thermal history of the Universe}\label{sec:VI-A-1}

How does the Universe evolve thermally? We have seen that the abundance of $\nuc{4}{He}$ shows that the Universe must have gone through an early phase which was hot enough for the nuclear fusion of light elements. But was there thermal equilibrium? Thus, can we speak of the ``temperature of the Universe''?

From isotropy, we must conclude that the Universe expanded \emph{adiathermally}: no net heat can have flowed between any two volume elements in the Universe because any flow would have singled out a direction, which is forbidden by isotropy. An \emph{adiathermal} process is \emph{adiabatic} if it proceeds slow enough for equilibrium to be maintained. Then, it is also \emph{reversible} and \emph{isentropic}. Of course, irreversible processes must have occurred during the evolution of the Universe. However, as we shall see later, the entropy in the Universe is so overwhelmingly dominated by the photons of the microwave background radiation that no entropy production by irreversible processes can have added a significant amount of entropy. Thus, we assume that the Universe has in fact expanded \emph{adiabatically}.

As the Universe expands and cools, particles are diluted and interaction rates drop, so thermal equilibrium must break down at some point for any particle species because collisions become too rare. Very early in the Universe, however, the expansion rate was very high, and it is important to check whether thermal equilibrium can have been maintained \emph{despite} the rapid expansion. The collision probability between any two particle species will be proportional to their number densities squared, $\propto n^2$, because collisions are dominated by two-body encounters. The collision \emph{rate}, i.e.~the number of collisions experienced by an individual particle with others will be $\propto n$, which is $\propto a^{-3}$ for non-relativistic particles. Thus, the collision time scale was $\tau_\mathrm{coll}\propto a^3$.

According to Friedmann's equation, the expansion rate in the very early Universe was determined by the radiation density, and thus proportional to $\propto\dot a/a\propto a^{-2}$, and the expansion time scale was $\tau_\mathrm{exp}\propto a^2$. Equilibrium could be maintained as long as the collision time scale was sufficiently shorter than the expansion time scale,
\begin{equation}
  \tau_\mathrm{coll}\ll\tau_\mathrm{exp}\;,
\label{eq:06-1}
\end{equation} 
which turns out to be well satisfied in the early Universe when $a\ll1$. Thus, even though the expansion rate was very high in the early Universe, the collision rates were even higher, and thermal equilibrium could indeed be maintained.

The final assumption is that the components of the cosmic fluid behave as ideal gases. By definition, this requires that their particles interact only via short-ranged forces, which implies that partition sums can be written as powers of one-particle partition sums and that the internal energy of the fluids does not depend on the volume occupied. This is a natural assumption which holds even for charged particles because they shield opposite charges on length scales small compared to the size of the observable universe.

It is thus well justified to assume that there was thermal equilibrium between all particle species in the early universe, that the constituents of the cosmic ``fluid'' can be described as ideal gases, and that the expansion of the universe can be seen as an adiabatic process. In later stages of the cosmic evolution, particle species will drop out of equilibrium when their interaction rates fall below the expansion rate of the Universe. As long as all species in the Universe are kept in thermodynamic equilibrium, there is a single temperature characterising the cosmic fluid. Once particle species drop out of thermal equilibrium because their interaction rates decrease, their temperatures, if defined, may begin deviating. Even then, we characterise the thermal evolution of the Universe by the temperature of the photon background.

\subsubsection{Mean properties of the CMB}\label{sec:VI-A-2}

As discussed before, the CMB had been predicted in order to explain the abundance of the light elements, in particular of $\nuc{4}{He}$ \citep{GA48.1,AL49.1}. It was serendipitously discovered by Penzias and Wilson in 1965 \citep{PE65.1}. Measurements of the energy density in this radiation were mostly undertaken at long (radio) wavelengths, i.e.~in the Rayleigh-Jeans part of the CMB spectrum. To firmly establish that the spectrum is indeed the Planck spectrum expected for thermal black-body radiation, the FIRAS experiment was placed on-board the COBE satellite, where it measured the best realisation of a Planck spectrum ever observed \citep{MA94.1,FI02.1}.

The temperature of the Planck curve best fitting the latest measurement of the CMB spectrum by the COBE satellite is
\begin{equation}
  T_0=2.726\,\mathrm{K}\;,
\label{eq:06-2}
\end{equation}
which implies the mean number density $n_\mathrm{CMB}=405\,\mathrm{cm}^{-3}$ of CMB photons and the energy density $u_\mathrm{CMB}=4.17\times10^{-13}\,\mathrm{erg\,cm^{-3}}$ in the CMB, equivalent to a mass density of $\rho_\mathrm{r, CMB}=4.63\times10^{-34}\,\mathrm{g\,cm^{-3}}$. Using (\ref{eq:01-14}), the density parameter in radiation is
\begin{equation}
  \Omega_\mathrm{r0}=8.51\times10^{-5}\;,
\label{eq:06-3}
\end{equation} 
which shows that the scale factor at matter-radiation equality was
\begin{equation}
  a_\mathrm{eq}=\frac{\Omega_\mathrm{m0}}{\Omega_\mathrm{r0}}\approx
  \frac{1}{3280}\approx3.0\times10^{-4}\;.
\label{eq:06-4}
\end{equation}
This calculation includes three neutrino species besides the photons, which were relativistic at the time of matter-radiation equality.

The number density of baryons in the Universe is approximately
\begin{equation}
  n_\mathrm{B}\approx
  \frac{\Omega_\mathrm{B0}\rho_\mathrm{cr}}{m_\mathrm{p}}\approx
  2.3\times10^{-7}\,\mathrm{cm}^{-3}\;,
\label{eq:06-5}
\end{equation} 
confirming that the photon-to-baryon ratio is extremely high,
\begin{equation}
  \eta^{-1}\approx\frac{405}{2.3\times10^{-7}}\approx
  1.8\times10^9\;.
\label{eq:06-6}
\end{equation}

\subsubsection{Decoupling of the CMB}\label{sec:VI-A-3}

When and how was the CMB set free? While the Universe was sufficiently hot to keep electrons and protons separated (we neglect heavier elements here for simplicity), the photons scattered off the charged particles, their mean free path was short, and the photons could not propagate. When the Universe cooled below a certain temperature, electrons and protons combined to form hydrogen, free charges disappeared, the mean free path became virtually infinite and photons could freely propagate.

The recombination reaction
\begin{equation}
  p+e^-\leftrightarrow H+\gamma
\label{eq:06-7}
\end{equation}
can thermodynamically be described by Saha's equation for the ionisation fraction $x$,
\begin{equation}
  \frac{x^2}{1-x}=\frac{\sqrt{\pi}}{4\sqrt{2}\zeta(3)\eta}\left(
  \frac{m_\mathrm{e}c^2}{kT}
  \right)^{3/2}\e^{-\chi/kT}\approx
  \frac{0.26}{\eta}\left(
  \frac{m_\mathrm{e}c^2}{kT}
  \right)^{3/2}\e^{-\chi/kT}\;,
\label{eq:06-8}
\end{equation}
where $\chi$ is the ionisation energy of hydrogen, $\chi=13.6\,\mathrm{eV}$, and $\zeta$ is the Riemann Zeta function.

Notice that Saha's equation contains the inverse of the $\eta$ parameter (\ref{eq:06-6}), which is a huge number due to the high photon-to-baryon ratio in the Universe. This counteracts the
exponential which would otherwise guarantee that recombination happens when $kT\approx\chi$, i.e.~at $T\approx1.6\times10^5\,\mathrm{K}$. Recombination is thus delayed by the high photon number, which illustrates that newly formed hydrogen atoms are effectively reionised by sufficiently energetic photons until the temperature has dropped well below the ionsation energy. Putting $x\approx0.5$ in (\ref{eq:06-8}) yields a recombination temperature of
\begin{equation}
  kT_\mathrm{rec}\approx0.3\,\mathrm{eV}\;,\quad
  T_\mathrm{rec}\approx3000\,\mathrm{K}
\label{eq:06-9}
\end{equation}
and thus a recombination redshift of $z_\mathrm{rec}\approx1100$. Since this is in the early matter-dominated era, the age of the Universe was then
\begin{eqnarray}
  t&=&\int_0^{a_\mathrm{rec}}\frac{\d a}{aH(a)}=
  \frac{2a_\mathrm{rec}^{3/2}}{3H_0\sqrt{\Omega_\mathrm{m0}}}\left[
    \sqrt{1+\alpha}(1-2\alpha)+2\alpha^{3/2}
  \right]\nonumber\\
  &\approx& 374\,\mathrm{kyr}\;,
\label{eq:06-10}
\end{eqnarray}
cf.~Eq.~(\ref{eq:01-15}), where $\alpha:=a_\mathrm{eq}/a_\mathrm{rec}\approx0.33$.

Recombination does not proceed instantaneously. The ionisation fraction $x$ drops from $0.9$ to $0.1$ within a temperature range of approximately $200\,\mathrm{K}$, corresponding to a redshift range of
\begin{equation}
  \Delta z\approx\left.
    \frac{\d z}{\d T}
  \right|_{z_\mathrm{rec}}\Delta T
  \approx\left.
    \frac{\d}{\d T}\left(\frac{T}{T_0}-1\right)
  \right|_{z_\mathrm{rec}}\Delta T\approx
  \frac{\Delta T}{T_0}\approx75
\label{eq:06-11}
\end{equation} 
or a time interval of
\begin{equation}
  \Delta t\approx\frac{\Delta a}{aH}\approx
  \frac{\Delta z}{H_0\sqrt{\Omega_\mathrm{m0}}(1+z)^{5/2}}\approx
  50\,\mathrm{kyr}\;.
\label{eq:06-12}
\end{equation}
We are thus led to conclude that the CMB was released when the Universe was approximately 374,000 years old, during a phase that lasted approximately 50,000 years. We have derived this result merely using the present temperature of the CMB, the photon-to-baryon ratio, the Hubble constant and the matter density parameter $\Omega_\mathrm{m0}$. The cosmological constant or a possible curvature of the Universe do not matter here.

\subsection{Structures in the CMB}\label{sec:VI-B}

\subsubsection{The dipole}\label{sec:VI-B-1}

The Earth is moving around the Sun, the Sun is orbiting around the Galactic centre, the Galaxy is moving within the Local Group, which is falling towards the Virgo cluster of galaxies. We can thus not expect the Earth to be at rest with respect to the CMB. We denote the net velocity of the Earth with respect to the CMB rest frame by $v_\oplus$. Lorentz transformation shows that, to lowest order in $v_\oplus/c$, the Earth's motion imprints a dipolar intensity pattern on the CMB with an amplitude of
\begin{equation}
  \frac{\Delta T}{T_0}=\frac{v_\oplus}{c}\;.
\label{eq:06-13}
\end{equation} 
The dipole's amplitude has been measured to be $\approx1.24\,m\mathrm{K}$, from which the Earth's velocity is inferred to be \citep{FI96.1}
\begin{equation}
  v_\oplus\approx(371\pm1.5)\,\mathrm{km\,s^{-1}}\;.
\label{eq:06-14}
\end{equation}
This is the highest-order deviation from isotropy in the CMB, but it is irrelevant for our purposes since it does not allow any conclusions on the Universe at large.

\subsubsection{Expected amplitude of CMB fluctuations}\label{sec:VI-B-2}

It is reasonable to expect that density fluctuations in the CMB should match density fluctuations in the matter because photons were tightly coupled to baryons by Thomson scattering before recombination. For adiabatic fluctuations, the density contrast $\delta$ in matter is $3/4$ that in radiation. Since the radiation density is $\propto T^4$, a density contrast $\delta$ is thus expected to produce relative temperature fluctuations of order
\begin{equation}
  \delta=\frac{3}{4}\,\frac{4T^3\delta T}{T^4}\quad\Rightarrow\quad
  \frac{\delta T}{T}\approx\frac{\delta}{3}\;.
\label{eq:06-15}
\end{equation}
Obviously, there are large-scale structures in the Universe today whose density contrast reaches or even substantially exceeds unity. Assuming linear structure growth on scales where $\delta_0\approx1$, and knowing the scale factor at recombination, we can thus infer that relative temperature fluctuations of order
\begin{equation}
  \frac{\delta T}{T}\approx\frac{1}{3}\left(
    \frac{\delta_0}{D_+(a_\mathrm{rec})}
  \right)\approx\frac{1}{3a_\mathrm{rec}}\approx10^{-3}
\label{eq:06-16}
\end{equation}
should be seen in the CMB, i.e.~fluctuations of order $m\mathrm{K}$, similar to the dipole. Such fluctuations, however, were not observed, although cosmologists kept searching increasingly desperately for decades after 1965 (see \cite{US84.1} for an example).

Why do they not exist? The estimate above is valid only under the assumption that matter and radiation were tightly coupled. Should this not have been the case, density fluctuations did not need to leave such a pronounced imprint on the CMB. In order to avoid the tight coupling, the majority of matter must not interact electromagnetically. Thus, we conclude from the absence of $m\mathrm{K}$ fluctuations in the CMB that matter in the Universe must be dominated by something that does not interact with light \citep{PE82.1}. This is perhaps the strongest argument in favour of not-electromagnetically interacting dark matter.

\subsubsection{Expected CMB fluctuations}\label{sec:VI-B-3}

Before we come to the results of CMB observations and their significance for cosmology, let us summarise which physical effects we expect to imprint structures on the CMB \citep{PE70.1,SU70.1,BO84.1,HU96.1,SE96.1}. The basic hypothesis is that the cosmic structures that we see today formed via gravitational instability from seed fluctuations in the early Universe, whose inflationary origin is likely, but yet unclear. This implies that there had to be density fluctuations at the time when the CMB was released. Via Poisson's equation, these density fluctuations were related to fluctuations in the Newtonian potential. Photons released in a potential fluctuation $\delta\Phi$ lost energy if the fluctuation was negative, and gained energy when the fluctuation was positive. This energy change can be translated to the temperature change
\begin{equation}
  \frac{\delta T}{T}=\frac{1}{3}\frac{\delta\Phi}{c^2}\;,
\label{eq:06-17}
\end{equation}
which is called the Sachs-Wolfe effect \citep{SA67.1}. The factor $1/3$ in front is caused by the gravitational redshift being offset by time retardation in the gravitational field of the perturbation.

Let us briefly look into the expected statistics of the Sachs-Wolfe effect. We introduced the power spectrum of the density fluctuations in (\ref{eq:01-27}) as the variance of the density
contrast in Fourier space. Poisson's equation implies
\begin{equation}
  \delta\hat\Phi\propto-\frac{\hat\delta}{k^2}\;,
\label{eq:06-18}
\end{equation}
for the Fourier modes of the gravitational potential. Thus the power spectrum of the temperature fluctuations due to the Sachs-Wolfe effect is
\begin{equation}
  P_T\propto P_\Phi\propto
  \frac{\langle\hat\delta\hat\delta^*\rangle}{k^4}\propto
  \frac{P_\delta}{k^4}\propto
  \begin{cases}
    k^{-3} & k\ll k_\mathrm{eq}\\
    k^{-7} & k\gg k_\mathrm{eq}\\
  \end{cases}
\label{eq:06-19}
\end{equation} 
according to (\ref{eq:01-28}) with $n_\mathrm{s}=1$. This shows that the Sachs-Wolfe effect can only be important at small $k$, i.e.~on large scales, and dies off quickly towards smaller scales.

The main constituents of the cosmic fluid were dark matter, baryons, electrons and photons. Overdensities in the dark matter compressed the fluid due to their gravity until the rising pressure in the tightly coupled baryon-electron-photon fluid was able to counteract gravity and drive the fluctuations apart. In due course, the pressure sank and gravity won again, and so forth: the baryon-electron-photon fluid underwent acoustic oscillations, but not the dark matter, which was decoupled. Since the pressure was dominated by the photons, whose pressure is a third of their energy density, a good approximation to the sound speed of the tightly coupled photon-baryon fluid was
\begin{equation}
  c_\mathrm{s}\approx\sqrt{\frac{p}{\rho}}=\frac{c}{\sqrt{3}}\approx
  0.58\,c\;.
\label{eq:06-20}
\end{equation}
Only such density fluctuations could undergo acoustic oscillations which were small enough to be crossed by sound waves in the available time. The largest comoving length that could be travelled by sound waves was the \emph{comoving sound horizon}
\begin{equation}
  w_\mathrm{s}=\int_0^{t_\mathrm{rec}}\frac{c_\mathrm{s}\d t}{a}=
  \frac{2c_\mathrm{s}\sqrt{a_\mathrm{rec}}}{H_0\sqrt{\Omega_\mathrm{m0}}}\left(
    \sqrt{1+\alpha}-\sqrt{\alpha}
  \right)=163.3\,\mathrm{Mpc}\;.
\label{eq:06-21}
\end{equation}
Larger-scale density fluctuations could not oscillate. We saw in (\ref{eq:01-18}) and (\ref{eq:01-19}) that the \textit{comoving} angular-diameter distance from today to scale factor $a_\mathrm{rec}$ is $f_K[w(a_\mathrm{rec})]=w(a_\mathrm{rec})$ if we assume spatial flatness, $K=0$. In excellent approximation,
\begin{equation}
  f_K[w(a_\mathrm{rec})]=w(a_\mathrm{rec})=
  3.195\frac{c}{H_0}\left(\frac{\Omega_\mathrm{m0}}{0.3}\right)^{-0.4}\;.
\label{eq:06-22}
\end{equation}
Thus, the sound horizon sets an angular scale of
\begin{equation}
  \theta_\mathrm{s}=\frac{w_\mathrm{s}}{w(a_\mathrm{rec})}=0.66^\circ\;.
\label{eq:06-23}
\end{equation}
This angular scale can be read off the first acoustic peak in the CMB power spectrum (see \S~\ref{sec:VI-B-5} below). Its relation to the physical sound horizon (\ref{eq:06-21}) depends almost precisely on the \textit{sum} of $\Omega_\mathrm{m0}$ and $\Omega_\mathrm{\Lambda0}$ if all else remains fixed. Thus, the location of the first acoustic peak determines the spatial curvature of the cosmological model. When combined with measurements of $H_0$, latest data (cf.~Tab.~\ref{tab:1}) show that $K=0$ to high precision.

A third effect influencing structures in the CMB is caused by the fact that, as recombination proceeds, the mean-free path of the photons increases. If $n_\mathrm{e}=xn_\mathrm{B}$ is the electron number density and $\sigma_\mathrm{T}$ is the Thomson cross section, the comoving mean-free path is
\begin{equation}
  \lambda\approx(xn_\mathrm{B}\sigma_\mathrm{T})^{-1}\;.
\label{eq:06-24}
\end{equation}
As the ionisation fraction $x$ drops towards zero, the mean-free path approaches infinity. After $N$ scatterings, the photons will have diffused by
\begin{equation}
  \lambda_\mathrm{D}\approx\sqrt{N}\lambda
\label{eq:06-25}
\end{equation}
on average. The number of scatterings per unit time is
\begin{equation}
  \d N\approx xn_\mathrm{B}\sigma_\mathrm{T}c\d t\;,
\label{eq:06-26}
\end{equation}
and thus the diffusion scale is given by
\begin{equation}
  \lambda_\mathrm{D}^2\approx\int\lambda^2\d N\approx
  \int\frac{c\d t}{xn_\mathrm{B}\sigma_\mathrm{T}}\;.
\label{eq:06-27}
\end{equation}
Inserting $x\approx1/2$ as a typical value into the latter integrand, we can approximate
\begin{equation}
  \lambda_\mathrm{D}^2\approx
  \frac{2c\Delta t}{n_\mathrm{B}\sigma_\mathrm{T}}\;.
\label{eq:06-28}
\end{equation}
Numerical evaluation returns a comoving damping length of order $50\,\mathrm{Mpc}$, corresponding to angular scales of $\theta_\mathrm{D}\approx10'$ on the sky. The effect of this diffusion process is called \emph{Silk damping} \citep{SI68.1}. We thus expect three principal mechanisms to shape the appearance of the microwave sky: the Sachs-Wolfe effect on the largest scales, the acoustic oscillations on scales smaller than the sound horizon, and Silk damping on scales smaller than a few arc minutes.

\subsubsection{CMB polarisation}\label{sec:VI-B-4}

If the CMB does indeed arise from Thomson scattering, interesting effects must occur because the Thomson scattering cross section is polarisation sensitive, and can thus convert unpolarised into linearly polarised radiation. Suppose an electron is illuminated by unpolarised radiation from the left, then the radiation scattered towards the observer will be linearly polarised in the perpendicular direction. Likewise, unpolarised radiation incoming from the top will be linearly polarised horizontally after being scattered towards the observer. Thus, if the electron is irradiated by a quadrupolar intensity pattern, the scattered radiation will be partially linearly polarised. The polarised intensity is expected to be of order 10\% of the total intensity. The polarised radiation must reflect the same physical effects as the unpolarised radiation, and the two must be cross-correlated. Much additional information on the physical state of the early Universe is thus contained in the polarised component of the CMB, apart from the fact that the mere detection of the polarisation supports the physical picture of the CMB's origin.

\subsubsection{The CMB power spectrum}\label{sec:VI-B-5}

Fourier transformation is impossible on the sphere, but the analysis of the CMB proceeds in a completely analogous way by decomposing the relative temperature fluctuations into spherical harmonics, finding the spherical-harmonic coefficients
\begin{equation}
  a_{lm}=\int\d^2\theta\,\frac{\delta T}{T}\,Y_{lm}(\vec\theta)\;,
\label{eq:06-29}
\end{equation}
and from them the power spectrum
\begin{equation}
  C_l\equiv\frac{1}{2l+1}\sum_{m=-l}^l|a_{lm}|^2\;,
\label{eq:06-30}
\end{equation}
which is the equivalent on the sphere to the three-dimensional power spectra defined in (\ref{eq:01-27}). The average over $m$ expresses the expectation of statistical isotropy. If parts of the observed sky need to be masked, care has to be taken to orthonormalise the spherical harmonics on the remaining domain.

The shape of the CMB power spectrum reflects the three physical mechanisms identified above: at small $l$ (on large scales), the Sachs-Wolfe effect causes a feature-less plateau, followed by pronounced maxima and minima due to the acoustic oscillations, damped on the smallest scales (largest $l$) by Silk damping.

\begin{figure}[ht]
  \includegraphics[width=\hsize]{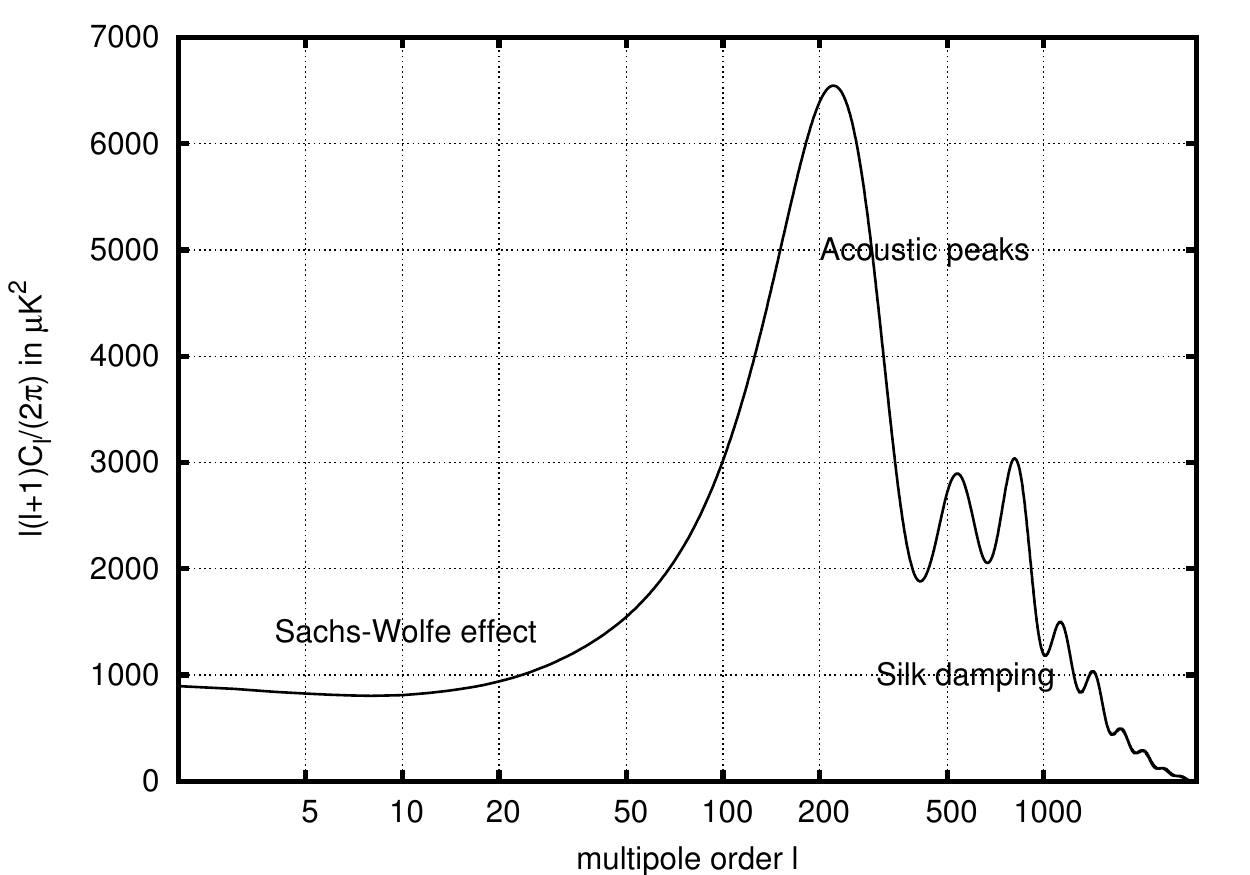}
\caption{Schematic appearance of the CMB power spectrum due to the three dominating physical effects defining its shape: the Sachs-Wolfe effect on largest scales, the Silk damping on smallest scales, and the acoustic oscillations in between.}
\label{fig:06-1}
\end{figure}

The detailed shape of the CMB power spectrum depends sensitively on the cosmological parameters, which can in turn be determined by fitting the theoretically expected to the measured $C_l$. This is the main reason for the detailed and sensitive CMB measurements pioneered by COBE, continued by ground-based and balloon experiments, and culminating recently in the spectacular results obtained by the WMAP satellite.

\subsubsection{Secondary anisotropies and foregrounds}\label{sec:VI-B-6}

By definition, the CMB is the oldest visible source of photons because all possible earlier sources could not shine through the hot cosmic plasma. Therefore, every source that produced microwave photons since, or that produced photons which became redshifted into the microwave regime by now, must appear superposed on the CMB. The CMB is thus covered by layers of foreground emission that have to be unveiled before the CMB can be observed.

Broadly, the CMB foregrounds can be grouped into point sources and diffuse sources. The most important among the point sources are infrared galaxies at high redshift and bodies in the Solar System such as the major planets, but even some of the asteroids.

The population of infrared sources at high redshift is poorly known, but the angular resolution of CMB measurements has so far been too low to be significantly contaminated by them. Future CMB observations will have to remove them carefully \citep{TO98.1}.

Microwave radiation from bodies in the Solar System has so far been used to calibrate microwave detectors. CMB observations at an angular resolution below $\sim10'$ are expected to detect hundreds of minor planets.

Diffuse CMB foregrounds are mainly caused by our Galaxy itself. There are three main components: synchrotron emission, emission from warm dust, and \emph{bremsstrahlung} \citep{HA82.1,SC98.1,DA01.1,FI03.1}.

Synchrotron radiation is emitted by relativistic electrons in the Galaxy's magnetic field. It is highly linearly polarised and has a power-law spectrum falling steeply from radio towards microwave frequencies. It is centred on the Galactic plane, but shows filamentary extensions from the Galactic centre towards the Galactic poles.

The dust in the Milky Way is also concentrated in the Galactic plane. It is between $~(10\ldots20)\,\mathrm{K}$ warm and thus substantially warmer than the CMB itself. It has a thermal spectrum modified by frequency-dependent self absorption. Due to its higher temperature, the warm dust has a spectrum rising with increasing frequency in the frequency window where the CMB is usually observed.

\emph{Bremsstrahlung} radiation is emitted by ionised hydrogen clouds (HII regions) in the Galactic plane. Its spectrum has the shape typical for thermal free-free radiation, falling exponentially at photon energies near and above the gas temperature, but is flat at CMB frequencies. Further sources of microwave radiation in the Galaxy are less prominent. Among them are line emission from CO molecules embedded in cool gas clouds.

The falling synchrotron spectrum, the flat spectrum of the free-free radiation, and the rising spectrum of the warm dust create a window for CMB observations between $\sim(100\ldots200)\,\mathrm{GHz}$. The different spectra of the foregrounds, and their non-Planckian character, are crucial for their proper removal from the CMB data. Therefore, CMB measurements have to be carried out in multiple frequency bands.

Secondary anisotropies are caused by propagation effects rather than photon emission. The most important are the integrated Sachs-Wolfe effect and the (thermal and kinetic) Sunyaev-Zel'dovich effects in galaxy clusters.

The integrated Sachs-Wolfe effect is caused by gravitational potential wells deepening while crossed by CMB photons. It is determined by the line-of-sight integral of the time derivative of the potential fluctuations caused by the density fluctuations between us and the CMB. By cross-correlating CMB temperature fluctuations with structures in the galaxy distribution, the integrated Sachs-Wolfe effect has indeed been detected \citep{GI06.1,GA06.1}. It constrains the ratio $D_+(a)/a$.

The Sunyaev-Zel'dovich effect \citep{SU72.1} is due to inverse Compton scattering of CMB photons off hot electrons in the intracluster plasma. On average, the photons gain energy and are thus moved from the low- to the high-frequency part of the spectrum. When observed through a galaxy cluster, the CMB therefore appears fainter at low and brighter at high frequencies, with the transition at 217~GHz. This gives galaxy clusters a peculiar spectral signature in the microwave regime as they cast shadows below, emit above, and vanish at 217~GHz. Once the angular resolution of CMB detectors will drop towards a few arc minutes, a large number of galaxy clusters are expected to show up in this way. Besides this thermal Sunyaev-Zel'dovich effect, there is the \emph{kinetic} effect caused by the bulk motion of the cluster as a whole, which causes CMB radiation to be scattered by the electrons moving with the cluster. Very few clusters have so far been detected in CMB data, but thousands are expected to be found in future missions.

\subsubsection{Measurements of the CMB}\label{sec:VI-B-7}

Wien's law (\ref{eq:05-3}) shows that the CMB spectrum peaks at $\lambda_\mathrm{max}\approx0.11\,\mathrm{cm}$, or at a frequency of $\nu_\mathrm{max}\approx282\,\mathrm{GHz}$. As we saw, Silk damping sets in below $\sim10'$ arc minutes, thus most of the structures in the CMB are resolvable for rather small telescopes. According to the formula
\begin{equation}
  \Delta\theta\approx1.44\,\frac{\lambda}{D}
\label{eq:06-31}
\end{equation}
relating the diffraction-limited angular resolution $\Delta\theta$ to the ratio between wavelength and telescope diameter $D$, we find that mirrors with
\begin{equation}
  D\lesssim1.44\,\frac{\lambda_\mathrm{max}}{\theta_\mathrm{D}}\approx
  55\,\mathrm{cm}
\label{eq:06-32}
\end{equation}
are sufficient to achieve sufficient angular resolution up to the Silk-damping scale $\theta_\mathrm{D}$. Detectors are needed which are sensitive to millimetre and sub-mm radiation and reach $\mu\mathrm{K}$ sensitivity, while the telescope optics can be kept rather small and simple.

Two types of detector are commonly used. The first are bolometers, which measure the energy of the absorbed radiation by the temperature increase it causes. They have to be cooled to very low temperatures, typically in the $m\mathrm{K}$ regime. The second are so-called \emph{high electron mobility detectors} (HEMTs), in which the currents caused by the incoming electromagnetic field are measured directly. They measure amplitude and phase of the waves and are thus polarisation-sensitive by construction, which bolometers are not. Polarisation measurements with bolometers are possible with suitably shaped so-called \emph{feed horns} guiding the radiation into the detectors.

Since water vapour in the atmosphere both absorbs and emits microwave radiation through molecular lines, CMB observations need to be carried out either at high, dry and cold sites on the ground (e.g.~in the Chilean Andes or at the South Pole), or from balloons rising above the troposphere, or from satellites in space.

After the breakthrough achieved with COBE \citep{BO92.1}, progress was made with balloon experiments and with ground-based interferometers. The balloons covered a small fraction of the sky (typically $\sim1\%$) at frequencies between 90 and 400~GHz, while the interferometers observe even smaller fields at somewhat lower frequencies (typically around 30~GHz).

The first discovery of the CMB polarisation and its cross-correlation with the CMB temperature was achieved with the \emph{DASI} interferometer \citep{KO02.2}. The existence, location and height of the first acoustic peak had been firmly established \citep{WA03.2} before the NASA satellite \emph{Wilkinson Microwave Anisotropy Probe} (WMAP for short) was launched, but the increased sensitivity and the full-sky coverage of WMAP produced breath-taking results \citep{SP07.1}. WMAP is still operating, measuring the CMB temperature at frequencies between 23 and 94~GHz with an angular resolution of $\gtrsim15'$. The sensitivity of WMAP is barely high enough for polarisation measurements.

By now, data from the first five years of operation have been published, and cosmological parameters have been obtained fitting theoretically expected to the measured temperature-fluctuation power spectrum and the temperature-polarisation power spectrum. Results were given in Tab.~\ref{tab:1}, adapted from \citep{KO08.1}.

\begin{figure}[ht]
  \includegraphics[width=\hsize]{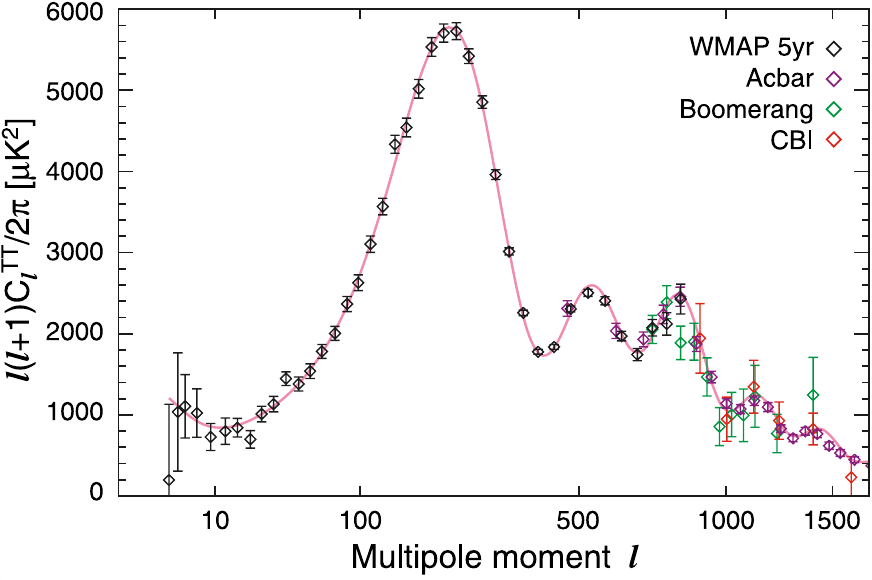}
\caption{Power spectrum of CMB temperature fluctuations as measured from the 5-year data of WMAP and several additional ground-based experiments. (Courtesy of the WMAP team)}
\label{fig:06-2}
\end{figure}

Although the CMB provides insight into the cosmological parameters on its own, it is most powerful when combined with other data sets. In particular, the Hubble constant is \emph{not} an independent measurement from the CMB alone. Only by \emph{assuming} a flat $\Lambda$CDM universe, it can be inferred from the location of the first acoustic peak in the CMB power spectrum to be $H_0=70.1\pm1.3\,\mathrm{km\,s^{-1}\,Mpc^{-1}}$, which agrees perfectly with the results of the Hubble Key Project and at least one interpretation of the gravitational-lens time delays.

A European CMB satellite mission is under way: ESA's \emph{Planck} satellite has been launched in May 2009. It will observe the microwave sky in nine frequency bands between 30 and 857 GHz with about ten times higher sensitivity than WMAP, and an angular resolution of $\gtrsim5'$ \citep{TA04.1}. Its wide frequency coverage will be very important for substantially improved foreground subtraction. It will also have sufficient sensitivity to precisely measure the CMB polarisation in some of its frequency bands \citep{LA03.1,VA07.1}. Moreover, it is expected that \emph{Planck} will detect a large number of galaxy clusters through their thermal Sunyaev-Zel'dovich effect.

\section{Cosmic Structures}\label{sec:VII}

\subsection{Quantifying structures}\label{sec:VII-A}

\subsubsection{Introduction}\label{sec:VII-A-1}

We have seen before that there is a very specific prediction for the power spectrum of cold dark matter density fluctuations in the Universe, characterised by (\ref{eq:01-28}). Recall that its shape was inferred from the simple assumption that the \textit{rms} mass of density fluctuations entering the horizon should be independent of the time when they enter the horizon, and from the fact that perturbation modes entering during the radiation era are suppressed until matter begins dominating. Given the simplicity of the argument, and the corresponding strength of the prediction, it is very important for cosmology to find out whether the actual power spectrum of matter density fluctuations does in fact have the expected \emph{shape}, and if it has, to determine the only remaining parameter, i.e.~the \emph{normalisation} of the power spectrum.

Since the comoving wave number $k_\mathrm{eq}$ of the peak in the power spectrum is set by the comoving horizon radius at matter-radiation equality (\ref{eq:01-26}),
\begin{equation}
  k_\mathrm{eq}\propto\frac{2\pi}{w_\mathrm{eq}}\propto
  \left(\frac{\Omega_\mathrm{m0}}{a_\mathrm{eq}}\right)^{1/2}=
  \frac{\Omega_\mathrm{m0}}{\sqrt{\Omega_\mathrm{r0}}}\;,
\label{eq:07-1}
\end{equation}
it is proportional to the matter-density parameter $\Omega_\mathrm{m0}$. A measurement of $k_\mathrm{eq}$ would thus provide an independent and very elegant determination of $\Omega_\mathrm{m0}$.

Since the power spectrum is defined in Fourier space, it is not obvious how it should be measured. The correlation function is related to the power spectrum by
\begin{equation}
  \xi(x)=\frac{1}{2\pi^2}\int_0^\infty 
  P(k)\frac{\sin kx}{kx}k^2\d k\;,
\label{eq:07-2}
\end{equation}
whose inverse transform is
\begin{equation}
  P(k)=4\pi\int_0^\infty\xi(x)\frac{\sin kx}{kx}x^2\d x\;.
\label{eq:07-3}
\end{equation}
The power spectrum can thus in principle be determined from a measured correlation function $\xi(x)$.

\subsubsection{Measuring the correlation function}\label{sec:VII-A-2}

The correlation function of the three-dimensional density field cannot be measured directly. Gravitational lensing by large-scale structures comes close by measuring the weighed line-of-sight projection of the matter power spectrum; see \S~\ref{sec:VIII} below. Starting from the assumption that galaxies may be tracers of the underlying density field, we can use their correlation function as an estimate for that of the matter.

Suppose we divide space into cells of volume $\d V$ small enough to contain at most a single galaxy. Then, the probability of finding one galaxy in $\d V_1$ and another in $\d V_2$ is
\begin{equation}
  \d P=\langle n(\vec x_1)n(\vec x_2)\rangle\d V_1\d V_2\;,
\label{eq:07-4}
\end{equation} 
where $n$ is the number density of the galaxies as a function of position. If we introduce a density contrast for the galaxies in analogy to the density contrast for the matter,
\begin{equation}
  \delta n\equiv\frac{n}{\bar n}-1\;,
\label{eq:07-5}
\end{equation}
and assume for now that $\delta n=\delta$, we find from (\ref{eq:07-4}) with $n=\bar n(1+\delta)$
\begin{equation}
  \d P=\bar n^2\langle(1+\delta_1)(1+\delta_2)\rangle\d V_1\d V_2=
  \bar n^2[1+\xi(x)]\d V_1\d V_2\;,
\label{eq:07-6}
\end{equation}
where $x$ is the comoving distance between the two volume elements. This shows that the correlation function quantifies the excess probability above random for finding galaxy pairs at a given distance.

Thus, the correlation function can be measured by counting galaxy pairs and comparing the result to the Poisson expectation, i.e.~to the pair counts expected in a \emph{random} point distribution. Symbolically,
\begin{equation}
  1+\xi=\frac{\langle DD\rangle}{\langle RR\rangle}\;,
\label{eq:07-7}
\end{equation}
where $D$ and $R$ represent the data and the random point set, respectively. Several other estimators for $\xi$ have been proposed which are all equivalent in the ideal situation of an infinitely extended point distribution. For finite point sets, their noise properties differ \citep{LA93.1,HA93.1}. The recipe for measuring $\xi(x)$ is thus to count pairs separated by $x$ in the data $D$ and in the random point set $R$, or between the data and the random point set, and to use one of the estimators proposed.

The obvious question is then how accurately $\xi$ can be determined. The simple expectation in the absence of clustering is
\begin{equation}
  \langle\xi\rangle=0\;,\quad
  \langle\xi^2\rangle=\frac{1}{N_\mathrm{p}}\;,
\label{eq:07-8}
\end{equation}
where $N_\mathrm{p}$ is the number of pairs found. Thus, the Poisson error on the correlation function is
\begin{equation}
  \frac{\Delta\xi}{1+\xi}=\frac{1}{\sqrt{N_\mathrm{p}}}\;.
\label{eq:07-9}
\end{equation}
This is a lower limit to the actual error, however, because the galaxies are in fact correlated. It turns out that the result (\ref{eq:07-9}) should be multiplied with $1+4\pi\bar nJ_3$, where $J_3$ is the volume integral over $\xi$ within the galaxy-survey volume \citep{PE73.1}. The true error bars on $\xi$ are therefore hard to estimate.

Having measured the correlation function, it would in principle suffice to carry out the Fourier transform (\ref{eq:07-3}) to find $P(k)$, but this is difficult in reality because of the inevitable sample limitations. Consider (\ref{eq:07-2}) and an underlying power spectrum of CDM shape, falling off $\propto k^{-3}$ for large $k$, i.e.~on small scales. For fixed $x$, the integrand in (\ref{eq:07-2}) falls off very slowly, which means that a considerable amount of small-scale power is mixed into the correlation function. Since $\xi$ at large $x$ is small and most affected by measurement errors, any uncertainty in the large-scale correlation function is propagated to the power spectrum even on small scales.

A further problem is the uncertainty in the mean galaxy number density $\bar n$. Since $1+\xi\propto\bar n^{-1}$ according to (\ref{eq:07-6}), the uncertainty in $\xi$ due to an uncertainty in $\bar n$ is
\begin{equation}
  \frac{\Delta\xi}{1+\xi}\approx\Delta\xi=
  \frac{\Delta\bar n}{\bar n}\;,
\label{eq:07-10}
\end{equation} 
showing that $\xi$ cannot be measured with an accuracy better than the relative accuracy of the mean galaxy density.

\subsubsection{Measuring the power spectrum}\label{sec:VII-A-3}

Given these problems with real data, it seems appropriate to estimate the power spectrum directly. The function to be transformed is the density field sampled by the galaxies, which can be represented by a sum of Dirac delta functions centred on the locations of the $N$ galaxies,
\begin{equation}
  n(\vec x)=\sum_{i=1}^N\delta_\mathrm{D}(\vec x-\vec x_i)\;.
\label{eq:07-11}
\end{equation}
The Fourier transform of the galaxy density contrast is then
\begin{equation}
  \hat\delta_\mathrm{g}(\vec k)=\frac{1}{N}\sum_{i=1}^N\e^{\ii\vec k\vec x_i}\;.
\label{eq:07-12}
\end{equation} 
In the absence of correlations, the Fourier phases of the individual terms are independent, and the variance of the Fourier amplitude for a single mode becomes
\begin{equation}
  \langle\hat\delta_\mathrm{g}(\vec k)\hat\delta_\mathrm{g}^*(\vec k)\rangle=
  \frac{1}{N^2}\sum_{i=1}^N
    \e^{\ii\vec k\vec x_i}\e^{-\ii\vec k\vec x_i}=
  \frac{1}{N}\;.
\label{eq:07-13}
\end{equation}
This is the so-called \emph{shot noise} present in the power spectrum due to the discrete sampling of the density field. The shot-noise contribution needs to be subtracted from the power spectrum of the real, correlated galaxy distribution,
\begin{equation}
  P(k)=\frac{1}{m}\sum|\hat\delta_\mathrm{g}(\vec k)|^2-\frac{1}{N}\;,
\label{eq:07-14}
\end{equation}
where the sum extends over all $m$ Fourier modes with wave number $k$ contained in the survey.

This is not the final result yet, because any real survey typically covers an irregularly shaped volume from which parts need to be excised because they are overshone by stars or unusable for any other reason. The combined effect of mask and irregular survey volume is described by a window function $f(\vec x)$ which multiplies the galaxy density,
\begin{equation}
  n(\vec x)\to f(\vec x)n(\vec x)\;,\quad
  (1+\delta_\mathrm{g})\to f(\vec x)(1+\delta_\mathrm{g})\;,
\label{eq:07-15}
\end{equation}
implying that the Fourier transform of the mask needs to be subtracted.

Moreover, the Fourier convolution theorem says that the Fourier transform of the \emph{product} $f(\vec x)\delta_\mathrm{g}(\vec x)$ is the \emph{convolution} of the Fourier transforms $\hat f(\vec k)$ and $\hat\delta_\mathrm{g}(\vec k)$,
\begin{equation}
  \widehat{f\delta}=\hat f*\hat\delta_\mathrm{g}\equiv
  \int\hat f(\vec k')\hat\delta_\mathrm{g}(\vec k'-\vec k)\d^3k'\;.
\label{eq:07-16}
\end{equation}
If the Fourier phases of $\hat f$ and $\hat\delta_\mathrm{g}$ are uncorrelated, which is the case if the dimensions of the survey are large enough compared to the size $2\pi/k$ of the density mode, this translates to a convolution of the power spectrum,
\begin{equation}
  P_\mathrm{obs}=P_\mathrm{true}*|\hat f(\vec k)|^2\;.
\label{eq:07-17}
\end{equation}
This convolution smoothes the observed compared to the true power spectrum and changes its amplitude.

If the Poisson error dominates in the survey, the different modes $\hat\delta_\mathrm{g}(\vec k)$ can be shown to be uncorrelated, and the standard deviation after summing over the $m$ modes with wave number $k$ is $\sqrt{2m}/N$, which yields the minimal error bar to be attached to the power spectrum.

Thus, the shot noise contribution and the Fourier transform of the window function need to be subtracted, and the window function needs to be deconvolved, and the amplitude needs to be corrected for the effective volume covered by the window function before the measured power spectrum can be compared to the theoretical expectation.

\subsubsection{Biasing}\label{sec:VII-A-4}

What we have determined so far is the power spectrum of the \emph{galaxy number-density contrast} $\delta n$ rather than that of the \emph{matter density contrast} $\delta$. Both are related by a possibly scale-dependent \emph{bias factor} $b(k)$, such that
\begin{equation}
  \widehat{\delta n}(\vec k)=b(k)\hat\delta_\mathrm{g}(\vec k)\;.
\label{eq:07-22}
\end{equation}
Clearly, different types of objects sample the underlying matter density field in different ways. Galaxy clusters, for instance, are much more rare than galaxies and are thus expected to have a substantially higher bias factor than galaxies. Obviously, the bias factor enters the power spectrum squared, e.g.
\begin{equation}
  P_\mathrm{gal}=b^2_\mathrm{gal}(k)\,P(k)\;.
\label{eq:07-23}
\end{equation}
It constitutes a major and possibly systematic uncertainty in the determination of the matter power spectrum from the galaxy power spectrum.

\subsubsection{Redshift-space distortions}\label{sec:VII-A-5}

For any quantification of three-dimensional structures in galaxy surveys, the three-dimensional positions $\vec x_i$ of the galaxies in the survey need to be known. Distances can be inferred only from the galaxy redshifts and thus from galaxy velocities. These, however, are composed of the Hubble velocities, from which the distances can be determined, and the peculiar velocities,
\begin{equation}
  v=v_\mathrm{Hubble}+v_\mathrm{pec}\;,
\label{eq:07-24}
\end{equation}
which are caused by local density perturbations and are unrelated to the galaxy densities. Since observations of individual galaxies do not allow any separation between the two velocity components, distances are inferred from the total velocity $v$ rather than the Hubble velocity as they should be,
\begin{equation}
  D=\frac{v}{H_0}=\frac{v_\mathrm{Hubble}+v_\mathrm{pec}}{H_0}=
  D_\mathrm{true}+\Delta D\;,
\label{eq:07-25}
\end{equation}
giving rise to a distance error $\delta D=v_\mathrm{pec}/H_0$, the so-called \emph{redshift-space distortion}.

The redshift-space distortions create a peculiar pattern through which they can be corrected \citep{KA87.1,HA93.2}. Consider a matter overdensity such as a galaxy cluster, containing galaxies moving with random virial velocities in it. The virial velocities of order $1000\,\mathrm{km\,s^{-1}}$ scatter around the systemic cluster velocity and thus broaden the redshift distribution of the cluster galaxies. In redshift space, therefore, the cluster appears stretched along the line-of-sight, which is called the \emph{finger-of-god} effect.

In addition, the cluster is surrounded by an infall region where the galaxies are not virialised yet, but move in an ordered, radial pattern towards the cluster. Galaxies in front of the cluster thus have higher, and galaxies behind the cluster have lower recession velocities compared to the Hubble velocity, leading to a flattening of the infall region in redshift space.

A detailed analysis shows that the redshift-space power spectrum $P_z$ is related to the real-space power spectrum $P$ by
\begin{equation}
  P_z(k)=P(k)\left(1+\beta\mu^2\right)^2\;,
\label{eq:07-26}
\end{equation}
where $\mu$ is the direction cosine between the line-of-sight and the wave vector $\vec k$, and $\beta$ is related to the bias parameter $b$ through
\begin{equation}
  \beta\equiv\frac{f(\Omega_\mathrm{m})}{b}\;,
\label{eq:07-27}
\end{equation}
and $f(\Omega_\mathrm{m})$ is the logarithmic derivative of the growth factor $D_+(a)$,
\begin{equation}
  f(\Omega_\mathrm{m})\equiv\frac{\d\ln D_+(a)}{\d\ln a}\approx
  \Omega_\mathrm{m}^{0.6}\;.
\label{eq:07-28}
\end{equation}
Thus, the characteristic pattern of the redshift-space distortions around overdensities allows a measurement of the bias factor \citep{HA03.1}. Another way of measuring $b$ is based upon gravitational lensing \citep{HO02.1,SI07.1}. Measurements of $b$ show that it is in fact only weakly scale-dependent, near unity for ``ordinary'' galaxies, but depends mildly on galaxy luminosity and type \citep{NO01.1,LA02.1,CO05.1}.

\subsubsection{Baryonic acoustic oscillations}\label{sec:VII-A-6}

As we have seen in the discussion of the CMB, the cosmic fluid underwent acoustic oscillations on comoving scales smaller than the sound horizon (\ref{eq:06-21}) $w_\mathrm{s}=163.3\,\mathrm{Mpc}$, corresponding to a comoving wave number $k_\mathrm{s}=2\pi/w_\mathrm{s}=0.038\,\mathrm{Mpc}^{-1}$. When the CMB decoupled, the oscillations ceased, leaving structures in the cosmic matter distribution with a fundamental wavelength of $w_\mathrm{s}$ and its overtones. The other scale characterising the cosmic structures is the horizon radius at matter-radiation equality (\ref{eq:01-26}), which was responsible to set the peak location of the matter-fluctuation power spectrum at $k_\mathrm{eq}=0.01\,\mathrm{Mpc}^{-1}$.

Thus, at wave numbers $\approx3.8$ times the peak scale and above, we expect the wave-like imprint of these baryonic acoustic oscillations (BAOs) on top of the otherwise smooth dark-matter power spectrum. Mode coupling due to non-linear evolution of cosmic structures must have stretched and distorted this pattern to some degree. As we shall see below, the BAOs have indeed been discovered in the largest galaxy surveys. They play an important role in attempts to recover the history of the cosmic expansion rate (see \cite{EI05.1} for a recent review).

\subsection{Measurements and results}\label{sec:VII-B}

\subsubsection{The power spectrum}\label{sec:VII-B-1}

Spectacularly successful measurements of the galaxy power spectrum became recently possible with the two largest galaxy surveys to date, the \emph{Two-Degree Field Galaxy Redshift Survey} (2dFGRS, \cite{CO99.1}) and the \emph{Sloan Digital Sky Survey} (SDSS, \cite{YO00.1}). As anticipated in the preceding discussion, an enormous effort had to be made to identify galaxies, measure their redshifts, select homogeneous galaxy subsamples by luminosity and colour as a function of redshift so as not to compare and correlate apples with oranges, estimate the window function of the survey, determine the average galaxy number density, correct for the convolution with the window function and for the bias, and so forth.

Moreover, numerical calibration experiments were carried out in which all measurement and correction techniques were applied to simulated data in the same way as to the real data to determine reliable error estimates and to test whether the full sequence of analysis steps ultimately yields an unbiased result.

Based on $221,414$ galaxies, the 2dFGRS consortium derived a power spectrum of superb quality \citep{CO05.1}. First and foremost, it agrees well with the power-spectrum shape expected for cold dark matter on the small-scale side of the peak. This is a highly remarkable result on its own. The 2dFGRS power spectrum also clearly shows a turn-over towards larger scales, signalling the presence of the peak. The survey is still not quite large enough to clearly show the peak, but the peak location can be estimated from the flattening of the power spectrum. Its proportionality to $\Omega_\mathrm{m0}$ allows an independent determination of the matter density parameter. Finally, and most spectacularly, the power spectrum shows the baryonic acoustic oscillations, whose amplitude allows an independent determination of the ratio between the density parameters of baryons and dark matter.

Apart from the fact that the galaxy power spectrum supports the CDM shape on small scales, the results obtained from the 2dFGRS can be summarised as follows:
\begin{eqnarray}
  \Omega_\mathrm{m0}&=&0.233\pm0.022\nonumber\\
  \Omega_\mathrm{b0}/\Omega_\mathrm{m0}&=&0.185\pm0.046\;.
\label{eq:07-29}
\end{eqnarray}
A Hubble constant of $h=0.72$ is assumed here. Indirectly, the baryon density is constrained to be $\Omega_\mathrm{b0}\approx0.04$, which is in perfect agreement with the value derived from primordial nucleosynthesis and the measured abundances of the light elements.

Based on $522,280$ galaxies, the power spectrum inferred from the SDSS \citep{PE07.2} is also well compatible with the CDM shape. The estimate for the matter density parameter overlaps with that from the 2dFGRS on large scales, $\Omega_\mathrm{m0}=0.22\pm0.04$, but increases when small scales are included.

\begin{figure}[ht]
  \includegraphics[width=\hsize]{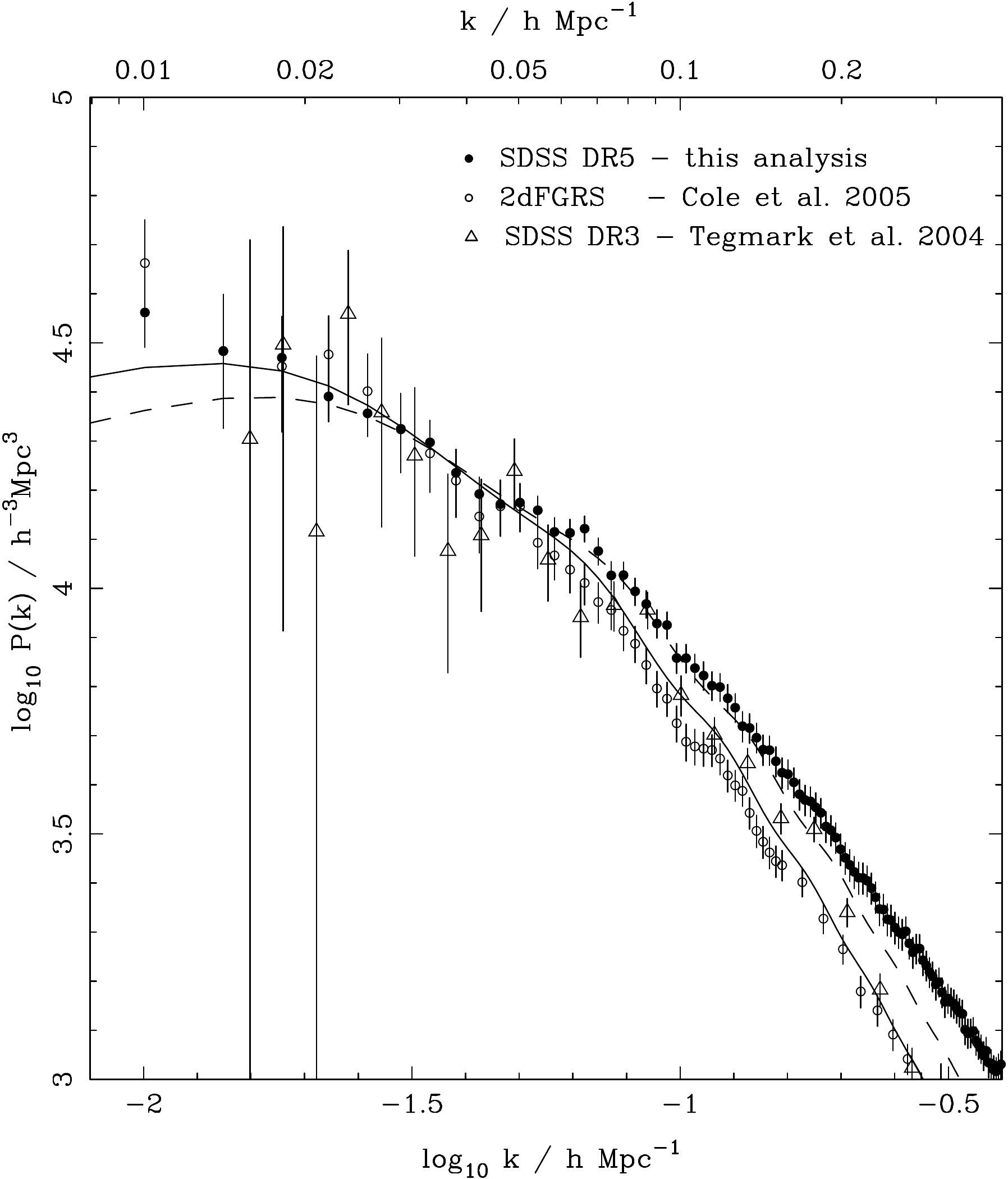}
\caption{Galaxy power spectra obtained from the 2dFGRS \citep{CO05.1} and two releases of the SDSS \citep{TE04.1,PE07.2}. The expected shape of the CDM power spectrum is well reproduced, and the difference in amplitudes may be attributed to scale-dependent galaxy biasing. (from \cite{PE07.2})}
\end{figure}

It should be kept in mind, however, that the assumption of linear biasing may turn out to be too simplistic for precise cosmological inferences to be drawn from galaxy power spectra. Nonetheless, power spectra for different types of galaxy will provide invaluable information on the physics of galaxy formation.

\section{Cosmological Weak Lensing}\label{sec:VIII}

\subsection{Cosmological light deflection}\label{sec:VIII-A}

\subsubsection{Deflection angle, convergence and shear}\label{sec:VIII-A-1}

Gravitational lensing was mentioned two times before: first in \S~\ref{sec:III-B} as a means for measuring the Hubble constant through the time delay caused by gravitational light deflection, and second as a means for measuring cluster masses in \S~\ref{sec:V-B-3}. For cosmology as a whole, gravitational lensing has also developed into an increasingly important tool (see \cite{RE03.1,BA01.1,ME99.1} for reviews).

Matter inhomogeneities deflect light. Working out this effect in the limit of a small Newtonian gravitational potential, $\Phi\ll c^2$, leads to the deflection angle
\begin{equation}
  \vec\alpha(\vec\theta)=\frac{2}{c^2}\int_0^w\d w'\,
  \frac{f_K(w-w')}{f_K(w)}\vec\nabla_\perp\Phi[f_K(w')\vec\theta]\;.
\label{eq:08-1}
\end{equation}
It is determined by a weighted integral over the gradient of the Newtonian gravitational potential $\Phi$ perpendicular to the line-of-sight into direction $\vec\theta$ on the observer's sky, and the weight is given by the comoving angular-diameter distance $f_K(w)$ defined in (\ref{eq:01-2}). The integral extends along the comoving radial distance $w'$ along the line-of-sight to the distance $w$ of the source.

Equation (\ref{eq:08-1}) can be intuitively understood. Light is deflected due to the pull of the dimension-less Newtonian gravitational field $\vec\nabla_\perp\Phi/c^2$ perpendicular to the otherwise unperturbed line-of-sight, and the effect is weighed by the ratio between the angular-diameter distances from the deflecting potential to the source, $f_K(w-w')$, and from the observer to the source, $f_K(w)$. Thus, a lensing mass distribution very close to the observer gives rise to a large deflection, while a lens near the source, $w'\approx w$, has very little effect. The factor of two is a relic from general relativity and is due to space-time curvature, which is missing in Newtonian gravity. Assuming $K=0$, we can replace $f_K(w)$ by $w$.

It is important to realise that the deflection itself is not observable. If all light rays emerging from a source were deflected by the same angle on their way to the observer, no noticeable effect remained. What is important, therefore, is not the deflection angle itself, but its change from one light ray to the next. This is quantified by the derivative of the deflection angle with respect to the direction $\vec\theta$,
\begin{equation}
  \frac{\partial\alpha_i}{\partial\theta_j}=
  \frac{2}{c^2}\int_0^w\d w'\,\frac{(w-w')w'}{w}\,
  \frac{\partial^2\Phi}{\partial x_i\partial x_j}(w'\vec\theta)\;.
\label{eq:08-2}
\end{equation}
The additional factor $w'$ in the weight function arises because the derivative of the potential is taken with respect to comoving coordinates $x_i$ rather than the angular components $\theta_i$. Obviously, the complete weight function
\begin{equation}
  W(w',w)\equiv\frac{(w-w')w'}{w}
\label{eq:08-3}
\end{equation} 
vanishes at the observer, $w'=0$, and at the source, $w'=w$, and peaks approximately half-way in between.

For applications of gravitational lensing, it is important to distinguish the trace-free part of the matrix (\ref{eq:08-2}) from its trace,
\begin{equation}
  \mathrm{tr}\left(\frac{\partial\alpha_i}{\partial\theta_j}\right)=
  \frac{2}{c^2}\int_0^w\d w'\,W(w',w)\,\left(
    \frac{\partial^2\Phi}{\partial x_1^2}+\frac{\partial^2\Phi}{\partial x_2^2}
  \right)(w'\vec\theta)\;.
\label{eq:08-4}
\end{equation}
The derivatives of $\Phi$ can be combined to the two-dimensional Laplacian, which can be replaced by the three-dimensional Laplacian because the derivatives parallel to the line-of-sight do not contribute to the integral (\ref{eq:08-4}). Thus, we find
\begin{equation}
  \mathrm{tr}\left(\frac{\partial\alpha_i}{\partial\theta_j}\right)=
  \frac{2}{c^2}\int_0^w\d w'\,W(w',w)\,\Delta\Phi\;.
\label{eq:08-5}
\end{equation}

Next, we can use Poisson's equation to replace the Laplacian of $\Phi$ by the density. In fact, we have to take into account that light deflection is caused by density \emph{perturbations}, and that we need the Laplacian in terms of \emph{comoving} rather than \emph{physical} coordinates. Thus,
\begin{equation}
  a^{-2}\Delta\Phi=4\pi G\bar\rho\delta\;,
\label{eq:08-6}
\end{equation} 
where $\delta$ is the density contrast and
\begin{equation}
  \bar\rho=\bar\rho_0a^{-3}=\rho_\mathrm{cr}\Omega_\mathrm{m0}a^{-3}=
  \frac{3H_0^2}{8\pi G}\Omega_\mathrm{m0}a^{-3}
\label{eq:08-7}
\end{equation}
is the mean matter density. Thus, Poisson's equation reads
\begin{equation}
  \Delta\Phi=\frac{3}{2}H_0^2\Omega_\mathrm{m0}\frac{\delta}{a}\;,
\label{eq:08-8}
\end{equation}
 and (\ref{eq:08-5}) becomes
\begin{equation}
  \mathrm{tr}\left(\frac{\partial\alpha_i}{\partial\theta_j}\right)=
  \frac{3H_0^2\Omega_\mathrm{m0}}{c^2}
  \int_0^w\d w'\,W(w',w)\,\frac{\delta}{a}\equiv2\kappa\;,
\label{eq:08-9}
\end{equation}
where we have introduced the (effective) \emph{convergence} $\kappa$.

The trace-free part of the matrix (\ref{eq:08-2}) is
\begin{equation}
  \frac{\partial\alpha_i}{\partial\theta_j}-\frac{1}{2}\delta_{ij}
  \mathrm{tr}\left(\frac{\partial\alpha_i}{\partial\theta_j}\right)=
  \frac{\partial\alpha_i}{\partial\theta_j}-\delta_{ij}\kappa\equiv
  \left(\begin{array}{cc}
    \gamma_1 &  \gamma_2 \\
    \gamma_2 & -\gamma_1 \\
        \end{array}\right)\;,
\label{eq:08-10}
\end{equation}
which defines the so-called \emph{shear} components $\gamma_i$. Specifically,
\begin{eqnarray}
  \gamma_1&=&\frac{1}{c^2}\int_0^w\d w'\,W(w',w)\,\left(
    \frac{\partial^2\Phi}{\partial x_1^2}-
    \frac{\partial^2\Phi}{\partial x_2^2}
  \right)\;,\nonumber\\
  \gamma_2&=&\frac{2}{c^2}\int_0^w\d w'\,W(w',w)\,\left(
    \frac{\partial^2\Phi}{\partial x_1\partial x_2}
  \right)\;.
\label{eq:08-11}
\end{eqnarray}

Combining results, we can write the matrix of deflection-angle derivatives as
\begin{equation}
  \left(\frac{\partial\alpha_i}{\partial\theta_j}\right)=\left(
  \begin{array}{cc}
    \kappa+\gamma_1 & \gamma_2 \\
    \gamma_2 & \kappa-\gamma_1 \\
  \end{array}\right)\;.
\label{eq:08-12}
\end{equation}
This matrix contains the important information on how an image is magnified and distorted. In the limit of weak gravitational lensing, the \emph{size} of a lensed image is changed by the relative magnification
\begin{equation}
  \delta\mu=2\kappa\;,
\label{eq:08-13}
\end{equation}
while the image \emph{distortion} is given by the shear components. In fact, an originally circular source with radius $r$ will appear as an ellipse with major and minor axes
\begin{equation}
  a=\frac{r}{1-\kappa-\gamma}\;,\quad
  b=\frac{r}{1-\kappa+\gamma}\;,
\label{eq:08-14}
\end{equation}
where $\gamma\equiv(\gamma_1^2+\gamma_2^2)^{1/2}$. The \emph{ellipticity} of the observed image of a circular source thus provides an estimate for the shear,
\begin{equation}
  \epsilon\equiv\frac{a-b}{a+b}=\frac{\gamma}{1-\kappa}\approx
  \gamma\;.
\label{eq:08-15}
\end{equation}

\subsubsection{Power spectra}\label{sec:VIII-A-2}

Of course, the exact light deflection expected along a particular line-of-sight cannot be predicted because the mass distribution along that light path is unknown. However, we can predict the \emph{statistical properties} of weak lensing from those of the density-perturbation field. We are thus led to the following problem: Suppose the power spectrum $P(k)$ of a Gaussian random density-perturbation field $\delta$ is known, what is the power spectrum of any weighed projection of $\delta$ along the line-of-sight? The answer is given by Limber's equation. Suppose the weight function is $q(w)$ and the projection is
\begin{equation}
  g(\vec\theta)=\int_0^w\d w'\,q(w')\delta(w'\vec\theta)\;.
\label{eq:08-16}
\end{equation} 
If $q(w)$ is smooth compared to $\delta$, i.e.~if the weight function changes on scales much larger than typical scales in the density contrast, then the power spectrum of $g$ is
\begin{equation}
  P_g(l)=\int_0^w\d w'\,\frac{q^2(w')}{w'^2}\,P\left(\frac{l}{w'}\right)\;,
\label{eq:08-17}
\end{equation}
where $\vec l$ is a two-dimensional wave vector which is the Fourier conjugate variable to the two-dimensional position $\vec\theta$ on the sky.

Strictly speaking, Fourier transforms are inappropriate because the sky is not an infinite, two-dimensional plane. Instead, the appropriate set of orthonormal base functions are the spherical harmonics. However, lensing effects are usually observed in areas whose solid angle is very small compared to the full sky. If this is so, the survey area can be approximated by a section of the plane locally tangent to the sky, and Fourier transforms can be used in this so-called flat-sky approximation.

Equation~(\ref{eq:08-9}) is clearly of the form (\ref{eq:08-16}) with the weight function
\begin{equation}
  q(w')=\frac{3}{2}\Omega_\mathrm{m0}\frac{H_0^2}{c^2}\,
  \frac{W(w',w)}{a}\;,
\label{eq:08-18}
\end{equation}
thus the power spectrum of the convergence is, according to Limber's equation,
\begin{equation}
  P_\kappa(l)=\frac{9\Omega_\mathrm{m0}^2}{4}\frac{H_0^4}{c^4}\,
  \int_0^w\d w'\,\bar W^2(w',w)\,P\left(
    \frac{l}{f_k(w')}
  \right)\;,
\label{eq:08-19}
\end{equation}
 with a new weight function
\begin{equation}
  \bar W(w',w)\equiv\frac{W(w',w)}{aw'}\;.
\label{eq:08-20}
\end{equation}

While it requires huge data sets and extremely careful data analysis to observe the differential magnification $\delta\mu$ or the convergence $\kappa$ \citep{SC05.1}, image distortions can in principle be measured in a more straightforward way. With a brief excursion through Fourier space, it can easily be shown that the power spectrum of the shear is exactly identical to that of the convergence,
\begin{equation}
  P_\gamma(l)=P_\kappa(l)\;.
\label{eq:08-21}
\end{equation}
Thus, the statistics of the image distortions caused by cosmological weak lensing contains integral information on the power spectrum of the matter fluctuations.

Since the shear is defined on the two-sphere (the observer's sky), its power spectrum is related to its correlation function $\xi_\gamma$ through the two-dimensional Fourier transform
\begin{equation}
  \xi_\gamma(\phi)=\int\frac{\d^2l}{(2\pi)^2}\,
  P_\gamma(l)\e^{\ii\vec\phi\vec l}=
  \int_0^\infty\frac{l\d l}{2\pi}\,P_\gamma(l)\mathrm{J}_0(l\phi)\;,
\label{eq:08-22}
\end{equation}
where $\mathrm{J}_\nu$ is the ordinary Bessel function of order $\nu$.

\subsubsection{Correlation functions}\label{sec:VIII-A-3}

In principle, shear correlation functions are measured by comparing the ellipticity of one galaxy with the ellipticity of other galaxies at an angular distance $\phi$ from the first. Ellipticities are oriented, of course, and one has to specify against what other direction the direction of, say, the major axis of a given ellipse is to be compared with. Since correlation functions are measured comparing the shear of galaxy pairs, a preferred direction is defined by the line connecting the two galaxies of the pair under consideration.

Let $\alpha$ be the angle between this direction and the major axis of the ellipse, then the \emph{tangential} and \emph{cross} components of the shear are defined by
\begin{equation}
  \gamma_+\equiv\gamma\cos2\alpha\;,\quad
  \gamma_\times\equiv\gamma\sin2\alpha\;.
\label{eq:08-23}
\end{equation}
The factor two is important because it accounts for the fact that an ellipse is mapped onto itself when rotated by an angle $\pi$. This illustrates that the shear is a spin-2 field: It returns into its original orientation when rotated by $\pi$ rather than $2\pi$.

The correlation functions of the tangential and cross components of the shear are
\begin{equation}
  \xi_{++}(\phi)=\langle\gamma_+(\theta)\gamma_+(\theta+\phi)\rangle=
  \frac{1}{2}\int_0^\infty\frac{l\d l}{2\pi}P_\kappa(l)\left[
    \mathrm{J}_0(l\phi)+\mathrm{J}_4(l\phi)
  \right]
\label{eq:08-24}
\end{equation}
and
\begin{equation}
  \xi_{\times\times}(\phi)=
  \langle\gamma_\times(\theta)\gamma_\times(\theta+\phi)\rangle=
  \frac{1}{2}\int_0^\infty\frac{l\d l}{2\pi}P_\kappa(l)\left[
    \mathrm{J}_0(l\phi)-\mathrm{J}_4(l\phi)
  \right]\;,
\label{eq:08-25}
\end{equation}
while the cross-correlation between the tangential and cross components must vanish, $\xi_{+\times}(\phi)=0$. This suggests to define the correlation functions $\xi_\pm=\xi_{++}\pm\xi_{\times\times}$, which are related to the power spectrum through $\xi_+=\xi_\gamma$ and
\begin{equation}
  \xi_-=\int_0^\infty\frac{l\d l}{2\pi}P_\kappa(l)\mathrm{J}_4(l\phi)\;.
\label{eq:08-27}
\end{equation}

The principle of all measures for cosmic shear is the same: They are integrals of the weak-lensing power spectrum times filter functions describing the specific response of the measurement to the underlying power spectrum of density fluctuations. The width of the filter functions controls the range of density-perturbation modes $\vec k$ contributing to one specific mode $\vec l$ of weak-lensing on the sky.

We can now estimate typical numbers for the cosmological weak-lensing effect. The power $\Delta_\kappa$ in the weak-lensing quantities such as the cosmic shear is given by the power spectrum $P_\kappa(l)$ found in (\ref{eq:08-19}), times the volume in $l$-space,
\begin{equation}
  \Delta_\kappa(l)\approx l^2P_\kappa(l)\;.
\label{eq:08-30}
\end{equation}
Assuming a cosmological model with $\Omega_\mathrm{m0}=0.3$ and $\Omega_{\Lambda0}=0.7$, the CDM power spectrum and a reasonable source redshift distribution, $\Delta_\kappa^{1/2}(l)$ is found to peak on scales $l$ corresponding to angular scales $2\pi/l$ of $2'\ldots3'$, and the peak reaches values of $0.04\ldots0.05$. This shows that cosmological weak lensing will typically cause source ellipticities of a few per cent, and correlations have a typical angular scale of a few arc minutes. Details depend on the measure chosen through the filter function.

\subsection{Cosmic-shear measurements}\label{sec:VIII-B}

\subsubsection{Typical scales and requirements}\label{sec:VIII-B-1}

How can cosmic gravitational lensing effects be measured? As shown in (\ref{eq:08-15}), the ellipticity of a hypothetic circular source is an \emph{unbiased estimator} for the shear. However, typical sources are not circular, but to first approximation elliptical themselves. Thus, measuring their ellipticities yields their intrinsic ellipticities in the first place.

Let $\epsilon^\mathrm{(s)}$ be the intrinsic source ellipticity. It is a two-component, spin-2 quantity. The cosmic shear adds to that ellipticity, such that the observed ellipticity is
\begin{equation}
  \epsilon\approx\epsilon^\mathrm{(s)}+\gamma
\label{eq:08-31}
\end{equation}
in the weak-lensing approximation. What is observed is therefore the sum of the signal, $\gamma$, and the intrinsic noise component $\epsilon^\mathrm{(s)}$.

On sufficiently deep observations, $\lesssim20$ galaxies per square arc minute are routinely detected. Since the full moon has half a degree diameter, it covers a solid angle of $15^2\pi=700$ square arc minutes, or $\lesssim14,000$ of such distant, faint galaxies! From this point of view, the sky is covered by densely patterned ``wall paper'' of distant galaxies. Thus, it is possible to average observed galaxy ellipticities. Assuming their shapes are intrinsically independent, the intrinsic ellipticities will average out, and the shear will remain,
\begin{equation}
  \langle\epsilon\rangle\approx
  \langle\epsilon^\mathrm{(s)}\rangle+\langle\gamma\rangle\approx
  \langle\gamma\rangle\;.
\label{eq:08-32}
\end{equation}

It is a fortunate coincidence that the typical angular scale of cosmic lensing, which we found to be of order a few arc minutes, is large compared to the mean distance between background galaxies, which is of order $\sqrt{1/20}\approx0.2'$. This enables averages over background galaxies without cancelling the cosmic shear signal. If $\gamma$ varied on scales comparable to or smaller than the mean galaxy separation, any average over galaxies would remove the lensing signal as well.

The intrinsic ellipticities of the faint background galaxies have a distribution with a standard deviation of $\sigma_\epsilon\approx0.3$. Averaging over $N$ of them, and assuming Poisson statistics, yields expectation values of
\begin{equation}
  \langle\epsilon^\mathrm{(s)}\rangle=0\;,\quad
  \delta\epsilon=\langle(\epsilon^\mathrm{(s)})^2\rangle^{1/2}=
  \frac{\sigma_\epsilon}{\sqrt{N}}
\label{eq:08-33}
\end{equation}
for the mean and its intrinsic fluctuation.

A rough estimate for the signal-to-noise ratio of a cosmic shear measurement can proceed as follows. Suppose the correlation function $\xi$ is measured by counting pairs of galaxies with a separation within $\delta\theta$ of $\theta$. As long as $\theta$ is small compared to the side length of the survey area $A$, the number of pairs will be
\begin{equation}
  N_\mathrm{p}=\frac{1}{2}\times n^2A\times2\pi\theta\,\delta\theta\;,
\label{eq:08-34}
\end{equation}
and thus the Poisson noise due to the intrinsic ellipticities will be
\begin{equation}
  \mbox{noise}\approx
  \frac{2\sigma_\epsilon}{n\sqrt{\pi A\theta\delta\theta}}\;,
\label{eq:08-35}
\end{equation}
where the factor of two arises because of the two galaxies involved in each pair. The signal is the square root of the correlation function $\xi$, which we can approximate as
\begin{equation}
  \xi\approx l^2P_\kappa(l)\delta\ln l\approx
  l^2P_\kappa(l)\frac{\delta l}{l}\approx
  l^2P_\kappa(l)\frac{\delta\theta}{\theta}\;,
\label{eq:08-36}
\end{equation}
where we have used in the last step that $\theta=2\pi/l$. Thus, the signal-to-noise ratio is estimated to be
\begin{equation}
  \frac{S}{N}\approx\frac{\sqrt{\xi}}{\mbox{noise}}\approx
  \frac{\ln\delta\theta\sqrt{\pi AP_\kappa}}{2\sigma_\epsilon}=
  \frac{n\sqrt{\pi^3AP_\kappa}}{\sigma_\epsilon}
  \frac{\delta\theta}{\theta}\;.
\label{eq:08-37}
\end{equation}
Evidently, the signal-to-noise ratio, and thus the significance of any cosmic-lensing detection, grows with the survey area and decreases with the intrinsic ellipticity of the source galaxies. In evaluating (\ref{eq:08-37}) numerically, we have to take into account that $l^2P_\kappa(l)$ must be a dimension-less number, which implies that the power spectrum $P_\kappa$ must have the dimension steradian. Therefore, either the survey area $A$ and the number density $n$ in (\ref{eq:08-37}) must be converted to steradians, or $P_\kappa$ must be converted to square arc minutes first.

The signal-to-noise ratio increases approximately linearly with scale. Assuming $\delta\theta/\theta=0.1$, it is $S/N\approx5$ on a scale of $0.5'$ for a survey of one square degree area. This shows that, if the cosmic shear should be measured on such small scales with an accuracy of, say, five per cent, a survey area of $A\approx(20/5)^2\approx16$ square degrees is needed since the signal-to-noise ratio scales like the square root of the survey area. On such an area, the ellipticities of $16\times3600\times30\approx2\times10^6$ background galaxies have to be accurately measured.

Matters are more complicated in reality, but the orders-of-magnitude are well represented by this rough estimate. Bearing in mind that typical fields-of-view of telescopes large enough for detecting sufficiently many faint background galaxies reach one tenth of a square degree up to one square degree, and that typical exposure times are of order one hour for that purpose, the total amount of telescope time for a weak-lensing survey of about 100 square degrees is estimated to be several hundred telescope hours. With perhaps five hours of telescope time per night, and perhaps half of the nights per year usable, it is easy to see that the time needed for such surveys is measured in months.

Since typical sizes of the faint background galaxies measure fractions of arc seconds, shape measurements require a pixel resolution of, say, $0.1''$. A total survey area of $100$ square degrees must therefore be resolved into $100\times3600\times3600/0.1^2\approx1.3\times10^{11}$ pixels.

These estimates neglect all sources of noise other than the shot noise caused by the finite number of galaxies. On angular scales below a few arc minutes, the cosmic variance caused by field-to-field variations in the shear signal due to the large-scale cosmic structures must be added \citep{KI04.1}.

\subsubsection{Ellipticity measurements}\label{sec:VIII-B-2}

The determination of image ellipticities is straightforward in principle, but difficult in practice \citep{KA95.1}. Often, the surface-brightness quadrupole
\begin{equation}
  Q_{ij}=\frac{\int I(\vec x)x_ix_j\d^2x}{\int I(\vec x)\d^2x}
\label{eq:08-38}
\end{equation}
is measured, from whose principal axes the ellipticity can be read off. Real galaxy images, however, are typically far from ideally elliptical. They are structured or otherwise irregular. In addition, if they are small, they are coarsely resolved into just a few pixels, so that only a crude approximation to the integral in (\ref{eq:08-38}) can be found.

How the image of a point-like source, such as a star, appears on the detector is described by the \emph{point-spread function} (PSF). The PSF may be anisotropic if the telescope optics is slightly astigmatic, and this anisotropy may, and will in general, depend on the location in the focal plane. The image is a convolution of the ideal image shape prior to any distortion by the atmosphere and the telescope optics. Any accurate measurement of image ellipticities requires a PSF correction or deconvolution, for which the PSF must of course be known. It is commonly measured off the images of stars in the field.

Many other effects may distort the PSF and thus the images in systematic ways. For instance, if the CCD chips are not exactly perpendicular to the optical axis of the telescope, or if the individual chips of a CCD mosaic are not exactly coplanar, or if the telescope is slightly out of focus, systematic image deformations may result which typically vary across the focal plane. They have to be measured and corrected. This is commonly achieved by fitting the measured PSF by low-order, two-dimensional polynomials on the focal plane. Since part of the image distortions may depend on time due to thermal deformation, changing atmospheric conditions and such, PSF corrections will also depend on time and have to be determined and applied with much care.

Even if the surface-brightness quadrupole of the image on the detector can be accurately determined, the image appears affected by imperfections of the telescope optics and by the turbulence in the atmosphere, the so-called \emph{seeing}. Due to the wave nature of light and the finite size of the telescope mirror, the telescope will have finite resolution. The angular resolution limit is given by (\ref{eq:06-31}). With $\lambda\approx6\times10^{-5}\,\mathrm{cm}$ and $D=400\,\mathrm{cm}$, the angular resolution is $\Delta\theta\approx0.04''$, much smaller than needed for our purposes.

The turbulence of the Earth's atmosphere effectively convolves images with a Gaussian whose width depends on the site, the weather and other conditions. Typical seeing ranges around $1''$. Under very good conditions, it can shrink to $\sim0.5''$ or even less. Clearly, if an image of sub-arc second size is convolved with a Gaussian of similar width, any ellipticity is substantially reduced.

Systematic effects may remain which need to be detected and quantified. Any coherent image distortions caused by gravitational lensing must be describable by the tidal gravitational field, i.e.~by second-order derivatives of a scalar potential. In analogy to the $\vec E$-field in electromagnetism, such distortion patterns are called $E$-modes. Similarly, distortion patterns which are described as the curl of a vector field are called $B$-modes. They cannot be due to gravitational lensing and thus signal systematic effects remaining in the data. Such $B$-mode contaminations could recently be strongly reduced or suppressed by improved algorithms for PSF correction \citep{HO04.1}. Absence of $B$-mode contamination does not allow the implication that the results are free of systematics, though, because optical distortions also tend to create spurious $E$-modes.

\subsubsection{Results}\label{sec:VIII-B-3}

Despite the smallness of the effect and the many difficulties in measuring it, much progress in cosmic-shear observations has been achieved in the past few years \citep{HE06.1,MA07.1}. Current and ongoing surveys, in particular the Canada-France-Hawaii Legacy Survey \citep{SE06.1,FU08.1}, combined with well-developed, largely automatic data-analysis pipelines, have succeeded in producing cosmic-shear correlation functions with very small error bars covering angular scales from below an arc minute to several degrees. The best correlation functions could be shown to be at most negligibly contaminated by $B$-modes.

The power spectrum $P_\kappa(l)$ depends crucially on the non-linear evolution of the dark-matter power spectrum. This, and the exact redshift distribution of the background galaxies, are the major uncertainties now remaining in the interpretation of cosmic-shear surveys. Apart from that, the measured cosmic-shear correlation functions agree very well with the theoretical expectation from CDM density fluctuations in a spatially-flat, low-density universe.

As (\ref{eq:08-19}) shows, the weak-lensing power spectrum $P_\kappa(l)$ depends on the product of a factor $\Omega_\mathrm{m0}^2$ due to the Poisson equation, times the amplitude $A$ of the matter power spectrum. An additional weak dependence on cosmological parameters is caused by the geometric weight function $\bar W(w',w)$. The cosmic-shear correlation function thus measures approximately the product $A\Omega_\mathrm{m0}^{2\alpha}$, $\alpha\lesssim1$, which means that the amplitude of the power spectrum is nearly degenerate with the matter density parameter. Only if it is possible to constrain $\Omega_\mathrm{m0}$ or $A$ in any other way can the degeneracy be broken.

We shall see later how this may work. The amplitude of the power spectrum $A$ is conventionally described by a parameter $\sigma_8^2$ which will be defined and described in more detail in \S~\ref{sec:X}. Weak lensing thus constrains the product $\sigma_8\Omega_\mathrm{m0}^\alpha$, and latest measurements find $\sigma_8(\Omega_\mathrm{m0}/0.25)^{0.64}\approx0.784\pm0.043$ \citep{FU08.1}.

Weak gravitational lensing is a fairly new field of cosmological research. Within a few years, it has considerably matured and returned cosmologically interesting constraints. Considerable potential is attributed to weak lensing in wide-area surveys in particular when combined with photometric redshift information, because this is expected to allow constraints on the growth of cosmic structures.

\begin{figure}[ht]
  \includegraphics[width=\hsize]{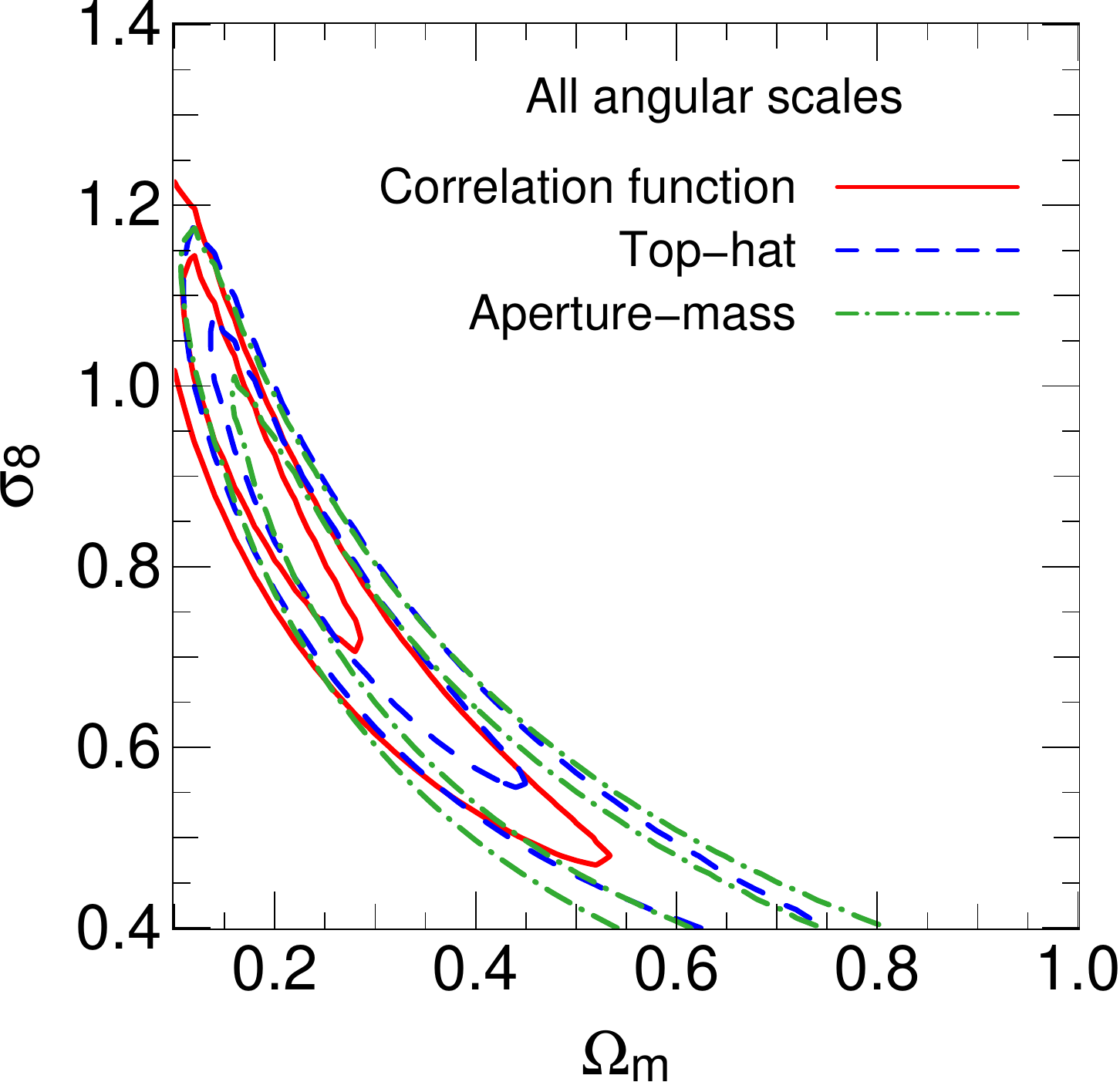}
\caption{Constraints in the $\Omega_\mathrm{m0}-\sigma_8$ plane from weak lensing on large angular scales in the CFHTLS. The Universe is assumed spatially flat here. (from \cite{FU08.1})}
\end{figure}

\section{Supernovae of Type Ia}\label{sec:IX}

\subsection{Standard candles and distances}\label{sec:IX-A}

\subsubsection{The principle}\label{sec:IX-A-1}

Before discussing supernovae of type Ia and their cosmological relevance, let us set the stage with a few illustrative considerations. Suppose we had a standard candle whose luminosity, $L$, we knew precisely. Then, according to the definition of the luminosity distance in (\ref{eq:01-20}), the distance can be inferred from the measured flux, $S$, through
\begin{equation}
  D_\mathrm{lum}=\sqrt{\frac{L}{4\pi S}}\;.
\label{eq:09-1}
\end{equation} 
Besides the redshift $z$, the luminosity distance will depend on the cosmological parameters,
\begin{equation}
  D_\mathrm{lum}=D_\mathrm{lum}
  (z;\Omega_\mathrm{m0},\Omega_{\Lambda0},H_0,\ldots)\;,
\label{eq:09-2}
\end{equation}
which can in principle be used to determine cosmological parameters from a set of distance measurements from a class of standard candles.

For this to work, the standard candles must be at a suitably high redshift for the luminosity distance to depend on the cosmological model. As we have seen in (\ref{eq:01-21}), all distance measures share the low-$z$ limit
\begin{equation}
  D\approx\frac{cz}{H_0}
\label{eq:09-3}
\end{equation}
and lose their sensitivity to all cosmological parameters except $H_0$.

In reality, we rarely know the absolute luminosity $L$ even of cosmological standard candles. The problem is that they need to be calibrated first, which is only possible from a flux measurement once the distance is known by other means, such as from parallaxes in case of the Cepheids. Supernovae, however, which are the subject of this chapter, are typically found at distances which are way too large to allow direct distance measurements. Therefore, the only way out is to combine distant supernovae with local ones, for which the approximate distance relation (\ref{eq:09-3}) holds.

Any measurement of flux $S_i$ and redshift $z_i$ of the $i$-th standard candle in a sample then yields an estimate for the luminosity $L$ in terms of the squared inverse Hubble constant,
\begin{equation}
  L=4\pi S_i\left(\frac{cz_i}{H_0}\right)^2\;.
\label{eq:09-4}
\end{equation}
Since all cosmological distance measures are proportional to the Hubble length $c/H_0$, the dependence on $H_0$ on both sides of (\ref{eq:09-1}) cancels, and the determination of cosmological parameters \emph{other than} the Hubble constant becomes possible. Thus, the first lesson to learn is that cosmology from distant supernovae requires a sample of nearby supernovae for calibration.

Of course, this nearby sample must satisfy the same criterion as the distance indicators used for the determination of the Hubble constant: their redshifts must be high enough for the peculiar velocities to be negligible, thus $z\gtrsim0.02$. If the redshifts are low enough for the linear approximation (\ref{eq:09-3}) to hold, the interpretation of the nearby sample is independent of the cosmological model.

It is important to note that it is not necessary to know the absolute luminosity $L$ even up to the uncertainty in $H_0$. If $L$ is truly independent of redshift, cosmological parameters could still be determined through (\ref{eq:09-1}) from the \emph{shape} of the measured relation between flux and redshift even though its precise \emph{amplitude} may be unknown. It is only important that the objects used \emph{are} standard candles, but not how \textit{bright} they are.

\subsubsection{Requirements and degeneracies}\label{sec:IX-A-2}

Let us now collect several facts about cosmological inference from standard candles. Since we aim at the determination of cosmological parameters, say $\Omega_\mathrm{m0}$, it is important to estimate the accuracy that we can achieve from measurements of the luminosity distance. Suppose we restrict the attention to spatially flat cosmological models, for which $\Omega_{\Lambda0}=1-\Omega_\mathrm{m0}$. Then, because the dependence on the Hubble constant was cancelled, $\Omega_\mathrm{m0}$ is the only remaining relevant parameter. We estimate the accuracy through first-order Taylor expansion,
\begin{equation}
  \Delta D_\mathrm{lum}\approx
  \frac{\d D_\mathrm{lum}}{\d\Omega_\mathrm{m0}}\,
  \Delta\Omega_\mathrm{m0}\;,
\label{eq:09-5}
\end{equation} 
about a fiducial model, such as a $\Lambda$CDM model with $\Omega_\mathrm{m0}=0.3$.

At a fiducial redshift of $z\approx0.8$, we find numerically
\begin{equation}
  \frac{\d\ln D_\mathrm{lum}}{\d\Omega_\mathrm{m0}}\approx-0.5\;,
\label{eq:09-6}
\end{equation}
which shows that a \emph{relative} distance accuracy of
\begin{equation}
  \frac{\Delta D_\mathrm{lum}}{D_\mathrm{lum}}\approx
  -0.5\Delta\Omega_\mathrm{m0}
\label{eq:09-7}
\end{equation}
is required to achieve an absolute accuracy of $\Delta\Omega_\mathrm{m0}$. For $\Delta\Omega_\mathrm{m0}\approx0.02$, say, distances thus need to be known to $\approx1\%$.

This accuracy requires sufficiently large supernova samples. Assuming Poisson statistics for simplicity and distance measurements to $N$ supernovae, the combined accuracy is
\begin{equation}
  |\Delta\Omega_\mathrm{m0}|\approx\frac{2}{\sqrt{N}}
  \frac{\Delta D_\mathrm{lum}}{D_\mathrm{lum}}\;.
\label{eq:09-8}
\end{equation}
That is, an accuracy of $\Delta\Omega_\mathrm{m0}\approx0.02$ can be achieved from $\approx100$ supernovae whose individual distances are known to $\approx10\%$.

Anticipating physical properties of type-Ia supernovae, their intrinsic peak luminosities in blue light are $L\approx3.3\times10^{43}\,\mathrm{erg\,s^{-1}}$, with a relative scatter of order $10\%$. \footnote{As we shall see later, type-Ia supernovae are \emph{standardisable} rather than standard candles, and the standardising procedure is currently not able to reduce the scatter further.} Given uncertainties $\Delta L$ in the luminosity $L$ and $\Delta S$ in the flux measurement $S$, error propagation on (\ref{eq:09-1}) yields the relative distance uncertainty
\begin{equation}
  \frac{\Delta D_\mathrm{lum}}{D_\mathrm{lum}}=\frac{1}{2}\left[
    \left(\frac{\Delta L}{L}\right)^2+
    \left(\frac{\Delta S}{S}\right)^2
  \right]^{1/2}\;.
\label{eq:09-9}
\end{equation}
Even if the flux could be measured precisely, the intrinsic luminosity scatter currently forbids distance determinations to better than $10\%$.

Fluxes have to be inferred from photon counts. For various reasons to be clarified later, supernova light curves should be determined until $\sim35$ days after the peak, when the luminosity has typically dropped to $\approx2.5\times10^{42}\,\mathrm{erg\,s^{-1}}$. The luminosity distance to $z\approx0.8$ is $\approx5\,\mathrm{Gpc}$, which implies fluxes $S\approx1.1\times10^{-14}\,\mathrm{erg\,s^{-1}\,cm^{-2}}$ at peak and $S\approx8.7\times10^{-16}\,\mathrm{erg\,s^{-1}\,cm^{-2}}$ 35 days later.

Dividing by an average photon energy of $5\times10^{-12}\,\mathrm{erg}$, multiplying with the area of a typical telescope mirror with $4\,\mathrm{m}$ diameter, and assuming a total quantum efficiency of $30\%$, we find detected photon fluxes of $S_\gamma\approx85\,\mathrm{s^{-1}}$ at peak and $S_\gamma\approx7\,\mathrm{s^{-1}}$ 35 days afterwards. These fluxes are typically distributed over a few CCD pixels.

Supernovae need to be identified against the background of their host galaxies and the sky brightness. For distant supernovae, the sky brightness on the area of the supernova image is typically 10--1000 times higher than the supernova itself, dominating the noise budget. Let $N_\mathrm{SN}$ and $N_\mathrm{sky}$ the number of photons received from the supernova and the sky, respectively, on the supernovae image, and assume for simplicity $N_\mathrm{sky}\approx100\,N_\mathrm{SN}$. Then, an estimate for the signal-to-noise ratio for the detection is
\begin{equation}
  \frac{\mbox{S}}{\mbox{N}}\approx\frac{N_\mathrm{SN}}{\sqrt{N_\mathrm{SN}+N_\mathrm{sky}}}
  \approx\frac{N_\mathrm{SN}}{\sqrt{100\,N_\mathrm{SN}}}\approx
  \frac{\sqrt{N_\mathrm{SN}}}{10}\;.
\label{eq:09-10}
\end{equation}
Signal-to-noise ratios of $\gtrsim10$ up to 35 days after the maximum thus require $N_\mathrm{SN}\approx10^4$ photons from the supernova. This implies exposure times of order $10^4/7\approx1400\,\mathrm{s}$, or about 20~minutes. Typical exposure times are thus of order $(15\ldots30)$ minutes to capture supernovae out to redshifts $z\sim1$. Then, the photometric error around peak luminosity is certainly less than the remaining scatter in the intrinsic luminosity, and relative distance accuracies of order $10\%$ are within reach.

However, a major difficulty is the fact that the identification of type-Ia supernovae requires spectroscopy. Sufficiently accurate spectra typically require exposures lasting 1--3 hours per spectrum on the world's largest telescopes, such as ESO's Very Large Telescope which consists of four individual mirrors with 8~m diameter each.

In order to see what we can hope to constrain by measuring luminosity distances, we form the gradient of $D_\mathrm{lum}$ in the $\Omega_\mathrm{m0}$-$\Omega_{\Lambda0}$ plane,
\begin{equation}
  \vec g\equiv\left(
    \frac{\partial D_\mathrm{lum}}{\partial\Omega_\mathrm{m0}},
    \frac{\partial D_\mathrm{lum}}{\partial\Omega_{\Lambda0}}
  \right)^T\;,
\label{eq:09-11}
\end{equation}
at a fiducial $\Lambda$CDM model with $\Omega_\mathrm{m0}=0.3$. When normalised to unit length, it turns out to point into the direction
\begin{equation}
  \vec g=\cvector{-0.76\\0.65}\;.
\label{eq:09-12}
\end{equation}

This vector rotated by $90^\circ$ then points into the direction in the $\Omega_\mathrm{m0}$-$\Omega_{\Lambda0}$ plane along which the luminosity distance does \emph{not} change. Thus, near the fiducial $\Lambda$CDM model, the parameter combination
\begin{equation}
  P\equiv\vec g\cdot \cvector{\Omega_\mathrm{m0}\\\Omega_{\Lambda0}}=
  -0.76\,\Omega_\mathrm{m0}+0.65\,\Omega_{\Lambda0}
\label{eq:09-13}
\end{equation}
is degenerate. The degeneracy direction, characterised by the vector $\mathcal{R}(\pi/2)\,\vec g=(0.65,0.76)^T$, encloses an angle of $\arctan(0.76/0.65)=49.5^\circ$ with the $\Omega_\mathrm{m0}$ axis, almost along the diagonal from the lower left to the upper right corner of the parameter plane. Thus, it is almost perpendicular to the degeneracy direction obtained from the curvature constraint due to the CMB. This illustrates how parameter degeneracies can very efficiently be broken by \emph{combining} suitably different types of measurement. Moreover, combining supernova data from a wide redshift range partially lifts the parameter degeneracy obtained from them already.

\subsection{Supernovae}\label{sec:IX-B}

\subsubsection{Types and classification}\label{sec:IX-B-1}

Supernovae are ``eruptively variable'' stars. A sudden rise in brightness is followed by a gentle decline. They are unique events which at peak brightness reach luminosities comparable to those of an entire galaxy, or $(10^{10}\ldots10^{11})\,L_\odot$. They reach their maxima within days and fade within several months.

Supernovae are traditionally characterised by their early spectra \citep{FI97.1}. If hydrogen lines are missing, they are of type I, otherwise of type II. Type-Ia supernovae show silicon lines, unlike type-Ib/c supernovae, which are distinguished by the prominence of helium lines. Normal type-II supernovae have spectra dominated by hydrogen. They are subdivided according to their lightcurve shape into type-IIL and type-IIP. Type-IIb supernova spectra are dominated by helium instead.

Except for type-Ia, supernovae arise due to the collapse of a massive stellar core, followed by a thermonuclear explosion which disrupts the star by driving a shock wave through it. Core-collapse supernovae of type-II arise from stars with masses between $(8\ldots30)\,M_\odot$, those of type-I (i.e.~types Ib/c) from more massive stars \citep{WO02.1}.

Type-Ia supernovae, which we are dealing with here, arise when a white dwarf is driven towards the Chandrasekhar mass limit by mass overflowing from a companion star. In a binary system, the more massive star evolves faster and can reach its white-dwarf stage before its companion leaves the main sequence and becomes a red giant. When this happens, and the stars are close enough, matter will flow from the expanding red giant on the white dwarf.

Electron degeneracy pressure can stabilise white dwarfs up to the Chandrasekhar mass limit of $\sim1.4\,M_\odot$ \citep{CH84.1,HI00.1}. Because the core material is degenerate, its pressure is independent of its temperature. The mass accreted from the companion star increases the pressure until nuclear burning can begin in isolated places somewhere in the core. The electron degeneracy is lifted, the temperature rises dramatically, and the thermonuclear runaway sets in. Neutrinos produced in inverse beta decays carry away much of the explosion energy unnoticed because they can leave the supernova essentially without further interaction.

This thermonuclear runaway destroys the white dwarf. Since this type of explosion (more precisely, deflagration) involves an approximately fixed amount of mass, it is physically plausible that the explosion releases a fixed amount of energy. Thus, the Chandrasekhar mass limit is the main responsible for type-Ia supernovae to be approximate standard candles.

The nuclear fusion processes in type-Ia supernovae converts the carbon and oxygen in the core of the white dwarf into $\nuc{56}{Ni}$, which later decays through $\nuc{56}{Co}$ into stable $\nuc{56}{Fe}$. According to detailed numerical explosion models, the nuclear fusion is started at random points near the centre of the white dwarf \citep{RO07.1}.

The presence of silicon lines in the type-Ia spectra indicates that not all of the white dwarf's material is converted into $\nuc{56}{Ni}$. This shows that there is no explosion, but a deflagration, in which the flame front propagates at velocities below the sound speed. The deflagration can burn the material fast enough if it is turbulent, because the turbulence dramatically increases the surface of the flame front and thus the amount of material burnt per unit time. Theoretical models predict that $\sim0.5\,M_\odot$ of $\nuc{56}{Ni}$ are typically produced.

The peak brightness is reached when the deflagration front reaches the former white dwarf's surface and drives it as a rapidly expanding envelope into the surrounding space. The $\gamma$ photons released in the nuclear fusion processes are redshifted by scattering off the expanding material and finally leave the explosion site as X-ray, UV, optical and infrared photons.

Once the thermonuclear fusion has ended, additional energy is released by the $\beta$ decay of $\nuc{56}{Co}$ into $\nuc{56}{Fe}$ with a half life of $77.12$ days. The exponential nature of the radioactive decay causes the typical exponential decline phase in supernova light curves \citep{TR67.1,CO69.1}. Since the supernova light has to propagate through the expanding envelope before we can see it, the opacity of the envelope and thus its metalicity are important for the appearance of the supernova \citep{BL06.1}.

\subsubsection{Observations}\label{sec:IX-B-2}

Since supernovae are transient phenomena, they can only be detected by sufficiently frequent monitoring of selected areas in the sky. Typically, fields are selected by their accessibility for the telescope to be used and the least degree of absorption by the Galaxy. Since a type-Ia supernova event lasts for about a month, monitoring is required every few days.

Supernovae are then detected by differential photometry, in which the average of all preceding images is subtracted from the last image taken. Since the seeing varies, the images appear convolved with point-spread functions of variable width even if they are taken with identical optics, thus the objects on them appear more or less blurred. Before they can be meaningfully subtracted, they therefore have to be convolved with the same effective point-spread function. This causes several complications in the later analysis procedure, in particular with the photometry.

Of course, this detection procedure returns many variable stars and supernovae of other types, which are not standard candles and have to be removed from the sample. Pre-selection of type-Ia candidates is done by colour and the light-curve shape, but the identification of type-Ia supernovae requires spectroscopy in order to identify the decisive silicon lines at 6347~\AA{} and 6371~\AA{}. Since these lines move out of the optical spectrum at redshifts $z\gtrsim0.5$, near-infrared observations are crucially important for the high-redshift supernovae relevant for cosmology.

Nearby supernovae, which are needed for calibration, reveal that type-Ia supernovae are \emph{not} standard candles, but show a substantial scatter in luminosity. It turned out that there is an empirical relation between the duration of the supernova event and its peak brightness in that brighter supernovae last longer \citep{PH99.1}. This relation between the light-curve shape and the brightness can be used to \emph{standardise} type-Ia supernovae. It was seen as a major problem for their cosmological interpretation that the origin for this relation was unknown, and that its application to high-redshift supernovae was based on the untested assumption that the relation found and calibrated with local supernovae would also hold there. Recent simulations indicate that the relation is an opacity effect \citep{KA07.1}: brighter supernovae produce more $\nuc{56}{Ni}$ and thus have a higher metalicity, which causes the envelope to be more opaque, the energy transport through it to be slower, and therefore the supernova to last longer.

Thus, before a type-Ia supernova can be used as a standard candle, its duration must be determined, which requires the light-curve to be observed over sufficiently long time. It must be taken into account here that the cosmic expansion leads to a time dilation, due to which supernovae at redshift $z$ appear longer by a factor of $(1+z)$. We note in passing that the confirmation of this time dilation effect indirectly supports the cosmic expansion. \emph{After} the standardisation, the scatter in the peak brightnesses of nearby supernovae is substantially reduced. This encourages (and justifies) their use as \emph{standardisable} candles for cosmology.

The remaining relative uncertainty is now typically between $(10\ldots15)\%$ for individual supernovae. Since, as we have seen following (\ref{eq:09-7}), we require relative distance uncertainties at the per cent level, of order a hundred distant supernovae are needed before meaningful cosmological constraints can be placed, which justifies the remark after (\ref{eq:09-8}).

An example for the several currently ongoing supernova surveys is the \emph{Supernova Legacy Survey} (SNLS, \cite{AS06.1}) in the framework of the Canada-France-Hawaii Legacy Survey (CFHTLS), which is carried out with the 4-m Canada-France-Hawaii telescope on Mauna Kea. It monitors four fields of one square degree each five times during the 18 days of dark time between two full moons (lunations).

Differential photometry is performed to find out variables, and candidate type-Ia supernovae are selected by light-curve fitting after removing known variable stars. Spectroscopy on the largest telescopes (mostly ESO's VLT, but also the Keck and Gemini telescopes) is then needed to identify type-Ia supernovae. To give a few characteristic numbers, the SNLS has taken 142 spectra of type-Ia candidates during its first year of operation, of which 91 were identified as type-Ia supernovae.

The light curves of these objects are observed in several different filter bands. This is important to correct for interstellar absorption. Any dimming by intervening material makes supernovae appear fainter, and thus more distant, and will bias the cosmological results towards faster expansion. Since the intrinsic colours of type-Ia supernovae are characteristic, any deviation of the observed from the intrinsic colours signals interstellar absorption which is corrected by adapting the amount of absorption such that the observed is transformed back into the intrinsic colour.

This correction procedure is expected to work well unless there is material on the way which absorbs equally at all wavelengths, so-called ``grey dust'' \citep{AG99.1}. This could happen if the absorbing dust grains are large compared to the wavelength. Currently, it is quite difficult to conclusively rule out grey dust, although it is implausible based on the interstellar absorption observed in the Galaxy \citep{RI04.1,OS05.2}.

After applying the corrections for absorption and duration, each supernova yields an estimate for the luminosity distance to its redshift. Together, the supernovae in the observed sample constrain the evolution of the luminosity distance with redshift, which is then fit varying the cosmological parameters except for $H_0$, i.e.~typically $\Omega_\mathrm{m0}$ and $\Omega_\mathrm{\Lambda0}$. This yields an ``allowed'' region in the $\Omega_\mathrm{m0}$-$\Omega_\mathrm{\Lambda0}$ plane compatible with the measurements which is degenerate in the direction calculated before.

\begin{figure}[ht]
  \includegraphics[width=\hsize]{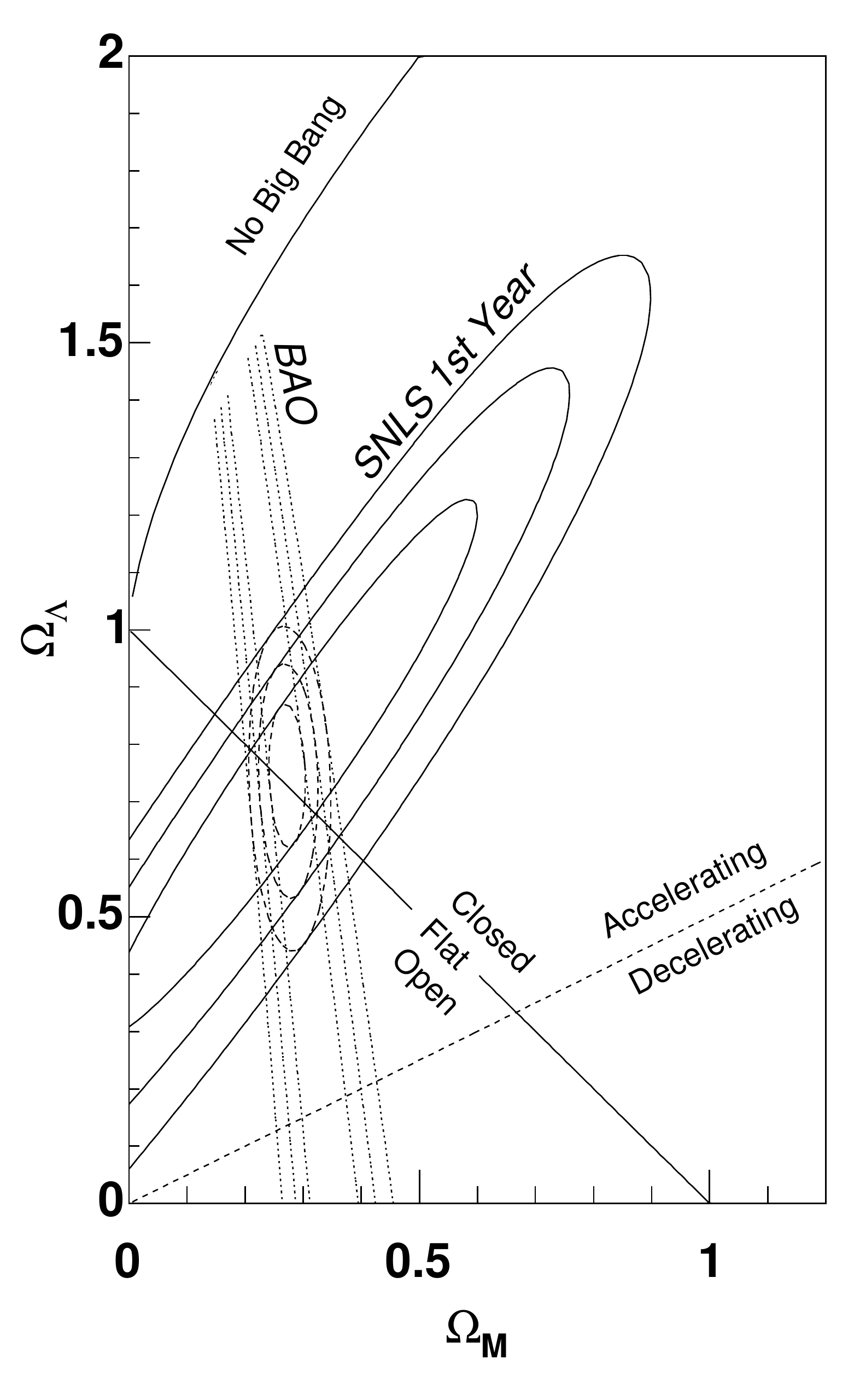}
\caption{Constraints in the $\Omega_\mathrm{m0}$-$\Omega_\mathrm{\Lambda0}$ plane obtained from type-Ia supernovae in the SNLS. (from \cite{AS06.1}).}
\end{figure}

More information or further assumptions are necessary to break the degeneracy. The most common assumption, justified by the CMB measurements, is that the Universe is spatially flat. Based upon it, the SNLS data yield a matter density parameter of
\begin{equation}
  \Omega_\mathrm{m0}=0.263\pm0.037\;.
\label{eq:09-14}
\end{equation}
This is a remarkable result. First of all, it confirms the other independent measurements we have already discussed, which were based on kinematics, cluster evolution and the CMB. Second, it shows that, in the assumed spatially flat universe, the dominant contribution to the total energy density must come from something else than matter, possibly the cosmological constant.

It is important for the later discussion to realise in what way the parameter constraints from supernovae differ from those from the CMB. The fluctuations in the latter show that the Universe is at least nearly spatially flat, and the density parameters in dark and baryonic matter are near $0.24$ and $0.04$, respectively. The rest must be the cosmological constant, or the dark energy. Arising \emph{early} in the cosmic history, the CMB itself is almost insensitive to the cosmological constant, and thus it can only constrain it indirectly.

Type-Ia supernovae, however, measure the angular-diameter distance during the \emph{late} cosmic evolution, when the cosmological constant is much more important. As (\ref{eq:09-13}) shows, the luminosity distance constrains the \emph{difference} between the two parameters,
\begin{equation}
  \Omega_{\Lambda0}=1.17\,\Omega_\mathrm{m0}+P\;,
\label{eq:09-15}
\end{equation} 
where the degenerate parameter $P$ is determined by the measurement. Assuming $\Omega_{\Lambda0}=1-\Omega_\mathrm{m0}$ as in a spatially-flat universe yields
\begin{equation}
  P=1-2.17\,\Omega_\mathrm{m0}\approx0.43
\label{eq:09-16}
\end{equation}
from the SNLS first-year result (\ref{eq:09-14}), illustrating that the survey has constrained the density parameters to follow the \emph{relation}
\begin{equation}
  \Omega_{\Lambda0}\approx1.17\,\Omega_\mathrm{m0}+0.43\;.
\label{eq:09-17}
\end{equation}

The relative \emph{acceleration} of the universe, $\ddot a/a$, is given by (\ref{eq:01-7}), which can be simplified to read
\begin{equation}
  \frac{\ddot a}{a}=H_0^2\left(
    \Omega_{\Lambda0}-\frac{\Omega_\mathrm{m0}}{2a^3}
  \right)
\label{eq:09-18}
\end{equation} 
if matter is pressure-less. Thus, the expansion of the universe accelerates today ($a=1$) if $\ddot a=H_0^2(\Omega_{\Lambda0}-\Omega_\mathrm{m0}/2)>0$, or $\Omega_{\Lambda0}>\Omega_\mathrm{m0}/2$. Given the measurement (\ref{eq:09-17}), the conclusion seems inevitable that the Universe's expansion does indeed accelerate today \citep{RI98.1,PE99.2,LE05.1,AS06.1}.

If the Universe is indeed spatially flat, then the transition between decelerated and accelerated expansion happened at
\begin{equation}
  1-0.263\approx\frac{0.263}{2a^3}\quad\Rightarrow\quad
  a=0.56\;,
\label{eq:09-19}
\end{equation} 
or at redshift $z\approx0.78$. Luminosity distances to supernovae at higher redshifts should show this transition, and in fact they do \citep{RI04.1}.

\subsubsection{Potential problems}\label{sec:IX-B-3}

The problem with possible grey dust has already been mentioned: While the typical colours of type-Ia supernovae allow the detection and correction of the reddening coming with typical interstellar absorption, grey dust would leave no trace in the colours and remain undetectable. However, grey dust would re-emit the absorbed radiation in the infrared and add to the infrared background, which is quite well constrained. It thus seems that grey dust is not an important contaminant, if it exists.

Gravitational lensing is inevitable for distant supernovae \citep{HO98.1}. Depending on the line-of-sight, they are either magnified or demagnified. Since high magnifications due to non-linear structures may occasionally happen, the magnification distribution must be skewed towards demagnification to keep the mean of unit magnification. Thus, the \emph{most probable} magnification experienced by supernova is \emph{below unity}. In other words, lensing may lead to a slight demagnification if lines-of-sight towards type-Ia supernovae are random with respect to the matter distribution. In any case, the \emph{rms} cosmic magnification adds to the intrinsic scatter of the supernova luminosities. It may become significant for redshifts $z\gtrsim1$.

It is a difficult and debated question whether supernovae at high redshifts are intrinsically the same as at low redshifts where they are calibrated. Should there be undetected systematic differences, cosmological inferences could be wrong. In particular, it may be natural to assume that metalicities at high redshifts are lower than at low redshifts. Since supernovae last longer if their atmospheres are more opaque, lower metalicity may imply shorter supernova events, leading to underestimated luminosities and overestimated distances. Simulations of type-Ia supernovae, however, seem to show that such an effect is probably not significant. For this and other systematic effects, see \cite{LE01.1}.

It was also speculated that distant supernovae may be intrinsically bluer than nearby ones due to their possibly lower metalicity. Should this be so, the extinction correction, which is derived from reddening, would be underestimated, causing intrinsic luminosities to be under- and luminosity distances to be overestimated. Thus, this effect would lead to an \emph{underestimate} of the expansion rate and counteract the cosmological constant. There is currently no indication of such a colour effect.

Supernovae of types Ib/c may be mistaken for those of type Ia if the identification of the characteristic silicon lines fails for some reason. Since they are typically fainter than type-Ia supernovae, they would contaminate the sample and bias results towards higher luminosity distances, and thus towards a higher cosmological constant. It seems, however, that the possible contamination by non-type-Ia supernovae is so small that it has no noticeable effect.

Several more potential problems exist. It has been argued for a while that, if the evidence for a cosmological constant was based exclusively on type-Ia supernovae, it would probably not be considered entirely convincing. However, since the supernova observations come to conclusions compatible with virtually all independent cosmological measurements, they add substantially to the persuasiveness of the cosmological standard model. Moreover, recent supernova simulations reveal good physical reasons why they should in fact be reliable, standardisable candles.

\section{The Normalisation of the Power Spectrum}\label{sec:X}

\subsection{Introduction}\label{sec:X-A}

We saw in \S~\ref{sec:VII-B} that the measured power spectrum of the galaxy distribution follows the CDM expectation in the range of wave numbers where current large surveys allow its measurement. This range can be extended to some degree towards smaller scales by measuring the autocorrelation of hydrogen absorption lines in the spectra of distant quasars. Such observations of the power spectrum of the so-called Lyman-$\alpha$ forest lines show that the power spectrum does indeed turn towards the asymptotic behaviour $\propto k^{-3}$ \citep{MD05.1,VI08.1}. In addition, we have seen that the peak location agrees with the expectation for a universe with $\Omega_\mathrm{m0}\approx0.3$ and $h\approx0.72$. This indicates that the CDM expectation for the dark-matter power spectrum is indeed at least very close to its real shape, which is a remarkable finding.

Although the shape of the power spectrum could thus be quite well established, its amplitude still poses a surprisingly obstinate problem. We shall see in this section why it is so difficult to measure. For this purpose, we shall discuss four ways of measuring $\sigma_8$; the amplitude of large-scale temperature fluctuations in the CMB, the cosmic-shear autocorrelation function, the abundance and evolution of the galaxy-cluster population, and the statistics of Lyman-$\alpha$ forest lines.

For historical reasons, the amplitude of the dark-matter power spectrum is characterised by the variance of the density fluctuations within spheres of $8\,h^{-1}\mathrm{Mpc}$ radius. More generally, one imagines placing spheres of radius $R$ randomly and measuring the density-contrast variance within them. Since the variance in Fourier space is characterised by the power spectrum, it can be written as
\begin{equation}
  \sigma_R^2=\int_0^\infty\frac{\d^3k}{(2\pi)^3}P_\delta(k)W_R^2(k)\;,
\label{eq:10-1}
\end{equation} 
where $W_R(k)$ is a \emph{window function} selecting the $k$ modes contributing to the variance within the spheres.

Imagining spheres of radius $R$ in real space, the window function should be the Fourier transform of a step function, which is inconvenient because it extends to infinite wave numbers. It is thus more common to use either Gaussians, since they Fourier transform to Gaussians, or step functions in Fourier space. For simplicity of the following illustrative calculations, we use the latter choice, thus
\begin{equation}
  W_R(k)=\Theta(k_R-k)=\Theta\left(\frac{2\pi}{R}-k\right)\;.
\label{eq:10-2}
\end{equation} 
This is a step function dropping to zero at $k=2\pi/R$. Inserting this into (\ref{eq:10-1}) gives
\begin{equation}
  \sigma_R^2=\int_0^{2\pi/R}\frac{k^2\d k}{2\pi^2}P_\delta(k)\;.
\label{eq:10-3}
\end{equation}
All modes \emph{larger} than $R$ contribute to the density fluctuations in spheres of radius $R$ because all smaller modes average to zero. The normalisation of the power spectrum is usually expressed in terms of $\sigma_8$, fixing $R$ to its historical value of $8\,h^{-1}\,\mathrm{Mpc}$.

\subsection{Fluctuations in the CMB}\label{sec:X-B}

\subsubsection{The large-scale fluctuation amplitude}\label{sec:X-B-1}

We saw in \S~\ref{sec:VI-B-3} that the long-wavelength (low-$k$) tail of the CMB power spectrum is caused by the Sachs-Wolfe effect, giving rise to relative temperature fluctuations (\ref{eq:06-17}) in terms of the Newtonian potential fluctuations $\delta\Phi$. The three-dimensional temperature-fluctuation power spectrum is then
\begin{equation}
  P_\tau(k)=\frac{P_\Phi(k)}{9c^4}\;.
\label{eq:10-5}
\end{equation} 
The Poisson equation in its form (\ref{eq:08-8}) implies that the power spectra of potential- and density fluctuations are related through
\begin{equation}
  P_\Phi(k)=\frac{9H_0^4}{4}\Omega_\mathrm{m0}^2\left(
    \frac{D_+(a_\mathrm{rec})}{a_\mathrm{rec}}
  \right)^2\frac{P_\delta(k)}{k^4}\;,
\label{eq:10-6}
\end{equation}
where the linear growth factor $D_+(a_\mathrm{rec})$ was introduced to relate the potential-fluctuation power spectrum at the time of decoupling to the \emph{present} density-fluctuation power spectrum $P_\delta(k)$.

Now we need to account for projection effects. A three-dimensional mode with comoving wave number $k$ and comoving wavelength $\lambda=2\pi/k$ appears under an angle $\theta=\lambda/w_\mathrm{rec}$, where $w_\mathrm{rec}\equiv w(a_\mathrm{rec})$ is the comoving angular-diameter distance (\ref{eq:06-22}) to the CMB. The \emph{angular} wave number under which the mode appears is thus
\begin{equation}
  l\approx\frac{2\pi}{\theta}\approx w_\mathrm{rec}k\;.
\label{eq:10-8}
\end{equation}

Expressing now the power spectrum (\ref{eq:10-5}) in terms of the \emph{angular} wave number $l$ yields, with (\ref{eq:10-6})
\begin{equation}
  P_\tau(l)\propto\left(\frac{H_0}{c}\right)^4\Omega_\mathrm{m0}^2\left(
    \frac{D_+(a_\mathrm{rec})}{a_\mathrm{rec}}
  \right)^2\frac{1}{w_\mathrm{rec}^2}\frac{w_\mathrm{rec}^4}{l^4}P_\delta\left(
    \frac{l}{w_\mathrm{rec}}
  \right)\;,
\label{eq:10-9}
\end{equation}
where the factor $w_\mathrm{rec}^{-2}$ arises because of the transformation from \emph{spatial} to \emph{angular} wave numbers $l$ [cf.~Limber's equation (\ref{eq:08-17})], and the factor $w_\mathrm{rec}^4/l^4$ expresses the factor $k^{-4}$ from the squared Laplacian. This shows that the angular power spectrum $P_\tau(l)$ of the large-scale CMB temperature fluctuations can only be translated into the amplitude of the dark-matter power spectrum $A$ if the cosmological parameters are already known well enough.

A further complication is added by the integrated Sachs-Wolfe effect introduced in \S~\ref{sec:VI-B-6}. It depends on $D_+(a)/a$ and adds secondary anisotropies to the CMB unless $D_+(a)=a$. The primordial CMB fluctuations are then lower than measured and need to be corrected by subtracting the integrated Sachs-Wolfe contribution, which adds a further dependence on the cosmological parameters.

\subsubsection{Translation to $\sigma_8$}\label{sec:X-B-2}

Two more complications arise in the translation of the large-scale amplitude $A$ to $\sigma_8$. Since structures with wave numbers larger than the peak location $k_\mathrm{eq}$ in the power spectrum contribute to $\sigma_8$, the dependence of $k_\mathrm{eq}$ on the cosmological parameters comes in. Finally, the spectral index $n_\mathrm{s}$ may affect the extrapolation from large to small scales substantially because of the long lever arm between the scales involved.

Of course, one could also use the small-scale part of the CMB power spectrum for normalising the dark-matter power spectrum. Due to the acoustic oscillations, however, this part depends in a much more complicated way on additional cosmological parameters, such as the baryon density. Reading $\sigma_8$ off the low-order multipoles is thus a safer, albeit intricate procedure.

Even if the cosmological parameters are now known well enough to translate the low-order CMB multipoles to $\sigma_8$, an additional uncertainty remains. We know that, although the Universe became neutral $\sim400,000$ years after the Big Bang, it must have been reionised after the first stars and other sources of UV radiation formed. Since then, CMB photons are travelling through ionised material again and experience Thomson (or Compton) scattering. The \emph{optical depth} for Thomson scattering is
\begin{equation}
  \tau=\int n_\mathrm{e}\sigma_\mathrm{T}\,c\d t\;,
\label{eq:10-15}
\end{equation}
where $n_\mathrm{e}$ is the number density of free electrons and $\sigma_\mathrm{T}$ is the Thomson scattering cross section. After propagating through the optical depth $\tau$, the CMB fluctuation amplitude is reduced by $\exp(-\tau)$.

Of course, the CMB photons cannot disappear through Thomson scattering, thus the CMB's overall intensity cannot change in this way, but the fluctuation amplitudes are lowered in this diffusion process. The optical depth $\tau$ depends on the path length through ionised material. In view of the CMB, this means that the degree of fluctuation damping depends on the \emph{reionisation redshift}, i.e.~the redshift after which the cosmic baryons were transformed back into a plasma. Unless the reionisation redshift is known, we cannot know by how much the CMB fluctuations were suppressed.

So far, the reionisation redshift can be estimated in two ways. First, as discussed in \S~\ref{sec:VI-B-4}, Thomson scattering creates linear polarisation. Of course, the polarisation due to \emph{reionised} material appears superposed on the primordial polarisation, but on different angular scales. The characteristic scale for \emph{secondary} polarisation is the horizon size at the reionisation redshift, which is much larger than the typical scales of the primordial polarisation. Thus, the reionisation redshift can be inferred from large-scale features in the CMB polarisation, provided the cosmological parameters are known well enough to translate angular scales into physical scales \citep{ZA97.1}. Modern cosmological parameter estimates aim at fitting the available data simultaneously for all relevant parameters mostly by Monte-Carlo Markov chain techniques, as explained in detail e.g.~in the textbook by \citealt{DU08.2}.

Unfortunately, this is aggravated by the polarised microwave radiation from the Milky Way. Synchrotron and dust emission can be substantially polarised and mask the CMB polarisation, which can only be measured reliably if the foregrounds of Galactic origin can be accurately subtracted \citep{PA07.1}. Thus, the degree to which the foreground polarisation is known directly determines the accuracy of the $\sigma_8$ parameter derived from the CMB fluctuations. This contributes considerably to the remaining uncertainty in the $\sigma_8$ derived from the 5-year WMAP data given in Tab.~\ref{tab:1}.

The other way to constrain the reionisation redshift uses the spectra of distant quasars. Light with wavelengths shorter than the Lyman-$\alpha$ wavelength cannot propagate through neutral hydrogen because it is immediately absorbed. Therefore, quasar spectra released before the reionisation redshift must be completely absorbed blueward of the Lyman-$\alpha$ emission line. The appearance of this so-called Gunn-Peterson effect \citep{GU65.1} at high redshift thus signals the transition from ionised into neutral material. Using this technique, the reionisation was estimated to end at redshifts $\sim6\ldots8$ \citep{FA06.1}, while the secondary polarisation of the CMB implies a reionisation redshift of $11.0\pm1.4$ if instantaneous reionisation is assumed \citep{DU08.1}. These apparently discrepant values do not contradict each other because a small admixture of neutral hydrogen is enough to produce the Gunn-Peterson effect, which therefore persists until reionisation has completed.

\subsection{Cosmological weak lensing}\label{sec:X-C}

Compared to the outlined procedure to obtain $\sigma_8$ from the CMB, it appears completely straightforward to derive it from the cosmic-shear measurements. As we have discussed in (\ref{sec:VIII-B-3}), the cosmic-shear power spectrum is proportional to $\Omega_\mathrm{m0}^{2\alpha}$ times the amplitude $A$ of the dark-matter power spectrum, which leads to the approximate degeneracy $\Omega_\mathrm{m0}^\alpha\sigma_8\approx\mbox{const.}$ between $\sigma_8$ and the matter-density parameter $\Omega_\mathrm{m0}$.

A more subtle dependence on $\Omega_\mathrm{m0}$ and to some degree also on other cosmological parameters is introduced by the geometrical weight function $\bar W(w',w)$ shown in (\ref{eq:08-20}), and by the growth of the power spectrum along the line-of-sight. This slightly modifies the form of the $\sigma_8$-$\Omega_\mathrm{m0}$ degeneracy, but does not lift it. However, knowing $\Omega_\mathrm{m0}$ well enough, we should be able to read $\sigma_8$ off the cosmic-shear correlation function. However, there are three problems associated with that.

First, the cosmic shear measured on angular scales below $\sim10'$ is heavily influenced by the onset of non-linear structure growth and the effect this has on the dark-matter power spectrum. While the linear growth factor can be straightforwardly calculated analytically, non-linear growth can only be quantified by means of large numerical simulations and recipes derived from them \citep{PE96.1,SM03.1}. Insufficient knowledge of the non-linear dark-matter power spectrum is a major uncertainty in the cosmological interpretation of cosmic shear.

Second, the amplitude of cosmological weak-lensing effects depends on the redshift distribution of the sources used for measuring ellipticities. Since these background galaxies are typically very faint, it is demanding to measure their redshifts. Two methods have typically been used. One adapts the known redshift distribution of sources in narrow, very deep observations such as the \emph{Hubble Deep Field} to the characteristics of the observation to be analysed. The other relies on photometric redshifts, i.e.~redshift estimates based on multi-band photometry. Yet, the precise redshift distribution of the background sources adds additional uncertainty to estimates of $\sigma_8$.

Third, it is possible that systematic effects remain in weak-lensing measurements because the effect is so small, and many corrections have to be applied to measured ellipticities before the cosmic shear can be extracted. Advanced correction methods have been developed which made the $B$-mode contamination almost or completely disappear. This is good news, but it does not yet guarantee the absence of other systematic effects in the data.

Nonetheless, cosmic lensing, combined with estimates of the matter-density parameter, is perhaps the most promising method for precisely determining $\sigma_8$. Table~\ref{tab:10-1} lists values of $\sigma_8$ derived from some cosmic-shear measurements under the assumption of $\Omega_\mathrm{m0}=0.30$ in a spatially-flat universe.

\begin{table}[ht]
\begin{center}
\begin{tabular}{|p{0.2\hsize}|p{0.35\hsize}|p{0.35\hsize}|}
\hline
$\sigma_8$ & data & reference \\
\hline
$0.86^{+0.09}_{-0.13}$ & RCS  & \cite{HO02.2} \\
$0.71^{+0.12}_{-0.16}$ & CTIO & \cite{JA03.1} \\
$0.72\pm0.09$ & Combo-17 & \cite{BR03.1} \\
$0.97\pm0.13$ & Keck-II  & \cite{BA03.1} \\
$1.02\pm0.16$ & HST/STIS & \cite{RH04.1} \\
$0.83\pm0.07$ & Virmos-Descart & \cite{WA05.1} \\
$0.68\pm0.13$ & GEMS & \cite{HE05.1} \\
$0.85\pm0.06$ & CFHTLS-wide & \cite{HO06.1} \\
$0.80\pm0.1$ & GaBoDS & \cite{HE07.1} \\
$0.866^{+0.085}_{-0.068}$ & COSMOS & \cite{MA07.2} \\
$0.74\pm0.04$ & 100 sq.~deg. combined & \cite{BE07.1} \\
$0.70\pm0.04$ & CFHTLS-wide & \cite{FU08.1} \\
\hline
\end{tabular}
\end{center}
\caption{Values for $\sigma_8$ derived from cosmic-shear measurements under the assumption of a spatially-flat universe with $\Omega_\mathrm{m0}=0.3$.}
\label{tab:10-1}
\end{table}

The systematic deviations between them are probably due to three different uncertainties. First, the redshift distribution of the background sources needs to be known; second, nonlinear evolution of the matter power spectrum need to be accurately modelled; and third, all kinds of PSF distortions need to be corrected.

\subsection{Galaxy clusters}\label{sec:X-D}

\subsubsection{The mass function}\label{sec:X-D-1}

Based on the assumption that the density contrast is a Gaussian random field and the spherical-collapse model, \cite{PR74.1} derived a mass function for dark-matter halos. It compares the standard deviation $\sigma_R$ of the density-fluctuation field to the linear density-contrast threshold $\delta_\mathrm{c}\approx1.686$ for collapse in the spherical-collapse model. The mean mass contained in spheres of radius $R$ sets the halo mass, which brings the mean (dark-) matter density $\bar\rho$ into the game.

The standard deviation $\sigma_R$ is related to the power spectrum. For convenience, we introduce an effective slope
\begin{equation}
  n=\frac{\d\ln P(k)}{\d\ln k}
\label{eq:10-16}
\end{equation} 
for the power spectrum, which will of course be scale-dependent. On large scales, $n\approx1$, while $n\to-3$ on small scales, i.e.~for small halo masses. For galaxy clusters, $n\approx-1$.

We introduce the \emph{non-linear} mass scale $M_*$ as the mass contained in spheres of radius $R$ chosen such that $\sigma_R=1$. Since $\sigma_R$ grows with the linear growth factor $D_+(a)$, the non-linear mass grows with time. It is convenient here to express the amplitude of the power spectrum, and thus $\sigma_8$, in terms of $M_*$. In sufficient approximation,
\begin{equation}
  \sigma_R=\left(\frac{M_*}{M}\right)^\alpha\;,\quad
  \alpha\equiv\frac{1}{2}\left(1+\frac{n}{3}\right)\;.
\label{eq:10-17}
\end{equation}

In terms of the dimensionless mass $m\equiv M/M_*$, the Press-Schechter mass function can be cast into the form
\begin{equation}
  N(m,a)\d m=\sqrt{\frac{2}{\pi}}
  \frac{\bar\rho\delta_\mathrm{c}}{M_*^2D_+(a)}\alpha m^{\alpha-2}
  \exp\left(
    -\frac{\delta_\mathrm{c}^2}{2D_+^2(a)}m^{2\alpha}
  \right)\d m\;.
\label{eq:10-19}
\end{equation}

The Press-Schechter mass function, and some improved variants of it \citep{JE01.1,SH02.1}, have been spectacularly confirmed by numerical simulations \citep{SP05.2}. It shows that the mass function is a power law with an exponential cut-off near the non-linear mass scale $M_*$. For galaxy clusters, $n\approx-1$, thus $\alpha\approx1/3$, and
\begin{equation}
  N(m,a)\d m\propto m^{-5/3}\exp\left(
    -\frac{\delta_\mathrm{c}^2}{2D_+^2(a)}m^{2/3}
  \right)\d m\;,
\label{eq:10-20}
\end{equation}
with an amplitude characterised by $M_*$, the mean dark-matter density $\bar\rho$, and the growth factor $D_+(a)$.

This opens a way to constrain cosmological parameters as well as $\sigma_8$ with galaxy clusters: if the abundance and evolution of the cluster mass function can be measured, they can be determined from the mass scale of the exponential cut-off and the amplitude of the power-law end. Today, the non-linear mass scale is a few times $10^{13}\,M_\odot$. Therefore, the exponential cut-off in the halo mass will not be seen in the galaxy mass function. Clusters, however, show a pronounced exponential cut-off \citep{RO02.1}, and thus their population is very sensitive to changes in $\sigma_8$. In principle, therefore, $\sigma_8$ should be very well constrained by the cluster population.

\subsubsection{What is a cluster's mass?}\label{sec:X-D-2}

The main problem here is how observable cluster properties should be related to quantities used in theory. Cluster \emph{masses}, as used in the theoretical mass function (\ref{eq:10-20}), are \emph{not} observables. Cluster observables are the X-ray temperature and flux, the optical luminosity and the velocity distribution of their galaxies, and their gravitational-lensing effects. Before we discuss their relation to mass, let us first see what the ``mass of a galaxy cluster'' could be.

It is easy to define masses of gravitationally bound, well localised objects, such as planets or stars. They have a well-defined boundary, e.g.~the planetary surfaces or the stellar photospheres. This is markedly different for objects like galaxies and galaxy clusters. As far as we know, their densities drop smoothly towards zero like power laws, $\propto r^{-(2\ldots3)}$. Thus, although they are gravitationally bound, it is less obvious what should be seen as their outer boundary. Strictly speaking, there is none.

The only way out is then to \emph{define} an outer boundary in such a way that it is well-defined in theory and identifiable in observational data. A common choice was introduced in \S~\ref{sec:V-A-2}: it defines the boundary by the mean overdensity it encloses. Although this is problematic as well, it may be as good as it gets. Three immediately obvious problems created by this definition are that objects like galaxy clusters are often irregularly shaped rather than spherical, that the overdensity of 200 is as arbitrary as any other, even if it is inspired by virial equilibrium in the spherical-collapse model, and that its measurement requires a sufficiently accurate density profile to be known or assumed.

How could standardised radii such as $R_{200}$ be measured? This could for instance be achieved applying equations such as (\ref{eq:05-37}) after measuring the slope $\beta$ and the core radius of the X-ray surface brightness profile together with the X-ray temperature, by calibrating an \emph{assumed} density profile with galaxy kinematics based on the virial theorem, or by constraining the cluster mass profile with gravitational lensing.

Obviously, all these measurements have their own problems. Being sensitive to all mass along the line-of-sight, gravitational lensing cannot distinguish between mass bound to a cluster or just projected onto it. Any measurement based on the virial theorem must of course rely on virial equilibrium, which takes time to be established and is often perturbed in real clusters because of merging and accretion. The common interpretation of X-ray measurements requires the assumption that the X-ray gas be in hydrostatic equilibrium with the host cluster's gravitational potential.

This illustrates that it may be fair to say that \emph{there is no such thing as the mass of a galaxy cluster}. Even if measurements of cluster ``radii'' were less dubious, it remained unclear whether they mean the same as those assumed in theory, which are related to the spherical-collapse model. Interestingly, but not surprisingly, cluster masses obtained from numerical simulations suffer from the same poor definition of the concept of a ``cluster radius''.

How can we make progress then? Observables such as the cluster temperature $T_X$ or its X-ray luminosity $L_X$ should be related to the depth of the gravitational potential they are embedded in, which should in turn be related to some measure of the total mass. If we can calibrate such expected temperature-mass or luminosity-mass relations, e.g.~using numerical simulations of galaxy clusters, a direct comparison between theory and observations seems possible. This is sometimes called an \emph{external calibration} of the required relations.

Internal or \textit{self-calibrations}, i.e.~calibrations based on cluster data alone, have become increasingly fashionable over the past years. Here, empirical temperature-mass and luminosity-mass relations are obtained based on one or more of the mass estimates sketched above.

\begin{table}[ht]
\begin{center}
\begin{tabular}{|p{0.2\hsize}|p{0.35\hsize}|p{0.35\hsize}|}
\hline
$\sigma_8$ & data & reference \\
\hline
$1.02\pm0.07$ & $M$-$T$ relation & \cite{PI01.1} \\
$0.77\pm0.07$ & $M$-$T$ relation & \cite{SE02.1} \\
$0.68^{+0.11}_{-0.09}$ & $M$-$L$ relation & \cite{RE02.1} \\
$0.75\pm0.16$ & lensing masses & \cite{SM03.2} \\
$0.79^{+0.06}_{-0.07}$ & luminosity function & \cite{PI03.1} \\
$0.77^{+0.05}_{-0.04}$ & temperature function & \cite{PI03.1} \\
$0.69\pm0.03$ & lensing masses & \cite{AL03.1} \\
$0.78\pm0.17$ & optical richness & \cite{EK06.1} \\
$0.67^{+0.04}_{-0.05}$ & lensing masses & \cite{DA06.1} \\
\hline
\end{tabular}
\end{center}
\caption{Values of $\sigma_8$ derived from the galaxy-cluster population based on different observational data.}
\label{tab:10-2}
\end{table}

The result of both procedures is qualitatively the same. It allows the conversion of observables to mass, and thus of the observed cluster temperature or luminosity functions to mass functions, which can then be compared to theory. The shape and amplitude of the power spectrum and the growth factor can then be adapted to optimise the agreement between observed and expected mass functions. Clusters at moderate or high redshift constrain the evolution of the mass function and allow an independent estimate of the matter-density parameter $\Omega_\mathrm{m0}$, as sketched in \S~\ref{sec:V-C} before.

In view of the many difficulties listed, it is an astonishing fact that, when applied to cluster samples rather than individual clusters, the determination of the cluster mass function and its evolution seems to work quite well. Values for $\sigma_8$ derived therefrom are given in Tab.~\ref{tab:10-2}. Systematic differences between these values are most likely due to the uncertainties of the calibration procedures applied to the relations between cluster masses and observables.

\subsection{The Lyman-$\alpha$ forest}\label{sec:X-E}

The Lyman-$\alpha$ forest lines mentioned before arise from absorption in neutral-hydrogen clouds. It is reasonable to assume that they are located where the dark matter is overdense, such that fluctuations in the neutral-hydrogen density are proportional to the dark-matter density contrast $\delta$. Then, observations of the Lyman-$\alpha$ forest lines, in particular their number per unit redshift, equivalent-width distribution and correlation properties, must contain information on the shape and the normalisation of the dark-matter power spectrum.

Retrieving this information would be straightforward if the biasing relation between the neutral-hydrogen and dark-matter densities was known and simple, the gas temperature was known, redshifts were unaffected by peculiar motions, and non-linear structure evolution could be neglected. While this is not the case in reality, these perturbing effects can be calibrated with numerical simulations and corrected. Two different methods have been proposed; one inverts the Lyman-$\alpha$ forest directly \citep{NU99.1}, while the other adapts power spectra of numerical simulations to those recovered from Lyman-$\alpha$ forest observations \citep{CR98.1}. Both methods have allowed interesting constraints which are summarised in Tab.~\ref{tab:10-3}.

\begin{table}[ht]
\begin{center}
\begin{tabular}{|p{0.2\hsize}|p{0.35\hsize}|p{0.35\hsize}|}
\hline
$\sigma_8$ & data & reference \\
\hline
$0.73\pm0.04$ & Keck spectra & \cite{CR02.1} \\
$0.94\pm0.08$ & LUQAS sample, Keck & \cite{VI04.1} \\
$0.91\pm0.07$ & SDSS spectra & \cite{VI06.1} \\
$0.92\pm0.09$ & LUQAS sample & \cite{ZA06.1} \\
\hline
\end{tabular}
\end{center}
\caption{Selection of $\sigma_8$ values obtained from Lyman-$\alpha$ forest data alone.}
\label{tab:10-3}
\end{table}

The method is young, but promising. Values of $\sigma_8$ obtained so far seem to be typically systematically higher than those derived from other observables. Possible sources of systematic error are the onset of nonlinear structure growth and, perhaps most importantly, the temperature of the hydrogen gas and its relation to the gas density (often called its equation of state).

\section{Inflation and Dark Energy}\label{sec:XI}

\subsection{Cosmological inflation}\label{sec:XI-A}

\subsubsection{Motivation}\label{subsec:XI-A-1}

In the preceding chapters, we have seen the remarkable success of the cosmological standard model, which is built upon the two symmetry assumptions underlying the class of Friedmann-Lema{\^\i}tre-Robertson-Walker models which experienced a Big Bang a finite time ago. We shall now discuss a fundamental problem of these models, and a possible way out. Historically, the problem was raised in a different way, but it intrudes with the very straightforward realisation that it is by no means obvious why the CMB should appear as isotropic as it is, and why there should be large coherent structures in it.

Let us begin with the so-called \emph{comoving particle horizon}, which is the comoving distance that light can travel between the Big Bang and time $t$. It is given by (\ref{eq:06-21}) with the sound speed $c_\mathrm{s}$ replaced by the light speed $c$,
\begin{equation}
  w_\mathrm{H}=\int_0^{t_\mathrm{rec}}\frac{c\d t}{a}=
  \frac{2c\sqrt{a_\mathrm{rec}}}{H_0\sqrt{\Omega_\mathrm{m0}}}\left(
    \sqrt{1+\alpha}-\sqrt{\alpha}
  \right)=282.8\,\mathrm{Mpc}\;,
\end{equation} 
with $\alpha\approx0.33$ as defined below (\ref{eq:06-10}). On the other hand, we have seen in (\ref{eq:06-22}) that the comoving angular-diameter distance to the CMB is
\begin{equation}
  w(a_\mathrm{rec})=3.195\frac{c}{H_0}
  \left(\frac{\Omega_\mathrm{m0}}{0.3}\right)^{-0.4}\,\mathrm{Gpc}\;,
\label{eq:11-4}
\end{equation}
which implies that the \emph{angular} size of the particle horizon is
\begin{equation}
  \theta_\mathrm{rec}=\frac{w_\mathrm{H}}{w(a_\mathrm{rec})}\approx
  1.14^\circ\,\left(\frac{\Omega_\mathrm{m0}}{0.3}\right)^{-0.1}\;.
\label{eq:11-5}
\end{equation}

The physical meaning of the particle horizon is that no event between the Big Bang and recombination can exert any influence on a given particle if it is more than the horizon length away. Our simple calculation thus shows that we can understand how causal processes could establish identical physical conditions in patches of the sky with about one degree radius. Points on the CMB separated by larger angles were never causally connected before the CMB was released. It is therefore not at all plausible how the CMB could have attained almost the same temperature across the entire sky! The simple fact that the CMB is almost entirely isotropic across the sky thus poses a problem which the standard cosmological model is apparently unable to solve. Moreover, the formation of coherent structures larger than the particle horizon remains mysterious. This is one way to state the \emph{horizon problem}. It is sometimes called the \emph{causality problem}: How can coherent structures in the CMB be larger than the particle horizon was at recombination?

Another uncomfortable problem of the standard cosmological model is the flatness, or at least the near-flatness, of spatial hypersurfaces of our Universe. To see this, we write Friedmann's equation (\ref{eq:01-6}) in the form
\begin{equation}
  H^2(a)=H^2(a)\left[\Omega_\mathrm{total}(a)-\frac{Kc^2}{a^2H^2}\right]\;,
\label{eq:11-6}
\end{equation}
which is equivalent to
\begin{equation}
  \Omega_\mathrm{total}(a)-1=\frac{Kc^2}{a^2H^2}\;.
\label{eq:11-7}
\end{equation}
The right-hand side typically grows as some power of the time $t$,
\begin{equation}
  (a^2H^2)^{-1}=\dot a^{-2}\propto t^{\,\beta}\;;
\label{eq:11-8}
\end{equation}
e.g.~$\beta=2/3$ in a matter-dominated universe without cosmological constant. Then,
\begin{equation}
  \Omega_\mathrm{total}(a)-1\propto t^{\,\beta}\;.
\label{eq:11-10}
\end{equation}

This shows that any deviation of the total density parameter $\Omega_\mathrm{total}$ from unity tends to grow with time. Thus, (spatial) flatness is an unstable property. If it is not very precisely flat in the beginning, the Universe will develop away from flatness. Since we know that spatial hypersurfaces are now almost flat, $|\Omega_\mathrm{total}(a)-1|\lesssim1\%$ (cf.~Tab.~\ref{tab:1}), the deviation from flatness must have been of order
\begin{equation}
  |\Omega_\mathrm{total}(a_\mathrm{rec})-1|\lesssim1\%\times\left(
    \frac{t_\mathrm{rec}}{t_0}
  \right)^{\,\beta}\approx10^{-5}\;,
\label{eq:11-11}
\end{equation}
or ten parts per million at the time of recombination. Clearly, this requires enormous fine-tuning. This is called the \emph{flatness problem}: How can we understand flatness in the late universe without assuming an extreme degree of fine-tuning at early times?

\subsubsection{The idea of inflation}\label{subsec:XI-A-2}

Since $c/H$ is the Hubble radius, the quantity $r_\mathrm{H}\equiv c/(aH)$ is the \emph{comoving} Hubble radius. According to (\ref{eq:11-8}), it typically grows with time as $r_\mathrm{H}\propto t^{\,\beta/2}$. Since we can write (\ref{eq:11-7}) as
\begin{equation}
  \Omega_\mathrm{total}(a)-1=Kr_\mathrm{H}^2\;,
\label{eq:11-12}
\end{equation}
this is equivalent to the flatness problem.

This motivates the idea that at least the flatness problem may be solved if the comoving Hubble radius could, at least for some sufficiently long period, \emph{shrink} with time. If that could be arranged, any deviation of $\Omega_\mathrm{total}(a)$ from unity would be driven towards zero.

Conveniently, such an arrangement would also remove or at least alleviate the causality problem. Possibly, the Hubble radius, characterising the radius of the observable universe, could be driven inside the horizon and thus move the entire observable universe into a causally-connected region. When the hypothesised epoch of a shrinking comoving Hubble radius is over, it starts expanding again, but if the reduction was sufficiently large, it could remain within the causally-connected region at least until the present.

How could such a \emph{shrinking} comoving Hubble radius be arranged? Obviously, we require
\begin{equation}
  \frac{\d}{\d t}\frac{c}{aH}=-\frac{c}{(aH)^2}(\dot aH+a\dot H)=
  -\frac{c\ddot a}{(aH)^2}<0\;,
\label{eq:11-13}
\end{equation}
which is possible if and only if $\ddot a>0$, in other words, if the expansion of the Universe \emph{accelerates}. This appears counter-intuitive because the cosmic expansion is dominated by gravity, which should be attractive and thus necessarily \emph{decelerate} the expansion. Friedmann's equation (\ref{eq:01-7}) implies the matter condition
\begin{equation}
  \rho c^2+3p<0\quad\Rightarrow\quad
  p<-\frac{\rho c^2}{3}\;.
\label{eq:11-14}
\end{equation}
In other words, cosmic acceleration is possible if and only if the dominant ingredient of the cosmic fluid has sufficiently \emph{negative pressure}. Equation~(\ref{eq:01-8}) implies the density evolution
\begin{equation}
  \frac{\dot\rho}{\rho}=-3\frac{\dot a}{a}\left(1+\frac{p}{\rho c^2}\right)<-2\frac{\dot a}{a}
\label{eq:11-17}
\end{equation}
for a cosmic fluid satisfying (\ref{eq:11-14}), showing that its density would fall off flatter than $a^{-2}$, and thus less steeply than the matter or radiation densities. Thus, once a component of the cosmic fluid with sufficiently negative pressure reaches a density comparable to the densities of matter or radiation, it quickly starts dominating the cosmic expansion. In the limiting case $\dot\rho=0$ or $p=-\rho c^2$, (\ref{eq:01-7}) reduces to
\begin{equation}
  \frac{\ddot a}{a}=\frac{8\pi G}{3}\rho=\mathrm{const.}>0\;,
\label{eq:11-18}
\end{equation}
which implies the exponential expansion or inflation
\begin{equation}
  a\propto\exp\left[\left(\frac{8\pi G}{3}\rho\right)^{1/2}t\right]\;.
\label{eq:11-19}
\end{equation} 

\subsubsection{Slow roll, structure formation, and observational constraints}\label{subsec:XI-A-3}

We have seen that we need inflation to solve the flatness and causality problems, and inflation needs a form of matter with \emph{negative pressure}. What could that be? Consider a scalar field $\phi$ with a self-interaction potential $V(\phi)$. Then, field theory shows that pressure and density of the scalar field are related by the \emph{equation of state}
\begin{equation}
  p_\phi=w\rho_\phi c^2\quad\mbox{with}\quad
  w\equiv\frac{\frac{1}{2}\dot\phi^2-V}{\frac{1}{2}\dot\phi^2+V}\;.
\label{eq:11-20}
\end{equation}
Evidently, negative pressure is possible if the kinetic energy of the scalar field is sufficiently smaller than its potential energy. For the cosmological-constant case, $\dot\phi=0$, we have $w=-1$ or $p=-\rho c^2$, in agreement with the conclusion from (\ref{eq:11-17}). In other words, a suitably strongly self-interacting scalar field has exactly the properties we need.

Inflation, i.e.~accelerated expansion, broadly requires $\dot\phi^2$ to be sufficiently smaller than $V$. Moreover, we need inflation to operate long enough to drive the total matter density parameter sufficiently close to unity for it to remain there to the present day. These two conditions are conventionally cast into the form
\begin{equation}
  \epsilon\equiv\frac{1}{24\pi G}\left(
    \frac{V'}{V}
  \right)^2\ll1\quad\mbox{and}\quad
  \eta\equiv\frac{1}{8\pi G}\frac{V''}{V}\ll1
\label{eq:11-21}
\end{equation}
(see the standard text-book on inflation, \cite{LI00.1}). They are called the \emph{slow-roll conditions}. The first assures that inflation can set in, because if it is satisfied, the potential has a small gradient and cannot drive rapid changes of the scalar field. The second restricts the curvature of the potential and thus assures that the inflationary condition is satisfied long enough.

Estimates show that inflation needs to expand the Universe by $\sim50\ldots60$ e-foldings \citep{AL06.1} (i.e.~by a factor of $\e^{50\ldots60}$) for it to solve the causality and flatness problems. Inflation ends once the slow-roll conditions are violated. By then, the Universe will have become extremely cold. While the density of the inflaton field $\phi$ will be approximately the same as at the onset of inflation (as for the cosmological constant, this is a consequence of the negative pressure), all other matter and radiation fields will have their energy densities lowered by factors of $a^{-3\ldots4}$, i.e.~by $\lesssim100$ orders of magnitude.

Once $\epsilon$ approaches unity, the kinetic term $\dot\phi^2$ will dominate the potential, and the scalar field will start oscillating rapidly. It is assumed that the scalar field then decays into ordinary matter which fills or \emph{reheats} the Universe after inflation is over.

It is an extremely interesting aspect of inflation that it also provides a mechanism for seeding structure formation \citep{MU81.1}. As any other quantum field, the \emph{inflaton field} $\phi$ must have undergone vacuum oscillations because the zero-point energy of a quantum harmonic oscillator cannot vanish due to Heisenberg's uncertainty principle. These vacuum oscillations cause the spontaneous creation and annihilation of particle-antiparticle pairs. Once inflation sets in, vacuum fluctuation modes are quickly driven out of the horizon and lose causal connection. Then, they cannot decay any more and ``freeze in''. Thus, inflation introduces the breath-taking notion that density fluctuations in our Universe today may have been seeded by vacuum fluctuations of the inflaton field before inflation set in and enlarged them to cosmological scales.

This idea has precisely quantifiable consequences \citep{GU82.1,PA83.1,BR84.1,HA85.1}. First, linear density fluctuations created by the inflaton field must be adiabatic and, by the central limit theorem, they should form a Gaussian random field. This is because they arise from incoherent superposition of extremely many independent fluctuation modes whose amplitude and wave number are all drawn from the same probability distribution. Under these circumstances, the central limit theorem shows that the result, i.e.~the superposition of all these modes, must be a Gaussian random field.

Second, it implies that the statistics of density fluctuations in the Universe today must be explicable by the statistics of vacuum fluctuations in a scalar quantum field. This is indeed the case. The power spectrum resulting from this consideration is very close to the scale-free Harrison-Zel'dovich-Peebles shape introduced in \S~\ref{sec:I-B-2},
\begin{equation}
  P_\delta(k)\propto k^{n_\mathrm{s}}\;,
\label{eq:11-22}
\end{equation} 
with $n_\mathrm{s}\approx1$. The spectral index $n_\mathrm{s}$ would be precisely unity if inflation lasted forever. Since this was obviously not so, $n_\mathrm{s}$ must deviate slightly from unity, and detailed calculations show that it must be slightly smaller \citep{LI00.1,AL06.1},
\begin{equation}
  n_\mathrm{s}=1+2\eta-6\epsilon\;.
\end{equation}
The latest WMAP measurements \citep{KO08.1} do in fact show that
\begin{equation}
  n_\mathrm{s}=0.960^{+0.014}_{-0.013}\;;
\label{eq:11-23}
\end{equation}
cf.~Tab.~\ref{tab:1}. The completely scale-invariant spectrum, $n_\mathrm{s}=1$, is thus excluded at more than $3\sigma$. The measured deviation of $n_\mathrm{s}$ from unity also restricts the number $N$ of e-foldings completed by inflation. Under fairly general assumptions,
\begin{equation}
  N=54\pm7
\label{eq:11-24}
\end{equation}
\citep{AL06.1} based on the three-year WMAP data.

Another prediction of inflation is that it may excite not only scalar, but also tensor perturbations \citep{LI00.1}. Scalar perturbations lead to the density fluctuations, tensor perturbations correspond to gravitational waves. Vector perturbations do not play any role because they decay quickly as the universe expands. Simple models of inflation predict that the ratio $r$ between the amplitudes of tensor and scalar perturbations, taken in the limit of small wave numbers, is
\begin{equation}
  r=16\epsilon\;.
\label{eq:11-25}
\end{equation}
An inflationary background of gravitational waves is in principle detectable through the polarisation of the CMB \citep{SE97.1}. Limits of order $r\lesssim0.05$ are expected from the upcoming \emph{Planck} satellite \citep{KI98.1,KA01.1}. Together with the result $n_\mathrm{s}\ne1$ from WMAP, it will then be possible to constrain viable inflation models, i.e.~to constrain the shape of the inflaton potential.

\subsection{Dark energy}\label{sec:XI-B}

\subsubsection{Motivation}\label{subsec:XI-B-1}

The CMB shows us that the Universe is at least nearly spatially flat. Constraints from kinematics, from cluster evolution and from the CMB show that the matter density alone cannot be responsible for flattening space, and primordial nucleosynthesis and the CMB show that baryons contribute at a very low level only. Something is missing, and it even dominates today's cosmic fluid.

From structure formation, we know that this remaining constituent cannot clump on the scales covered by the galaxy surveys and below. It is thus different from dark matter. The terms \emph{dark energy}, \textit{kosmon} or \textit{quintessence} have been coined for it \citep{PE88.1,WE88.1,PE03.2}. The type-Ia supernovae tell us that it behaves at least very similar to a cosmological constant in the recent cosmic past. Maybe the dark energy \emph{is} a cosmological constant? Nothing currently indicates any deviation from this ``simplest'' assumption (e.g.~\citealt{KO08.1}). So far, the cosmological constant is a perfectly viable description for all observational evidence we have. However, this is deeply dissatisfactory from the point of view of theoretical physics. The problem is the value of $\Omega_{\Lambda0}$. As we have seen above, a self-interacting scalar field with negligible kinetic energy behaves like a cosmological constant. Then, its density should simply be given by its potential $V$. Simple arguments suggest that $V$ should be the Planck mass divided by the third power of the Planck length, which turns out to be 120 \emph{orders of magnitude} larger than the cosmological constant derived from observations. Since this fails, it seems natural to expect that the cosmological constant should vanish, but it does not. The main problem with the cosmological constant is therefore, why is it not zero if it is so small?

Anthropic arguments suggest that we observe the Universe as it is because it needs to be that way for us to exist. The Universe needs to be old enough for stars to have produced carbon, on which life must be based to the best of our knowledge. Yet, it must not be too old for the majority of stars to have evolved past their main-sequence life for the stability of benign conditions for life. If the cosmological constant was much larger, inflationary expansion would have prohibited structure formation and thus also the formation of life. It may thus be that the cosmological parameters have the values we measure because otherwise we would not exist to measure them.

The explanation of inflation by means of an inflaton field suggests another way out. As we have seen there, accelerated expansion can be driven by a self-interacting scalar field while its potential energy dominates. Moreover, it can be shown that if the potential $V$ has an appropriate shape, the dark energy has attractor properties in the sense that a vast range of initial density values can evolve towards the same value today \citep{ZL99.1}. At this point at the latest, we leave the realm of what may be considered the established cosmological standard model, and enter into the very controversial discussion of dark energy.

\subsubsection{Observational constraints?}\label{subsec:XI-B-2}

If the dark energy is indeed dynamical and provided by a self-interacting scalar field, how can we find out more about it? Reviewing the cosmological measurements we have discussed so far, it becomes evident that they are all derived from constraints on

\begin{itemize}

\item cosmic time, as in the age of the Galaxy or of globular clusters, or in primordial nucleosynthesis;

\item distances, as in the spatial flatness derived from the CMB, the type-Ia supernovae or the geometry of cosmological weak lensing; or

\item the growth of cosmic structures, as in the acoustic oscillations in the CMB, the evolution of the cluster population, the structures in the galaxy distribution or the source of cosmological weak-lensing effects.

\end{itemize}

We must therefore seek to constrain the dark energy by measurements of distances, times, and structure growth. Since they can all be traced back to the expansion behaviour of the universe as described by Friedmann's equation, we must see how the dark energy enters there, and what effects it can seed through it.

Let us therefore assume that the dark energy is a suitably self-interacting, homogeneous scalar field. Then, its pressure can be described by
\begin{equation}
  p=w(a)\rho c^2\;,
\label{eq:11-26}
\end{equation} 
where the equation-of-state parameter $w$ is some function of $a$. According to (\ref{eq:11-14}), accelerated expansion needs $w<-1/3$, and the cosmological constant corresponds to $w=-1$. Since all cosmological measurements to date are in agreement with the assumption of a cosmological constant, we need to arrange things such that $w\to-1$ today. Suppose we had some function $w(a)$, which could either be obtained from a phenomenological choice, a model for the self-interaction potential $V(\phi)$ through (\ref{eq:11-20}) or from a suitable \emph{ad-hoc} parameterisation. Then, (\ref{eq:11-17}) implies
\begin{equation}
  \frac{\dot\rho}{\rho}=-3(1+w)\frac{\dot a}{a}\;,
\label{eq:11-27}
\end{equation}
or
\begin{equation}
  \rho(a)=\rho_0\exp\left\{
    -3\int_1^a[1+w(a')]\frac{\d a'}{a'}
  \right\}\equiv\rho_0f(a)\;.
\label{eq:11-28}
\end{equation}
For constant $w$, this simplifies to
\begin{equation}
  \rho(a)=\rho_0\exp\left[-3(1+w)\ln a\right]=\rho_0a^{-3(1+w)}\;.
\label{eq:11-29}
\end{equation}
If $w=-1$, we recover the cosmological-constant case $\rho=\rho_0=\mbox{const.}$, for pressure-less material, $w=0$ and $\rho\propto a^{-3}$, and for radiation, $w=1/3$ and $\rho\propto a^{-4}$.

Therefore, we can take account of the dynamical dark energy by replacing the term $\Omega_{\Lambda0}$ in the Friedmann equation (\ref{eq:01-10}) by $\Omega_\mathrm{DE0}f(a)$, and the expansion function $E(a)$ turns into
\begin{equation}
  E(a)=\left[
    \Omega_\mathrm{r0}a^{-4}+\Omega_\mathrm{m0}a^{-3}+
    \Omega_\mathrm{DE0}f(a)+\Omega_\mathrm{K0}a^{-2}
  \right]^{1/2}\;,
\label{eq:11-30}
\end{equation}
where $\Omega_\mathrm{K0}=1-\Omega_\mathrm{r0}-\Omega_\mathrm{m0}- \Omega_\mathrm{DE0}$ is a density parameter assigned to the spatial curvature.

We thus see that the equation-of-state parameter enters the expansion function in integrated form. Since all cosmological observables are integrals over the expansion function, including the growth factor $D_+(a)$, this implies that cosmological observables measure integrals over the integrated equation-of-state function $w(a)$. Note that in cosmological models with dynamical dark energy, the growth factor is more complicated than described by Eq.~(\ref{eq:01-23}) because dark-energy clustering needs to be taken into account \cite[e.g.][]{KU07.1}. Needless to say, the dependence of cosmological measurements on the exact form of $w(a)$ will be extremely weak, which in turn implies that extremely accurate measurements will be necessary for constraining the nature of the dark energy.

\begin{figure}[ht]
  \includegraphics[width=\hsize]{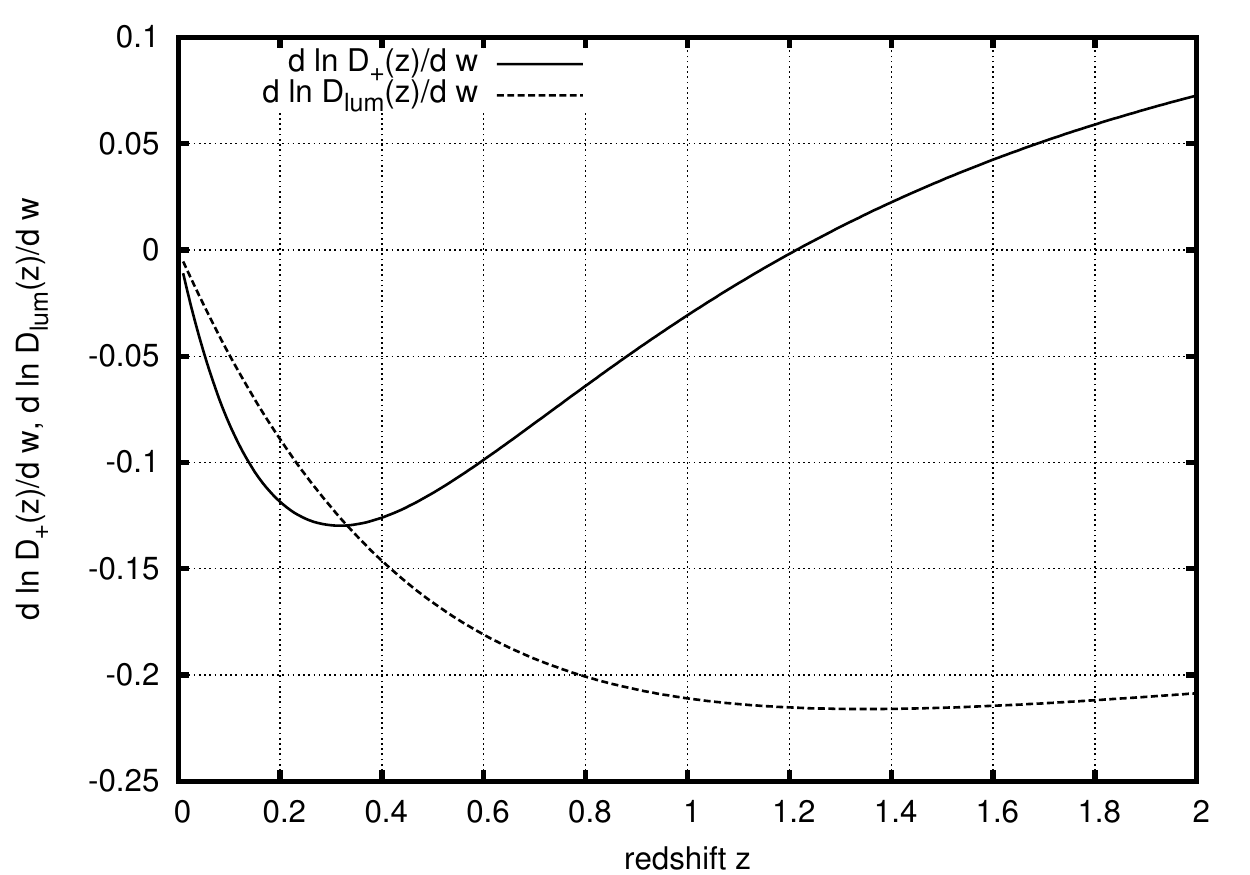}
\caption{Relative changes of the growth factor $D_+$ and the luminosity distance $D_\mathrm{lum}$ with $w$ as functions of redshift, assuming constant $w$.}
\end{figure}

In order to illustrate the required accuracies, let us consider by how much the angular-diameter distance and the growth factor change compared to $\Lambda$CDM upon changes in $w$ away from $-1$,
\begin{equation}
  \frac{\d\ln D_\mathrm{ang}(z)}{\d w}\;,\quad
  \frac{\d\ln D_+(z)}{\d w}\;,
\label{eq:11-31}
\end{equation} 
as a function of redshift $z$. Assuming $\Omega_\mathrm{m0}=0.3$ and $\Omega_{\Lambda0}=0.7$, we find typical values between $-0.1$ and $-0.2$ at most. Since we currently expect deviations of $w$ from $-1$ at most at the $\sim10\%$ level, accurate constraints on the dark energy require relative accuracies of distances and the growth factor at the per-cent level.

Clearly, all models for dynamical dark energy proposed so far are purely phenomenological. The most straightforward question to be addressed in this situation may be whether a cosmological constant can be ruled out in some way. All suitable cosmological information will need to be combined in order to progress.

Currently, the largest hope is put on the BAOs (see \S~\ref{sec:VII-A-6}) and on so-called \emph{tomographic} measurements. BAOs have a characteristic physical scale whose angular scale depends on the angular-diameter distance, which is in turn sensitive to the dark energy. Tomography attempts to trace the evolution of structures throughout cosmic history \citep{HU02.1,HU02.2}. An example is given by weak gravitational lensing: Since its geometrical sensitivity peaks approximately half-way between the sources and the observer, sources at higher redshift also probe more distant, and thus less evolved, cosmic structures. If lensing effects can be measured for sub-samples of sources in different redshift shells, the growth factor can be probed differentially. First examples for this technique have been published. They give rise to the expectation that clarifying the nature of the dark energy may indeed be feasible in the near future.

\subsection*{Acknowledgements}

I am most grateful to many colleagues for inspiring and clarifying discussions, in particular to Peter Schneider, Achim Wei{\ss} and Simon White, whose detailed comments helped improving this review substantially. Three anonymous referees contributed numerous thoughtful and constructive comments. This work was supported in part by the German Science Foundation (DFG) through the Collaborative Research Centres SFB 439 and TRR 33.

\bibliographystyle{apsrmp}
\bibliography{./main.bib}

\end{document}